\documentclass[twocolappendix]{emulateapj}
\usepackage{graphicx}
\usepackage{amssymb}
\usepackage{amsmath}
\usepackage{natbib}
\usepackage{color}
\usepackage[flushleft]{threeparttable}

\newcommand\be{\begin{equation}}
\newcommand\ee{\end{equation}}

\newcommand\ba{\begin{eqnarray}}
\newcommand\ea{\end{eqnarray}}
\begin{document}

\title{Gaps and Rings in Protoplanetary Disks with Realistic Thermodynamics: The Critical Role of In-Plane Radiation Transport}

\author{Ryan Miranda\altaffilmark{1,3} and Roman R. Rafikov\altaffilmark{1,2}}

\altaffiltext{1}{Institute for Advanced Study, Einstein Drive, Princeton, NJ 08540}
\altaffiltext{2}{Centre for Mathematical Sciences, Department of Applied Mathematics and Theoretical Physics, University of Cambridge, Wilberforce Road, Cambridge CB3 0WA, UK}
\altaffiltext{3}{miranda@ias.edu}
\begin{abstract}
Many protoplanetary disks exhibit annular gaps in dust emission, which may be produced by planets. Simulations of planet-disk interaction aimed at interpreting these observations often treat the disk thermodynamics in an overly simplified manner, which does not properly capture the dynamics of planet-driven density waves driving gap formation. Here we explore substructure formation in disks using analytical calculations and hydrodynamical simulations that include a physically-motivated prescription for radiative effects associated with the planet-induced density waves. For the first time, our treatment accounts not only for cooling from the disk surface, but also for radiation transport along the disk midplane. We show that this in-plane cooling, with a characteristic timescale typically an order of magnitude shorter than the one due to surface cooling, plays a critical role in density wave propagation and dissipation (we provide a simple estimate of this timescale). We also show that viscosity, at the levels expected in protoplanetary disks ($\alpha \lesssim 10^{-3}$), has a negligible effect on density wave dynamics. Using synthetic maps of dust continuum emission, we find that the multiplicity and shape of the gaps produced by planets are sensitive to the physical parameters---disk temperature, mass, and opacity---that determine the damping of density waves. Planets orbiting at $\lesssim 20$ au produce the most diverse variety of gap/ring structures, although significant variation is also found for planets at $\gtrsim 50$ au. By improving the treatment of physics governing planet-disk coupling, our results present new ways of probing the planetary interpretation of annular substructures in disks.
\end{abstract}


\section{Introduction}
\label{sect:introduction}

Recent observations of protoplanetary disks in dust continuum emission using ALMA have revealed a variety of disk substructures, among them numerous axisymmetric gaps and rings \citep{Huang2018}. A number of mechanisms have been invoked to explain the formation of these substructures. However, planet-disk interaction appears to provide the most promising quantitative description of these features (e.g., \citealt{Zhang2018}). Recent indirect detections of planets embedded in such disks using gas kinematics \citep{Pinte2019,Pinte2020} appear to lend support to the planet hypothesis.

In the planet-disk interaction scenario, one or more planets within the disk excite density waves in the surrounding gas via their gravity \citep{GT79}. Dissipation of the waves leads to the transfer of their angular momentum to the bulk disk material, resulting in the formation of axisymmetric structures in the disk surface density. The resulting pressure maxima are capable of trapping dust, leading to the appearance of rings and gaps in dust continuum emission.

In the classical picture of gap opening by massive planets (e.g., \citealt{LinPap1986}), damping of the planet-driven waves produces a wide gap around the planetary orbit. For lower mass planets, at least initially and provided that the disk viscosity is low enough, a planet is expected to produce a pair of narrow gaps situated on either side of its orbit  \citep{R02b,Duffell,Zhu2013}, as a result of nonlinear wave dissipation via shock formation \citep{GR01,R02}. Recent numerical simulations have shown that the formation of several additional gaps interior to the orbit of the planet is also possible \citep{Bae2017,DongGaps2017,DongGaps2018}. Such gaps are associated with the splitting of a planet-driven density wave into multiple spiral arms in the inner disk (which is a purely linear process, see \citealt{AR18} and \citealt{Miranda-Spirals}), and subsequent shocking and dissipation of each arm due to the nonlinear effects \citep{Bae2017,BZ18a}. Therefore, the multiple rings and gaps seen in many disks may require only a single planet to produce. For example, \citet{Zhang2018} demonstrated that the positions of the five gaps in the AS 209 disk can potentially be accounted for by a single planet.

Until recently, numerical efforts to understand the formation of gaps and rings by planets have largely relied upon a rather simple, locally isothermal treatment of disk thermodynamics, in which the disk temperature is assumed to be a fixed function of the distance from the central star. Recently, we described the shortcomings of this approximation and argued in favor of the more realistic treatment of thermal physics in numerical simulations \citep{Miranda-ALMA}.

The passage of a density wave generates inhomogeneous compressional heating and cooling of the disk fluid, taking it locally out of equilibrium with the heating (e.g., stellar irradiation) and cooling (e.g., radiative losses from the disk surface) processes operating in an unperturbed disk. Radiative transport associated with these temperature perturbations acts to re-establish thermodynamic equilibrium, thus reducing the restoring force due to pressure and resulting in wave damping. In \citet{Miranda-Cooling} we presented the first study examining the role of such cooling processes in planet-disk interaction using linear perturbation theory, demonstrating their strong impact on the wave propagation and wave-driven disk evolution. Most importantly, when cooling is neither very slow nor very rapid, it leads to linear damping of density waves, which is strongest when $\beta$, the ratio cooling timescale to the local orbital timescale, is comparable to the disk aspect ratio $h$ (see Section \ref{sect:expect} for more details). In this case the multiple gap structure expected to be produced by a planet in a low-viscosity disk is suppressed, in favor of a single wide gap. Another important result of this work was that the cooling timescale must be very short, typically $10^2-10^3$ times smaller than the orbital timescale, for the conventional locally isothermal approximation to provide an accurate description of density wave dynamics.

These findings have been recently corroborated by several numerical studies. In particular, \citet{Zhang2020} and \citet{Ziampras2020b} have also found the damping of density waves and suppression of multiple gap structure for intermediate cooling timescales, whereas \citet{Facchini2020} noted the importance of cooling for predicting the number and positions of rings in their modeling of the LkCa 15 disk.

In \citet{Miranda-Cooling}, we used a cooling prescription in which the dimensionless cooling timescale $\beta$ is constant across the disk; \citet{Zhang2020} and \citet{Facchini2020} took the same approach. While appropriate for a theoretical study of the effects of cooling, this is not a good approximation for real disks. In this paper, we go beyond the constant-$\beta$ assumption, and provide a realistic, radially-dependent estimate of cooling due to radiative losses from the surface of the disk, as has also been done in \citet{Ziampras2020b}. 

\begin{figure}
\begin{center}
\includegraphics[width=0.49\textwidth,clip]{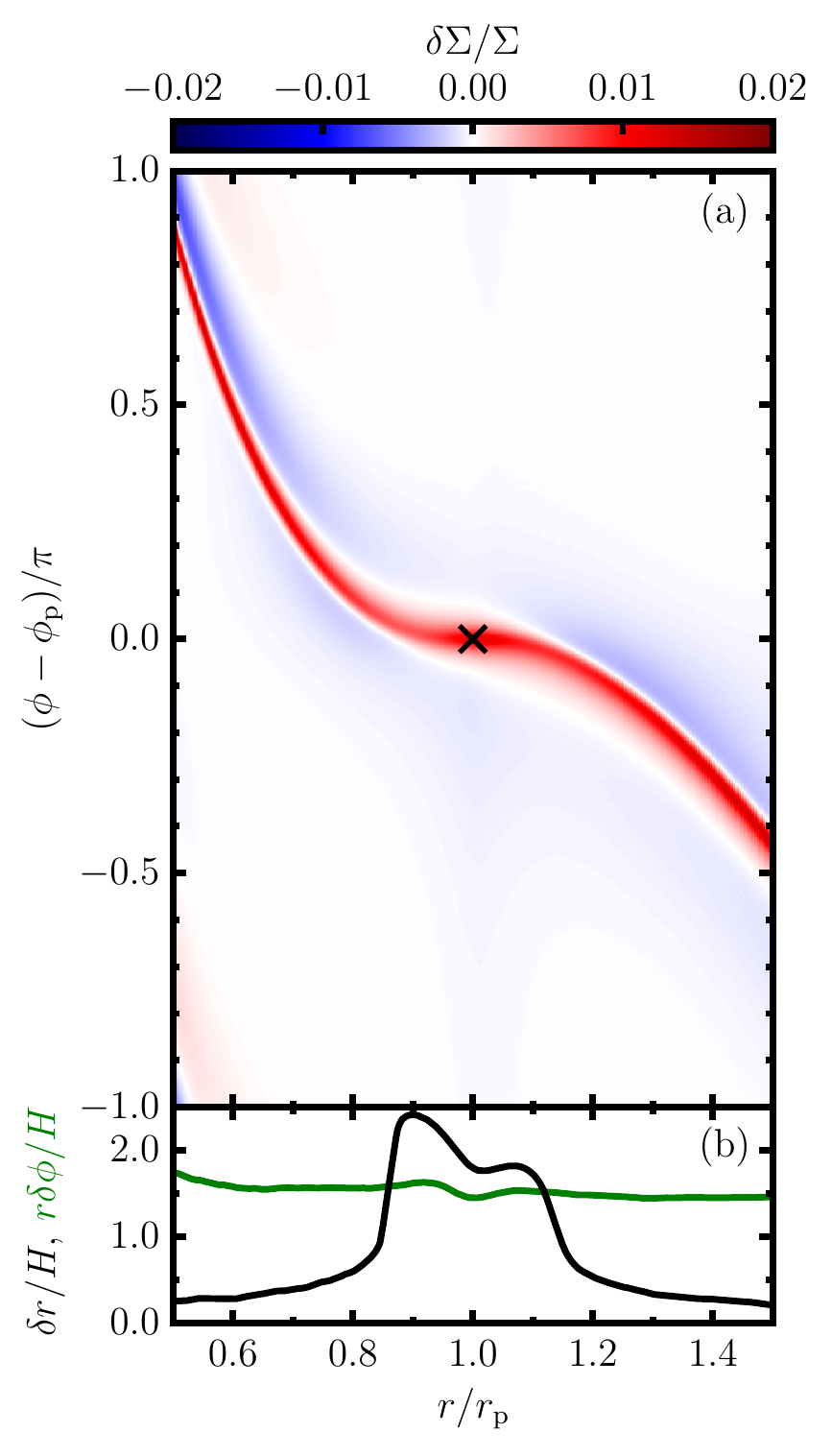}
\caption{(a) Fractional surface density perturbation $\delta\Sigma/\Sigma$ in polar coordinates $r$ and $\phi$ (relative to the azimuthal position of the planet $\phi_\mathrm{p}$) for a planet with mass $M_\mathrm{p} = 0.01 M_\mathrm{th}$ and orbital radius $r_\mathrm{p} = 50$ au, using the Fiducial disk model (see Section \ref{sect:basic-setup}). Note that this figure is stretched along the radial direction (by a factor of a few); the real spiral arm is more tightly wrapped than it appears in this figure. The marker indicates the position of the planet. Note the emergence of the secondary spiral arm \citep{Miranda-Spirals} for $r/r_\mathrm{p} \lesssim 0.7$. (b) The radial width $\delta r$ and azimuthal width $r\delta\phi$ of the primary spiral arm, defined using the full width at half maximum of $\delta\Sigma/\Sigma$, as a fraction of the local scale height $H$. Note that the radial scale of the density wave $\delta r\ll H$ for $|r-r_\mathrm{p}|\gtrsim H$.}
\label{fig:spiral_width}
\end{center}
\end{figure}

However, surface cooling, which relies on vertical energy transport towards the disk surface, is only one way in which radiative effects can affect planet-driven density waves. Localized disk regions experiencing transient compressional heating/cooling by the wave can also radiatively exchange energy in the horizontal direction, i.e., parallel to the disk midplane. Far from the planet, density waves evolve into tightly-wrapped, sharp structures with large radial temperature gradients, as shown in Fig.~\ref{fig:spiral_width}, where a perturbation due to a typical density wave is displayed together with its characteristic radial and azimuthal scales. One can see that outside the wave launching zone close to the planet, the radial width of the wave is much smaller than the disk scale height $H$---the characteristic scale of radiative cooling perpendicular to the disk midplane. For that reason, \citet{GR01} suggested that cooling of the spiral arms along the midplane may be much more efficient than cooling from the disk surface. The role of such {\it in-plane} cooling, as we refer to it in this work, has not been explored in detail previously in the context of planet-disk interaction. 

In our present study we fully account for the effects of both surface and in-plane cooling on the propagation of density waves and wave-driven disk evolution. Including in-plane cooling is a rather non-trivial exercise, as it is sensitive to the spatial structure of the perturbation, i.e., its wavelength. Here we develop a realistic in-plane cooling prescription that accounts for such details and use it both for semi-analytical calculations and in 2D hydrodynamical simulations (we also propose an approximate analytical expression for the cooling timescale, which can be used for simple estimates, see Section~\ref{sect:beta-mstar}). Our calculations demonstrate that in-plane cooling plays the dominant role in thermodynamics of the density waves and thus cannot be ignored. 

Using this approach, we explore the sensitivity of the gap and ring structures produced by a planet to various parameters of the underlying disk---including temperature, surface density, and opacity---which affect the efficiency of cooling. We do this using the results of hydrodynamical simulations, from which we produce simulated dust continuum emission maps.

The plan for this paper is as follows. We first present our basic setup in Section~\ref{sect:basic-setup}. In Section~\ref{sect:cooling}, we derive a prescription for the cooling timescale appropriate for planet-driven density waves. We describe the setup and details of our numerical simulations in Section~\ref{sect:setup}. We present our main results, the gas surface density profiles and dust emission maps for disks with realistic cooling for a variety of disk models, in Section~\ref{sect:results}. In particular, in Section~\ref{sect:nu-var} we explore the role of viscosity in density wave propagation and disk evolution. In Section~\ref{sect:results-iso}, we compare the results of our simulations to the results of locally isothermal simulations. We discuss and contextualize our results in Section~\ref{sect:disc}, and summarize and conclude in Section~\ref{sect:summary}.

\begin{table*}
\caption{Summary of disk models and their parameters}
\begin{center}
\begin{tabular}{ccccccccccc}
\hline
\hline
Disk Model & $h_{50\mathrm{au}}$ & $q$ & $\Sigma_{50\mathrm{au}}~(\mathrm{g}~\mathrm{cm}^{-2})$ & $p$ & $\bar{\kappa}_{{\rm d},0}$ & $\alpha$ & $M_{\rm p}/M_{\rm th}$ & $r_\mathrm{p}$~(au) & Resolution & In-plane Cooling \\ 
\hline
\hline
Fiducial & $0.1$ & $1/2$ & $10$ & $1$ & $1$ & $0$ & $0.1,0.3,1$ & $10,20,50,100$ & $1276 \times 2048$ & Yes \\
Low Mass & $0.1$ & $1/2$ & $3$ & $1$ & $1$ & $0$ & $0.3$ & $10,20,50,100$ & $1276 \times 2048$ & Yes \\
High Mass & $0.1$ & $1/2$ & $30$ & $1$ & $1$ & $0$ & $0.3$ & $10,20,50,100$ & $1276 \times 2048$ & Yes \\
$\Sigma \propto r^{-1/2}$ & $0.1$ & $1/2$ & $10$ & $1/2$ & $1$   & $0$ & $0.3$ & $10,20,50,100$ & $1276 \times 2048$ & Yes \\
Cold & $0.07$ & $1/2$ & $10$ & $1$ & $1$ & $0$ & $0.3$ & $10,20,50,100$ & $1914 \times 3072$ & Yes \\
Hot & $0.15$ & $1/2$ & $10$ & $1$ & $1$ & $0$ & $0.3$ & $10,20,50,100$ & $958 \times 1536$ & Yes \\
$T \propto r^{-1/4}$ & $0.1$ & $1/4$ & $10$ & $1$ & $1$  & $0$ & $0.3$ & $10,20,50,100$ & $1276 \times 2048$ & Yes \\
Low Opacity & $0.1$ & $1/2$ & $10$ & $1$  & $0.3$ & $0$ & $0.3$ & $10,20,50,100$ & $1276 \times 2048$ & Yes \\
High Opacity & $0.1$ & $1/2$ & $10$ & $1$  & $3$ & $0$ & $0.3$ & $10,20,50,100$ & $1276 \times 2048$ & Yes \\
$\alpha = 10^{-4}$ & $0.1$ & $1/2$ & $10$ & $1$ & $1$  & $10^{-4}$ & $0.3$ & $10,20,50,100$ & $1276 \times 2048$ & Yes \\
$\alpha = 10^{-3}$ & $0.1$ & $1/2$ & $10$ & $1$ & $1$  & $10^{-3}$ & $0.3$ & $10,20,50,100$ & $1276 \times 2048$ & Yes \\
Surface Cooling Only & $0.1$ & $1/2$ & $10$ & $1$ & $1$ & $0$ & $0.3$ & $10,20,50,100$ & $1276 \times 2048$ & No \\ 
\hline
\end{tabular}
\end{center}
\begin{tablenotes}
\item {\bf Notes:} (1) Name of the disk model, (2) $h_{50\mathrm{au}}$, the aspect ratio at $50$ au, (3) temperature power law index $q$, (4) $\Sigma_{50\mathrm{au}}$, the surface density at at $50$ au, (5) surface density power law index $p$, (6) dimensionless scaled opacity $\bar{\kappa}_{\mathrm{d,0}}$ (see equation~(\ref{eq:opac-law})), (7) viscosity parameter $\alpha$, (8) planet mass $M_\mathrm{p}$, in terms of the thermal mass $M_\mathrm{th}$ (see equation~(\ref{eq:Mth})), (9) planetary orbital radius $r_\mathrm{p}$, (10) numerical resolution $N_r \times N_\phi$, and (11) whether or not in-plane cooling is included in the simulation.
\end{tablenotes}
\label{tab:parameters}
\end{table*}

\section{Basic Setup}
\label{sect:basic-setup}

We consider a thin, two-dimensional gaseous disk around a star of mass $M_*$, that is described in polar coordinates $(r,\phi)$ by the surface density $\Sigma$, height-integrated pressure $P = c_\mathrm{s}^2 \Sigma$, where $c_\mathrm{s} = (k_\mathrm{B}T/\mu)^{1/2}$ is the (isothermal) sound speed, radial velocity $u_r$, and azimuthal velocity $u_\phi = r\Omega$. The rotation rate $\Omega$ differs from the Keplerian rate $\Omega_\mathrm{K} = (GM_*/r^3)^{1/2}$ by $\mathcal{O}(h^2)$, where $h = H/r$ is the disk aspect ratio and $H = c_\mathrm{s}/\Omega_\mathrm{K}$ is the pressure scale height. We neglect the disk self-gravity (its effect was explored by \citealt{Zhang2020}). We primarily consider a disk with a negligible viscosity, although we also carry out a subset of simulations with explicit viscosity, which we parameterize via the conventional dimensionless $\alpha$-parameter \citep{SS73}.

The disk temperature follows a power law profile with $T(r) \propto r^{-q}$. It is specified indirectly through the aspect ratio $h(r)$, parameterized by its value at $r = 50$ au, according to
\be
\label{eq:T-profile}
h(r) = h_{50\mathrm{au}} r_{50}^{(1-q)/2},
\ee
where $r_{50} = r/\left(50~\mathrm{au}\right)$. The surface density profile is similarly parameterized in terms of its value at $50$ au and the power law index $p$, 
\be
\label{eq:surfdens-profile}
\Sigma(r) = \Sigma_{50\mathrm{au}} r_{50}^{-p}.
\ee

For computing the thermodynamic properties of the disk, we adopt the opacity (due to dust grains) of \citet{Bell}, appropriate for the cold ($T < 170$ K) outer regions of protoplanetary disks,
\be
\label{eq:opac-law}
\begin{aligned}
\kappa_\mathrm{d}  = \kappa_{\mathrm{d},0} T^2  = 0.38~M_{*,1}^2 \bar{\kappa}_{\mathrm{d},0} h_{0.1}^4 r_{50}^{-2}~\mathrm{cm}^2\mathrm{g}^{-1},
\end{aligned}
\ee
where $M_{*,1} = M_*/M_\odot$, $h_{0.1} = h(r)/0.1$. We have also introduced $\bar{\kappa}_{\mathrm{d},0}$, the constant $\kappa_{\mathrm{d},0}$ scaled by its fiducial value $2 \times 10^{-4}~\mathrm{cm}^2~\mathrm{g}^{-1}~\mathrm{K}^{-2}$, which is used to scale the overall opacity of the disk material (which may depend on metallicity). As a result of our adopted opacity law, the optical depth (from the midplane to the surface) of the disk is
\be
\label{eq:tau}
\begin{aligned}
\tau  = \frac{1}{2} \kappa_\mathrm{d} \Sigma 
 = 1.9~M_{*,1}^2 \bar{\kappa}_{\mathrm{d},0} h_{0.1}^4 \Sigma_{10} r_{50}^{-2},
\end{aligned}
\ee
where $\Sigma_{10} = \Sigma(r)/(10~\mathrm{g}~\mathrm{cm}^{-2})$. In the numerical estimates above, we have assumed a mean molecular weight $\mu = 2 m_\mathrm{p}$.

The temperature and surface density profiles, along with the opacity $\kappa_\mathrm{d}$, fully define our disk model for the purposes of computing the cooling timescale. The disk model is therefore fully specified by six parameters: $h_{50\mathrm{au}}$, $q$, $\Sigma_{50\mathrm{au}}$, $p$, $\bar{\kappa}_{\mathrm{d},0}$, and $\alpha$. In describing our results we consider twelve different disk models, which are summarized in Table~\ref{tab:parameters}. The Fiducial disk model has $h_{50\mathrm{au}} = 0.1$, $q = 1/2$, $\Sigma_{50\mathrm{au}} = 10~\mathrm{g}~\mathrm{cm}^{-2}$, $p = 1$, $\bar{\kappa}_{\mathrm{d},0} = 1$, and $\alpha = 0$. The disk mass is varied (by varying $\Sigma_{50\mathrm{au}}$) in the Low Mass and High Mass models. The temperature (through $h_{50\mathrm{au}}$) is varied in the Hot and Cold models. The opacity is varied by means of $\bar{\kappa}_{\mathrm{d},0}$ in the Low Opacity and High Opacity models. The temperature and surface density power law indices $q$ and $p$ are varied in the $T \propto r^{-1/4}$ and $\Sigma \propto r^{-1/2}$ models, respectively. The effects of a non-zero viscosity are explored in the $\alpha = 10^{-4}$ and $10^{-3}$ models. 

For each disk model, we consider planets with a range of different masses $M_\mathrm{p}$ and orbital radii $r_\mathrm{p}$. For $r_\mathrm{p}$ we choose $10$ au, $20$ au, $50$ au, and $100$ au, spanning the range of radii at which most of the rings and gaps in protoplanetary disks are observed \citep{Huang2018}. Note that for each disk model, $h(r)$ and $\Sigma(r)$ are defined as functions of $r$, and not $r/r_\mathrm{p}$. The aspect ratio and surface density at the location of the planet $h_\mathrm{p}$ and $\Sigma_\mathrm{p}$ are therefore $h_{50\mathrm{au}} r_{\mathrm{p},50\mathrm{au}}^{(1-q)/2}$ and $\Sigma_{50\mathrm{au}} r_{\mathrm{p},50\mathrm{au}}^{-p}$ (where $r_{\mathrm{p},50\mathrm{au}} = r_\mathrm{p}/50~\mathrm{au}$) respectively.

The planet mass $M_\mathrm{p}$ is parameterized in terms of the thermal mass \citep{GR01},
\be
\label{eq:Mth}
M_\mathrm{th} = h_\mathrm{p}^3 M_* = 1 M_\mathrm{J} \left(\frac{M_*}{M_\odot}\right) \left(\frac{h_\mathrm{p}}{0.1}\right)^3.
\ee
The ratio $M_\mathrm{p}/M_\mathrm{th}$ describes the nonlinearity of disk response. Note that since we choose the planet mass to be a fixed fraction of $M_\mathrm{th}$, the planet mass varies with $r_\mathrm{p}$, as $M_\mathrm{th} = h_{50\mathrm{au}}^3 r_{\mathrm{p},50\mathrm{au}}^{3(1-q)/2}$, for a fixed $M_\mathrm{p}/M_\mathrm{th}$.

\section{Cooling Prescription}
\label{sect:cooling}

In this section we provide a detailed description of the procedure that we use to compute the thermal relaxation timescale for density waves propagating through protoplanetary disks. In Appendix~\ref{sect:en-eq} we derive the following 2D form of the energy equation for a disk perturbed by the passage of a density wave, accounting for radiative effects:
\begin{align}  
\label{eq:energy-eq}
\frac{\mathrm{d}e}{\mathrm{d}t} + P\frac{\mathrm{d}}{\mathrm{d}t}\left(\frac{1}{\Sigma}\right) &=
\left(\frac{\partial e}{\partial t}\right)_\mathrm{surf} + \left(\frac{\partial e}{\partial t}\right)_\mathrm{mid}, \\
\label{eq:loss-surf}
\left(\frac{\partial e}{\partial t}\right)_\mathrm{surf} &= -2\frac{\sigma T_\mathrm{eff}^4}{\Sigma e_0}\delta e, \\
\label{eq:loss-mid}
\left(\frac{\partial e}{\partial t}\right)_\mathrm{mid} &= -\frac{\mathbf{\nabla} \cdot\delta\mathbf{F}}{\Sigma}.
\end{align}
Here $e = c_V T$ and $P = (\gamma-1)\Sigma e$ are the 2D (properly vertically averaged) internal energy (per unit mass) and pressure, and $\delta e$ is the wave-driven perturbation of $e$ from its equilibrium value $e_0$. The two terms in the right-hand side of (\ref{eq:energy-eq}) represent the energy source/sink terms\footnote{If these terms were set to zero, equation (\ref{eq:energy-eq}) would reduce to the energy equation (6) given in \citet{Miranda-Cooling} for an adiabatic disk.} associated with the transport of radiation caused by the passage of a density wave. The one given by equation (\ref{eq:loss-surf}) accounts for the radiative losses from the disk surface, with $T_\mathrm{eff}$ being the effective temperature of the disk in equilibrium. The term given by equation (\ref{eq:loss-mid}) accounts for the transport of radiation along the disk midplane, with $\delta\mathbf{F}$ being the density wave-induced perturbation of the (vertically integrated) component of the radiative flux along the plane of the disk.  

Using this form of the energy equation, we first provide a calculation of the cooling timescale due to radiative losses from the surface of the disk (\S \ref{sect:cooling-surf}), followed by the cooling timescale associated with radiative energy transfer along the disk midplane  (\S \ref{sect:cooling-mid}). We then present the cooling timescale that arises from the combination of both of these effects  (\S \ref{sect:cooling-full}). As we will show, this timescale depends explicitly on the azimuthal number $m$ of the perturbation  (\S \ref{sect:beta_m}). In the last part of this section, we present the calculation of an effective cooling timescale characterizing the effect of cooling on planet-driven density waves that are made up of many different Fourier harmonics (\S \ref{sect:tc-eff}).

\subsection{Surface Cooling}
\label{sect:cooling-surf}

The surface cooling term (\ref{eq:loss-surf}) can be directly re-written in the form
\be
\label{eq:surf_standard}
\frac{1}{\Omega}\left(\frac{\partial e}{\partial t}\right)_\mathrm{surf} = -\frac{\delta e}{\beta_\mathrm{surf}}.
\ee
Here we introduced the cooling timescale due to thermal continuum dust emission from the disk surface:
\be
t_\mathrm{surf} = \frac{\Sigma e}{2\sigma T_\mathrm{eff}^4},
\label{eq:surf_cool}
\ee
and the dimensionless cooling timescale $\beta_\mathrm{surf} = \Omega t_\mathrm{surf}$ (in the following we will often omit the word ``dimensionless'', and simply refer to $\beta$ as a cooling timescale). The factor of two in the denominator accounts for the fact that the disk cools from both its upper and lower surfaces. In the above, we do not distinguish between $e$ and $e_0$, since the cooling term (\ref{eq:surf_standard}) is already linear in a perturbed variable $\delta e$.

The effective temperature $T_\mathrm{eff}$ is related to the midplane temperature $T$ according to \citep{Hubeny1990}
\be
\label{eq:T-eff}
T_\mathrm{eff}^4 = \tau_\mathrm{eff}^{-1} T^4,
\ee
where
\be
\label{eq:tau-eff}
\tau_\mathrm{eff} = \frac{3\tau}{8} + \frac{\sqrt{3}}{4} + \frac{1}{4\tau}
\ee
is the effective optical depth. Note that the cooling timescale (\ref{eq:surf_cool}) combined with the relations (\ref{eq:T-eff})--(\ref{eq:tau-eff}) should be valid not only for self-luminous disks, but also for externally irradiated ones (at least approximately). In particular, 
our $t_\mathrm{surf}$ would reduce (up to factors of order unity) to the cooling time $t_\mathrm{cool}$ derived in \citet{Zhu2015}, namely their equation (8), in which one needs to set $T_\mathrm{c} = T_\mathrm{irr}$. However, the precise form of $t_\mathrm{surf}$ is not especially important, since, as we show later, surface cooling is not the dominant factor in density wave thermodynamics.  
With this consideration, we have
\be
\label{eq:beta-surf}
\beta_\mathrm{surf} = \beta_{\mathrm{surf},0} f(\tau),
\ee
where
\be
\label{eq:beta-surf-0}
\begin{aligned}
\beta_{\mathrm{surf},0} & = \frac{c_V \Omega}{4 \kappa_{\mathrm{d},0} \sigma T^5} \\
& = 8.7 \times 10^{-3}~M_{*,1}^{-9/2} \bar{\kappa}_{\mathrm{d},0}^{-1} h_{0.1}^{-10} r_{50}^{7/2}
\end{aligned}
\ee
is the surface cooling timescale in the optically thin regime ($\tau \ll 1$) and
\be
\label{eq:ftau}
f(\tau)  = 1 + \sqrt{3}\tau + \frac{3}{2}\tau^2.
\ee 
The numerical estimate in equation (\ref{eq:beta-surf-0}) assumes an adiabatic index $\gamma = 7/5$. Note that $h(r) \propto r^{(1-q)/2}$, so the radial dependence of $\beta_{\mathrm{surf},0}$ is not determined solely by the $r_{50}^{7/2}$ term. For a fiducial temperature profile with $q = 1/2$, $\beta_{\mathrm{surf},0} \propto r$.

Note that $f(\tau) \approx 1$ for $\tau \ll 1$, and that $f(\tau) > 1$ always. Therefore, $\beta_{\mathrm{surf},0}$ is a lower limit for the surface cooling timescale. 

\subsection{In-plane Cooling}
\label{sect:cooling-mid}

In addition to cooling from the disk surface, thermal relaxation also occurs through radiative energy transfer along the midplane of the disk. Its contribution to the energy evolution is described by equation~(\ref{eq:loss-mid}), in which we need to specify the explicit form of the (vertically integrated) flux perturbation $\delta\mathbf{F}$. The form of this perturbation, and hence of the resulting cooling term, depend on the ratio $l_e/l_\mathrm{rad}$, where $l_e = |\delta e|/|\mathbf{\nabla}\delta e|$ is the perturbation lengthscale, and 
\be
\label{eq:l-rad}
\begin{aligned}
l_\mathrm{rad} = \frac{1}{\kappa_\mathrm{d}\rho} 
= 3.3~M_{*,1}^{-2} \bar{\kappa}_{\mathrm{d},0} h_{0.1}^{-3} \Sigma_{10}^{-1} r_{50}^{3}~\mathrm{au}
\end{aligned}
\ee
is the radiative lengthscale or photon mean free path length. In the above estimate, we have taken the 3D density $\rho$ equal to the midplane density $\rho_\mathrm{mid} = \Sigma/(\sqrt{2\pi}H)$ (assuming the disk is vertically isothermal).

In general, computing $\delta\mathbf{F}$ from the full radiation transfer equation is highly non-trivial, but can be bypassed by using an approximate theory, e.g., flux-limited diffusion (FLD; \citealt{Levermore1981,Kley1989}). However, the behavior of $\delta\mathbf{F}$ can be understood in two important limits: the diffusion limit $l_e \gg l_\mathrm{rad}$ and the streaming limit $l_e \ll l_\mathrm{rad}$.

\subsubsection{Diffusion Limit}
\label{sect:diffusion-limit}

For $l_e \gg l_\mathrm{rad}$, we show in Appendix \ref{sect:flux} that the vertically integrated flux perturbation takes the form
\be
\label{eq:flux-diff}
\delta\mathbf{F} = -\Sigma\eta\mathbf{\nabla} \delta e,
\ee
where 
\be
\label{eq:diff-coef}
\begin{aligned}
\eta  = \frac{16\sigma T^3}{3\kappa_\mathrm{d}\rho^2 c_V}
\end{aligned}
\ee
is the radiative diffusion coefficient (here we again assume $\rho = \rho_\mathrm{mid}$). As a result of specifying $\delta\mathbf{F}$ via equation~(\ref{eq:flux-diff}), the in-plane cooling term (equation~(\ref{eq:loss-mid})) reads
\be
\label{eq:loss-mid-diff}
\left(\frac{\partial e}{\partial t}\right)_\mathrm{mid} = \frac{1}{\Sigma}\mathbf{\nabla} \cdot \left(\Sigma\eta\mathbf{\nabla}\delta e\right).
\ee
One can see that in this regime, the in-plane cooling contribution is indeed diffusive, depending on the specific form of $\delta e$ through its second-order spatial derivative. This raises a serious problem for trying to understand the density wave dynamics in the well-defined limit of linear perturbations, an approximation that has been successfully used for this purpose in the past \citep{GT80,Miranda-Spirals,Miranda-Cooling}. 

In linear theory, we assume an expansion of $\delta e$ in Fourier harmonics:
\be
\label{eq:fourier}
\delta e(r,\phi) = \sum_m \delta e_m(r) \exp(\mathrm{i}m\phi),
\ee
where $\delta e_m(r)$ is a complex quantity describing the amplitude and phase of the perturbation with azimuthal number $m$. As a result of this expansion, equation (\ref{eq:loss-mid-diff}) then describes the evolution of each harmonic $\delta e_m$, rather than the total perturbation $\delta e$.

An analogous Fourier expansion (\ref{eq:fourier}) is assumed for all other perturbed fluid variables. As demonstrated in \citet{Miranda-Cooling}, one can then manipulate the fluid continuity and momentum equations into a single master equation describing the global behavior of linear perturbations. If the cooling term depends linearly on $\delta e$, like in equation (\ref{eq:surf_standard}), then this master equation ends up being a second-order ordinary differential equation (ODE) in the radial coordinate, which has been studied in detail in \citet{Miranda-Cooling}.

However, with the in-plane cooling contribution in the form (\ref{eq:loss-mid-diff}) the perturbations of, e.g., pressure and surface density would necessarily involve {\it second-order radial derivatives} of the energy (or enthalpy) perturbation. Propagating this dependence through the linear perturbation analysis therefore results in a {\it fourth-order} master ODE rather than a second-order equation as in \citet{Miranda-Cooling}. This significantly increases the complexity of the analysis and numerical solutions of the master equation.

To avoid this complication in our present work, and to fully benefit from the mathematical framework and physical understanding developed in \citet{Miranda-Cooling}, we have resorted to a different, approximate approach. Instead of taking full spatial derivatives in the right-hand side of equation~(\ref{eq:loss-mid-diff}), we evaluate them using the local (WKB) approximation, assuming $l_{e,m} \ll r$ (which is a good approximation, as demonstrated by Fig.~\ref{fig:spiral_width}). Here $l_{e,m} = |\delta e_m|/|\mathbf{\nabla}\delta e_m|$ is the lengthscale associated with a particular Fourier harmonic of the perturbation. As a result of making this assumption, equation~(\ref{eq:loss-mid-diff}) for a Fourier harmonic of the energy perturbation $\delta e_m$ takes the simple form
\be
\left(\frac{\partial e}{\partial t}\right)_{\mathrm{mid},m} \approx -\frac{\delta e_m}{t_{\mathrm{diff},m}},
\label{eq:diff_cool}
\ee
where
\be
\label{eq:t-diff}
t_{\mathrm{diff},m} = \frac{l_{e,m}^2}{\eta}
\ee
is the in-plane cooling timescale for the $m$-th Fourier harmonic of the perturbation in the diffusive limit. Note that $t_{\mathrm{diff},m}$ explicitly depends on the radial structure of the perturbation due to the wave, i.e., on the wavenumber $m$. Nevertheless, we can still deal with cooling in the form (\ref{eq:diff_cool}) directly using the framework of \citet{Miranda-Cooling}, which allows one to describe density wave propagation using a single second-order ODE.\footnote{Even though the calculation in \citet{Miranda-Cooling}, by design, adopted the dimensionless cooling time for each Fourier harmonic to be independent of $m$.}

\subsubsection{Streaming Limit and Interpolation}
\label{sect:streaming-limit}

In the opposite streaming limit, for $l_e \ll l_\mathrm{rad}$, we show in Appendix \ref{sect:flux} that the vertically integrated flux perturbation approximately satisfies 
\be
\label{eq:flux_stream}
\mathbf{\nabla}\cdot\delta\mathbf{F} = \frac{16\kappa_\mathrm{d}\Sigma\sigma T^3}{c_V}\delta e.
\ee 
Using definitions (\ref{eq:l-rad}) and (\ref{eq:diff-coef}), we can then transform equation~(\ref{eq:loss-mid}) into
\be
\left(\frac{\partial e}{\partial t}\right)_\mathrm{mid} \approx -\frac{\delta e}{t_\mathrm{stream}},
\ee
where
\be
\label{eq:t-stream}
t_\mathrm{stream} = \frac{l_\mathrm{rad}^2}{3\eta}
\ee
is the timescale for in-plane cooling in the streaming limit \citep{LinYoudin2015,Malygin2017}. Note that $t_\mathrm{stream}$ is independent of the perturbation lengthscale.

In the general situation, following \citet{LinYoudin2015}, we express the midplane cooling timescale for the $m$-th Fourier harmonic of the perturbation as $t_{\mathrm{mid},m} \approx t_\mathrm{stream} + t_{\mathrm{diff},m}$, which captures the correct behavior in both the diffusive and streaming limits. The dimensionless timescale for in-plane cooling $\beta_{\mathrm{mid},m} = \Omega t_{\mathrm{mid},m}$ is then, using equations (\ref{eq:t-diff}) and (\ref{eq:t-stream}),
\be
\label{eq:beta-mid}
\beta_{\mathrm{mid},m} \approx \frac{\Omega}{\eta}\left(l_{e,m}^2 + \frac{1}{3} l_\mathrm{rad}^2\right).
\ee
Once again, the subscript $m$ denotes that the in-plane cooling timescale $\beta_{\mathrm{mid},m}$ explicitly depends on the azimuthal wavenumber of the perturbation through the lengthscale $l_{e,m}$, which we derive next. 

\subsubsection{Perturbation Lengthscale}

For a density wave with azimuthal number $m$, the (inverse) perturbation lengthscale is
\be
l_{e,m}^{-1} = \left(k_m^2 + \frac{m^2}{r^2}\right)^{1/2} \approx k_m,
\ee
where $k_m(r)$ is the radial wavenumber for the $m$-th Fourier harmonic (this notation is consistent with \citealt{Miranda-Spirals}). The second (approximate) equality above follows from the fact that the radial wavelength $k_m^{-1}$ is shorter than the azimuthal wavelength $m/r$ by a factor of $\approx h$ after the wave has propagated far enough away from the planet. Indeed, Fig.~\ref{fig:spiral_width}(b) shows that $\delta r \sim r\delta\phi \sim H$ only within about one scale height of the planet (where the in-plane cooling is not dominant anyway). At larger distances, beyond about $2H$ from the planet, the radial width becomes much smaller, with $\delta r \approx 0.2 H$. On the other hand, the azimuthal width $r\delta\phi$ remains roughly constant, so that $\delta r/(r\delta\phi) \ll 1$. The spiral arm is therefore much narrower in the radial direction than in the azimuthal direction for $|r - r_\mathrm{p}|\gtrsim H$, and so $l_{e,m}^{-1}$ can be approximated by $k_m$. 

For planet-driven density waves, the radial wave-number is well-approximated by its expression in the WKB limit,
\be
\label{eq:kr-wkb}
k_m^2 = \tilde{k}_m^2 - \frac{1}{H^2},
\ee
where
\be
\label{eq:kr-tilde}
\tilde{k}_m = \frac{m}{H} \left|\frac{\Omega_\mathrm{p}}{\Omega} - 1\right|
\ee
is the approximate wavenumber for $m \gg 1$ (e.g., \citealt{OL02,Miranda-Spirals}). From equation~(\ref{eq:kr-wkb}) we see that $k_m^2 > 0$ for $r > r_+$ and and $r < r_-$, where $r_\pm$ is the location of the outer/inner Lindblad resonance (LR), defined by $k_m(r_\pm) = 0$. Between the LRs, density waves are evanescent, with $k_m^2 < 0$. Estimating the perturbation lengthscale as $l_{e,m} \approx k_m^{-1}$ using equation~(\ref{eq:kr-wkb}) would then result in an unphysical negative timescale for diffusive in-plane cooling (equation~(\ref{eq:t-diff})) in the evanescent zone.

However, note that $\tilde{k}_m^2$ (equation~(\ref{eq:kr-tilde})) is positive everywhere. Furthermore, it differs from the actual (WKB) $k_m^2$ (equation~(\ref{eq:kr-wkb})) only by $\mathcal{O}(m^{-2})$ far from the LRs. For planet-driven waves, which are dominated by harmonics with $m \sim h_\mathrm{p}^{-1}$ \citep{Miranda-Spirals}, this difference is $\mathcal{O}(h_\mathrm{p}^2)$ (also note that the $m = 1$ harmonic is very weak). Therefore, $\tilde{k}_m$ is a good approximation for $k_m$ outside of the LRs. We will henceforth approximate the perturbation lengthscale of planet-driven density waves as $l_{e,m} \approx \tilde{k}_m^{-1}$. 

\subsection{Combined Cooling Timescale}
\label{sect:cooling-full}

Taking into account both cooling from the disk surface with a timescale $\beta_\mathrm{surf}$ (equation~(\ref{eq:beta-surf})) and in-plane cooling with a timescale $\beta_{\mathrm{mid},m}$ (equation~(\ref{eq:beta-mid})), the cooling/thermal relaxation part of the evolution equation for a Fourier harmonic of the specific internal energy perturbation $\delta e_m$ reads
\be
\frac{1}{\Omega}\left(\frac{\partial \delta e_m}{\partial t}\right)_\mathrm{cool} = -\frac{\delta e_m}{\beta_\mathrm{surf}} - \frac{\delta e_m}{\beta_{\mathrm{mid},m}}.
\ee
We can rewrite this as
\be
\label{eq:em-cool}
\frac{1}{\Omega}\left(\frac{\partial \delta e_m}{\partial t}\right)_\mathrm{cool} = -\frac{\delta e_m}{\beta_m}, 
\ee
where we have defined the cooling timescale resulting from the combined effects of both surface and in-plane cooling,
\be
\label{eq:beta-m-1}
\beta_m = \left(\frac{1}{\beta_\mathrm{surf}} + \frac{1}{\beta_{\mathrm{mid},m}}\right)^{-1}.
\ee

Making use of the fact that $\eta\beta_{\mathrm{surf},0}/(l_\mathrm{rad}^2\Omega) = 4/3$ (see equations~(\ref{eq:beta-surf-0}), (\ref{eq:l-rad}), and (\ref{eq:diff-coef})), the cooling timescale (equation~(\ref{eq:beta-m-1})) can be expressed as
\be
\label{eq:beta-m-2}
\beta_m = \beta_\mathrm{surf} \left[1 + \frac{4 f(\tau)}{1+3(\tilde{k}_m l_\mathrm{rad})^{-2}}\right]^{-1},
\ee
where
\be
\label{eq:kr-lrad}
\tilde{k}_m l_\mathrm{rad} = \left(\frac{\pi}{2}\right)^{1/2} \left|\frac{\Omega_\mathrm{p}}{\Omega} - 1\right| \frac{m}{\tau}
\ee
is the ratio of the radiative lengthscale (equation~(\ref{eq:l-rad})) to the (approximate) radial wavelength of the density wave Fourier harmonic (equation~(\ref{eq:kr-tilde})). Equations~(\ref{eq:beta-mid}) and (\ref{eq:kr-lrad}) indicate that the transition between the diffusive and streaming limits for the in-plane cooling (taking place when $\tilde{k}_m l_\mathrm{rad} \sim 1$) occurs at $\tau \approx m$ in the inner disk, provided that $|\Omega-\Omega_\mathrm{p}|\sim \Omega$. Also note that $\beta_m$ is, in general, a function of the orbital distance $r_\mathrm{p}$ of the planet through the factor $\Omega_\mathrm{p}/\Omega$, which also ensures that $\beta_m(r_\mathrm{p}) = \beta_\mathrm{surf}$.

Equation~(\ref{eq:beta-m-2}) represents our final working expression for the cooling timescale associated with a particular Fourier harmonic of the density wave perturbation.

\subsubsection{Limiting Behaviors}

The behavior of the cooling timescale (equation~(\ref{eq:beta-m-2})) depends on two key parameters: the optical depth $\tau$, and the ratio of the radiative and perturbation lengthscales $\tilde{k}_m l_\mathrm{rad}$. Except within the immediate vicinity of the planet, the value of $\tilde{k}_m l_\mathrm{rad}$ is mostly determined by $m/\tau$ (see equation~(\ref{eq:kr-lrad})). Therefore, $\beta_m$ is largely determined by $\tau$. Its behavior can be broken down into several different regimes, delineated by whether $\tau$ is $\gg m$, $\ll m$, $\gg 1$, or $\ll 1$.

For $\tau \gg m$, in the diffusive cooling limit, we have
\be
\label{eq:beta-m-tauggm}
\beta_m \approx \beta_\mathrm{surf}\left[1 + \pi m^2\left(\frac{\Omega_\mathrm{p}}{\Omega} - 1\right)^2\right]^{-1},
\ee
or roughly $\beta_m \sim \beta_\mathrm{surf}/m^2$ for $|\Omega-\Omega_\mathrm{p}|\sim \Omega$. This reduction of the cooling timescale by a factor of $m^2$ compared to the surface cooling timescale $\beta_\mathrm{surf}$ is easy to understand by examining equation~(\ref{eq:kr-tilde}) and noticing that the characteristic radial scale of the wave-like temperature perturbation far from the planet is $\sim m^{-1}H$. This is considerably shorter than the vertical disk thickness $H$ (see Fig.~\ref{fig:spiral_width}) that the radiation needs to diffuse through to reach the disk surface, leading to in-plane cooling being a factor of $m^2$ faster (in the diffusive regime) than surface cooling. 

For $\tau \ll m$, in the streaming limit, we have
\be
\label{eq:beta-m-taullm}
\beta_m \approx \frac{\beta_{\mathrm{surf}}}{1 + 4 f(\tau)}=\beta_{\mathrm{surf},0}\frac{f(\tau)}{1 + 4 f(\tau)},
\ee
which is independent of $m$. 

Furthermore, in the conventional optically thin limit $\tau \ll 1$, we have $f(\tau) \approx 1$ and $\beta_{\mathrm{surf}}\approx \beta_{\mathrm{surf},0}$, so that
\be
\label{eq:beta-m-taull1}
\beta_m \approx \frac{1}{5}\beta_{\mathrm{surf}} \approx \frac{1}{5}\beta_{\mathrm{surf},0}.
\ee
This is the result of an equivalence (to within a constant factor) between the timescale for cooling from the disk surface (equation~(\ref{eq:beta-surf})) in the optically thin limit, and for cooling in the midplane (equation~(\ref{eq:beta-mid})) in the streaming limit. The factor of $1/5$ indicates that the combined effect of surface and in-plane optically thin cooling results in faster cooling than either of these effects alone. Whereas a more careful, fully 3D treatment of the radiation transfer would likely find the numerical prefactor to deviate from $1/5$, the difference is not fundamental. It is a consequence of assumptions made about other approximate prefactors in our calculation of the cooling timescale.

Finally, when $\tilde{k}_m = 0$---which occurs only at $r = r_\mathrm{p}$, regardless of the optical depth---we have $\beta_m = \beta_\mathrm{surf}$. Therefore, $\beta_\mathrm{surf}$ sets an upper limit for $\beta_m$, i.e., we always have $\beta_m \leq \beta_\mathrm{surf}$.

\subsubsection{Radial Profiles of $\beta_m$}
\label{sect:beta_m}

\begin{figure}
\begin{center}
\includegraphics[width=0.49\textwidth,clip]{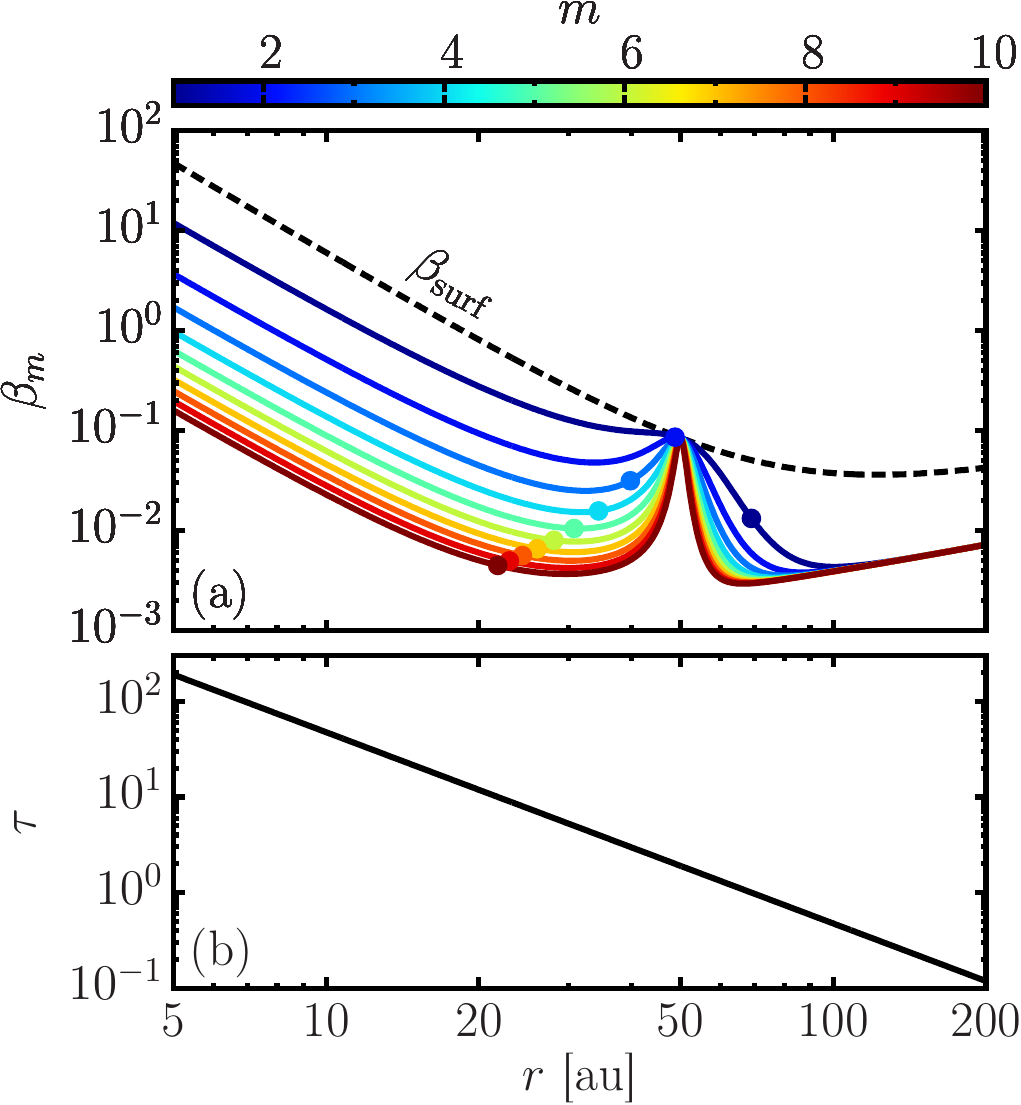}
\caption{Dimensionless cooling timescale $\beta_m$ (equation~(\ref{eq:beta-m-2})) for different Fourier harmonics of density waves with azimuthal wavenumber $m$ driven by a planet at $r_\mathrm{p} = 50$ au for the Fiducial disk model (a), and the optical depth $\tau(r)$ for that disk model (b). In (a), the dashed black curve shows the cooling timescale due to radiative cooling from the disk surface alone, $\beta_\mathrm{surf}$ (equation~(\ref{eq:beta-surf})). The filled point on each curve shows the location at which $\tau = m$, where in-plane cooling transitions from the diffusive regime to the streaming regime for the corresponding harmonic.}
\label{fig:beta-m}
\end{center}
\end{figure}

Fig.~\ref{fig:beta-m}(a) shows profiles of $\beta_m$ (equation~(\ref{eq:beta-m-2})) for a range of azimuthal numbers $m$, for the density waves launched by a planet at $50$ au, using our Fiducial disk model (see Section~\ref{sect:basic-setup} and Table~\ref{tab:parameters} for details). These serve to illustrate the different aspects of the behavior of $\beta_m$ outlined above.

The dashed line in Fig.~\ref{fig:beta-m}(a) shows the profile of the surface cooling timescale $\beta_\mathrm{surf}$ (equation~(\ref{eq:beta-surf})). Each $\beta_m$ (different colored solid lines) is equal to $\beta_\mathrm{surf}$ at the planetary orbital radius $r_\mathrm{p} = 50$ au. Everywhere else, $\beta_m < \beta_\mathrm{surf}$, as expected. This roughly indicates that surface cooling may play a role in the excitation of density waves (by determining the total torque absorbed by the density wave, see  \citealt{Miranda-Cooling}), which takes place within $(1$ -- $2)H$ from the planet where $m|\Omega-\Omega_\mathrm{p}|/\Omega\lesssim 1$. However, midplane cooling dominates the subsequent propagation and evolution of the waves.

Far from the planet, the behavior of the midplane cooling depends on the optical depth $\tau$, which is shown in Fig.~\ref{fig:beta-m}(b). For each $\beta_m$ shown in Fig.~\ref{fig:beta-m}(a), a filled point indicates the radius at which $\tau = m$, which marks a transition in the behavior of the in-plane cooling from diffusive (for $\tau > m$) to streaming ($\tau < m$). In the diffusive region, to the left of the $\tau = m$ radius, $\beta_m$ is described by equation~(\ref{eq:beta-m-tauggm}), and hence is smaller than $\beta_\mathrm{surf}$ by $\sim m^2$. In the streaming region, to the right of the $\tau = m$ radius, $\beta_m$ is described by equation~(\ref{eq:beta-m-taullm}), and hence should be independent of $m$. In practice, however, in the model presented in Fig.~\ref{fig:beta-m}, the interval $1 \lesssim \tau \lesssim m$ falls in the radial range where $\Omega$ is rather close to $\Omega_\mathrm{p}$. As a result, $\tilde{k}_m l_\mathrm{rad}$ ends up being not very large (see equation~(\ref{eq:beta-m-2})) and some sensitivity to $m$ gets retained in the expression (\ref{eq:kr-lrad}) for $\beta_m$, as we see in Fig.~\ref{fig:beta-m}(a).

At sufficiently large radii in the outer disk ($r \gtrsim 100$ au for the disk model in Fig.~\ref{fig:beta-m}), $\tau \ll 1$, and hence $\tau \ll m$ for all $m$. In this region, $\beta_m$ is in the streaming regime, and therefore independent of $m$, for all $m$, see equations (\ref{eq:beta-m-taullm})--(\ref{eq:beta-m-taull1}). Hence all of the $\beta_m$ converge to the same profile, as they are all proportional to $\beta_\mathrm{surf}$ with the same constant of proportionality of $1/5$ for $\tau\ll 1$.

We see that the different $\beta_m$ span a wide range of values in the model shown in Fig.~\ref{fig:beta-m}(a), from $\approx 10$ for $m = 1$ in the inner disk, to $\lesssim 0.01$ (for all $m$) in the outer disk. This indicates that the effect of cooling varies greatly in different regions of the disk and for different perturbation harmonics. The large spread in $\beta_m$ in the inner disk suggests that understanding the role of cooling on planet-driven density waves made up of many Fourier harmonics may be highly non-trivial. 

\subsection{Effective Cooling Timescale}
\label{sect:tc-eff}

Equation~(\ref{eq:beta-m-2}) describes the thermal relaxation timescale for a single Fourier harmonic with azimuthal number $m$. In general, this timescale is different for each harmonic. The specific internal energy perturbation $\delta e_m$ for each Fourier harmonic of a density wave is then damped on its own individual timescale $\beta_m$. In the numerical simulations in this work, we evolve each harmonic independently in this fashion, even when the density wave is non-linear, see \S \ref{sect:hydrosims} and Appendix \ref{sect:cooling-numerical}. However, to facilitate the interpretation of our results it is useful to define an effective cooling timescale $t_\mathrm{c,eff}$, or dimensionless $\beta_\mathrm{eff} = \Omega t_\mathrm{c,eff}$, which gives the approximate relaxation timescale for the  perturbation of the {\it total} internal energy, which is made up of many harmonics. 

First, using Parseval's theorem, we relate the total internal energy perturbation $\delta e$ to its Fourier harmonics $\delta e_m$:
\be
\label{eq:e-parseval}
\frac{1}{2\pi} \oint |\delta e|^2 \mathrm{d}\phi = \sum_m |\delta e_m|^2.
\ee
Differentiating (\ref{eq:e-parseval}) and making use of the cooling law for the individual harmonics (equation~\ref{eq:em-cool}), we have
\be
\frac{1}{4\pi\Omega} \frac{\partial}{\partial t} \oint |\delta e|^2 \mathrm{d}\phi = -\beta_\mathrm{eff}^{-1} \sum_m |\delta e_m|^2,
\label{eq:eff-cooling}
\ee
where we have defined the {\it effective cooling timescale}
\be
\label{eq:beta-eff}
\beta_\mathrm{eff}^{-1} = \sum_m \beta_m^{-1} |\delta e_m|^2 \left(\sum_m |\delta e_m|^2\right)^{-1}.
\ee
Using equation (\ref{eq:e-parseval}) once again, we can re-write equation (\ref{eq:eff-cooling}) as
\be
\oint |\delta e|\left(\frac{1}{\Omega} \frac{\partial |\delta e|}{\partial t}+\frac{|\delta e|}{\beta_\mathrm{eff}}\right)\mathrm{d}\phi = 0,
\label{eq:eff-cooling1}
\ee
which shows that $\beta_\mathrm{eff}$ indeed plays a role of a dimensionless cooling timescale for the total energy perturbation. 
\begin{figure*}
\begin{center}
\includegraphics[width=0.99\textwidth,clip]{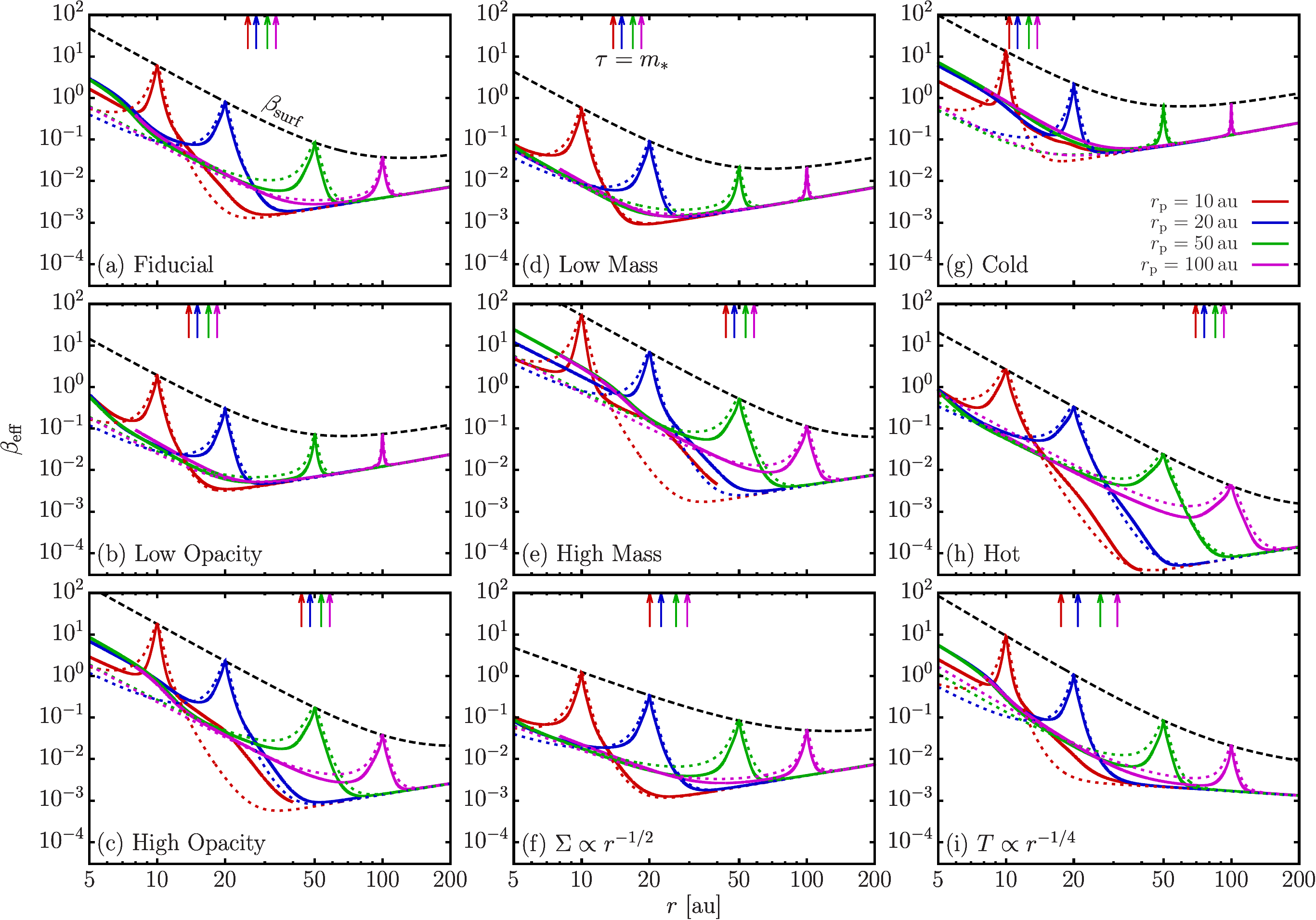}
\caption{Effective cooling timescale $\beta_\mathrm{eff}$ (equation~(\ref{eq:beta-eff})) for planet-driven density waves, for different disk models (different panels; see Table~\ref{tab:parameters}) and different planetary orbital radii $r_\mathrm{p}$ (solid curves in each panel). The dotted curves correspond to $\beta_{m*}$, the cooling timescale for the Fourier harmonic with the dominant azimuthal number $m = m_*$, which serves as an approximation for $\beta_\mathrm{eff}$ (see Section~\ref{sect:beta-mstar}). The black dashed curve in each panel shows the surface cooling timescale $\beta_\mathrm{surf}$ (equation~(\ref{eq:beta-surf})). The arrows at the top of each panel mark the locations at which the optical depth $\tau$ is equal to $m_*$ for each $r_\mathrm{p}$, corresponding to the transition from diffusive to streaming regimes of in-plane cooling.}
\label{fig:beta-eff}
\end{center}
\end{figure*}

Notice that $\beta_\mathrm{eff}(r)$ is a weighted harmonic mean of the cooling timescales $\beta_m$, with weights given by the squared Fourier amplitudes $|\delta e_m(r)|^2$. In order to calculate it, we need the specific form of $\delta e_m(r)$ for all of the harmonics of interest. As a benchmark, we determine these using numerical solutions of the {\it linear} perturbation equations (see Section~\ref{sect:setup-linear}) previously developed in \citet{Miranda-Cooling}, which is a well-defined way of fixing $\delta e_m(r)$. For massive planets $\delta e_m(r)$ is modified by nonlinear effects (which are fully captured in our numerical simulations). Nonetheless, computing $\beta_\mathrm{eff}$ using the linear $\delta e_m(r)$ provides a useful diagnostic for the role of cooling in the behavior of planet-driven density waves and reduces to the true effective cooling timescale in the limit of $M_\mathrm{p}\ll M_\mathrm{th}$.

We will not discuss in detail the dependence of the linear $\delta e_m$ on $r$. The most salient features of the behavior of $\delta e_m(r)$ can be roughly understood by examining the radial profiles of the $F_J^m$---contributions of the different Fourier harmonics to the angular momentum flux---derived in \citet{Miranda-Cooling}. While connecting $F_J^m(r)$ profiles to $\delta e_m(r)$ it should be remembered that $\delta e_m$ is a linear function of the enthalpy perturbation $\delta h_m$, whereas $F_J^m(r)$ depends on $\delta h_m$ quadratically. Also, the results in \citet{Miranda-Cooling} were obtained for a radially-constant and $m$-independent cooling time $\beta$, which is clearly different from $\beta_\mathrm{m}$ given by the equation (\ref{eq:beta-m-2}). Nevertheless, one can still use the $F_J^m(r)$ profiles in \citet{Miranda-Cooling} to get a qualitative idea of how the radial variation of $\delta e_m(r)$ might impact the calculation of  $\beta_\mathrm{eff}(r)$.

The effective cooling timescale $\beta_\mathrm{eff}$ for different disk models (see Table~\ref{tab:parameters}) is shown in Fig.~\ref{fig:beta-eff}. Profiles of $\beta_\mathrm{eff}(r)$ for waves driven by planets orbiting at four different radii $r_\mathrm{p}$, from $10$ au to $100$ au, are shown by the solid colored curves for each disk model.

We see that $\beta_\mathrm{eff}$ is equal to $\beta_\mathrm{surf}$ (dashed black curve in each panel of Fig.~\ref{fig:beta-eff}) at $r = r_\mathrm{p}$, and is strictly less than $\beta_\mathrm{surf}$ everywhere else. The factor by which $\beta_\mathrm{eff}$ is smaller than $\beta_\mathrm{surf}$ depends on whether in-plane cooling operates in the diffusive or streaming limit. Far from the planet in the outer, optically thin part of the disk, $\beta_\mathrm{eff}$ converges to the behavior given by equation~(\ref{eq:beta-m-taull1}), which is independent of $m$. Far inside of the planetary orbit, in the optically thick part of the disk, the $\beta_\mathrm{eff}(r)$ curves tend to show a universal behavior independent\footnote{This is not true in the diffusive limit for $r>r_\mathrm{p}$ since $\Omega_\mathrm{p}/\Omega$ is not small there.} of $r_\mathrm{p}$, which is due to $\Omega\gg\Omega_\mathrm{p}$ there, making $\beta_m$ independent of $r_\mathrm{p}$, see equation (\ref{eq:beta-m-tauggm}).

\subsubsection{Simple Estimate for $\beta_\mathrm{eff}$}
\label{sect:beta-mstar}

For a single Fourier harmonic, the character of the in-plane cooling depends on $\tau$ and $m$, but density waves are composed of multiple such harmonics. The density waves launched by the planet are initially dominated by perturbations with azimuthal wavenumber $m = m_* \approx (2h_\mathrm{p})^{-1}$ \citep{Miranda-Spirals}. In the Fiducial disk model, $m_* \approx 4$ -- $7$ for $r_\mathrm{p}$ in the range $10$ -- $100$ au (it is roughly in this range for our other disk models as well). 
To provide an idea of the behavior of $\beta_\mathrm{eff}(r)$, we can assume for simplicity that $\delta e_{m_*}$---the $m=m_*$ harmonic of internal energy---is the only non-zero $\delta e_m$ in equation~(\ref{eq:beta-eff}). In other words, we approximate $\beta_\mathrm{eff}$ with $\beta_{m*}$, defined as the value of $\beta_m$ (see equation~(\ref{eq:beta-m-2})) with $\tilde{k}_m l_\mathrm{rad}$ computed by setting $m = m_*$ in equation~(\ref{eq:kr-lrad}).

The dotted curves in Fig.~\ref{fig:beta-eff} show the behavior of $\beta_{m*}$ defined in such a way, and we see that in many cases $\beta_{m*}$ provides a reasonably good estimate for the full $\beta_\mathrm{eff}$. Typically the two timescales differ by at most factor of a few\footnote{More substantial deviations of $\beta_{m*}(r)$ from $\beta_\mathrm{eff}(r)$ visible in Fig.~\ref{fig:beta-eff}(c),(e),(g) are caused by the strong radiative damping of the $m = m_*$ harmonic of the perturbation, which invalidates our assumption of this Fourier mode dominating $\delta e_m$.}, which is small compared to the $3$ -- $4$ orders of magnitude over which $\beta_\mathrm{eff}$ intrinsically varies.

The approximation $\beta_\mathrm{eff} \approx \beta_{m*}$ simplifies our interpretation of the behavior of $\beta_\mathrm{eff}$ and allows one to make simple estimates of the role of in-plane cooling. When $\tau/m_* \gtrsim 1$, in-plane cooling is mostly diffusive, and far from the planet $\beta_{m*}$ reduces to
\be
\beta_{m*} \approx \frac{4}{\pi}h_\mathrm{p}^2 \beta_\mathrm{surf}\left(\frac{\Omega_\mathrm{p}}{\Omega} - 1\right)^{-2},
\label{eq:beta-mstar}
\ee
see equation (\ref{eq:beta-m-tauggm}). In other words, in this limit the cooling timescale $\beta_{m*}$ for a mode with $m \approx m_*$ is smaller than $\beta_\mathrm{surf}$ by a factor $\sim m_*^{-2}\sim h_\mathrm{p}^2$. Thus, $\beta_{m*}$ is least an order of magnitude smaller than $\beta_\mathrm{surf}$ in this regime, in agreement with Fig.~\ref{fig:beta-eff}. 

On the other hand, when $\tau/m_* \lesssim 1$, in-plane cooling operates in the streaming regime, hence $\beta_{m*}$ is given by equation (\ref{eq:beta-m-taullm}). The location at which $\tau = m_*$, where the transition between these two regimes occurs, is indicated by the arrows at the top of each panel in Fig.~\ref{fig:beta-eff}.

\subsubsection{Dependence of $\beta_\mathrm{eff}$ on Disk Parameters}
\label{sect:beta_eff-pars}

Next we discuss how the effective cooling timescale $\beta_\mathrm{eff}$ changes as we vary the parameters of the disk model away from their values in the Fiducial model. 

Examination of Fig.~\ref{fig:beta-eff} reveals that $\beta_\mathrm{eff}$ is very sensitive to the disk temperature (or aspect ratio $h$; see panels (g),(h)). For the Fiducial disk model ($h_{50\mathrm{au}} = 0.1$), the minimum value of $\beta_\mathrm{eff}$ is $\approx 10^{-3}$, while for the Cold model ($h_{50\mathrm{au}} = 0.07$), the minimum value is $\approx 0.05$, whereas for the Hot model ($h_{50\mathrm{au}} = 0.15$), the minimum value drops to $\approx 10^{-4}$. This dramatic variation stems from the fact that the minimum of $\beta_\mathrm{eff}$ is typically attained at the transition to free-streaming regime of the in-plane cooling (at $\tau \approx m_*$). In the free-streaming limit of in-plane cooling, according to equation (\ref{eq:beta-m-taullm}), $\beta_\mathrm{eff}$ depends primarily on the timescale for surface cooling in the optically thin limit $\beta_{\mathrm{surf},0}$ (with only weak additional dependence on $\tau$), which has a very steep scaling with the disk aspect ratio $h$, $\beta_{\mathrm{surf},0} \propto h^{-10}$, see equation (\ref{eq:beta-surf-0}). Thus, a factor of two variation in $h$ corresponds to a factor of $10^3$ variation of $\beta_\mathrm{eff}$, as we see in Fig.~\ref{fig:beta-eff}(g),(h). In the diffusive, optically thick limit, the temperature dependence is weaker with $\beta_\mathrm{surf} \sim \beta_{\mathrm{surf},0} \tau^2 \propto h^{-2}$, see equation (\ref{eq:beta-m-tauggm}).\footnote{Note that equation (\ref{eq:beta-mstar}) then predicts $\beta_{m*}$ to not depend on $h$, but Fig.~\ref{fig:beta-eff}(g),(h) does show that $\beta_\mathrm{eff}$ varies with $h$ in the diffusive regime. This shows that in these disk models the modes with $m = m_*$ are strongly damped radiatively and do not dominate $\delta e_m$.}

Varying the disk mass (normalization of $\Sigma$) also has an effect on the profile of $\beta_\mathrm{eff}$. Varying the disk mass shifts the $\tau = m_*$ radius relative to the Fiducial model. Also, there is an effect on $\beta_\mathrm{eff}\propto \beta_\mathrm{surf}$ through the $\Sigma$ dependence of $\tau$ in $\beta_\mathrm{surf}\propto f(\tau)$. In the optically thick region (inside of the $\tau = m_*$ radius), $\beta_\mathrm{eff}\propto \tau^2\propto \Sigma^2$, while in the optically thin region it is independent of $\Sigma$ (as $f(\tau)\to 1$). Hence, in the Low Mass model (Fig.~\ref{fig:beta-eff}(d)), the $\tau = m_*$ radius is shifted inwards relative to the Fiducial model, and the cooling timescale is reduced inside of this radius. Conversely, in the High Mass model (Fig.~\ref{fig:beta-eff}(e)), the $\tau = m_*$ radius is shifted outwards, and the cooling timescale is increased inside of this radius. This has the effect of increasing $\beta_\mathrm{eff}$ in most of the disk ($r \lesssim 100$ au).

Varying the opacity affects $\beta_\mathrm{eff}$ similarly to varying the disk mass. Note that $\beta_\mathrm{eff}\propto \beta_{\mathrm{surf}}$ scales as $\bar{\kappa}_{\mathrm{d},0}^{-1}$ in the optically thin region, whereas it is proportional to $\bar{\kappa}_{\mathrm{d},0}$ in the optically thick limit. In the Low Opacity model, (Fig.~\ref{fig:beta-eff}(b)), the $\tau = m_*$ radius is shifted inwards relative to the Fiducial model (since $\tau \propto \bar{\kappa}_{\mathrm{d},0}$). The cooling timescale is correspondingly reduced inside of this radius and increased outside of it. On the other hand, in the High Opacity model (Fig.~\ref{fig:beta-eff}(c)), the $\tau = m_*$ radius is shifted outwards, to about $100$ au. This has the effect of making most of the disk optically thick, hence increasing $\beta_\mathrm{eff}$ in this region.

Finally, varying the profile of either the surface density or temperature (Figs.~\ref{fig:beta-eff}(f),(i)) has a fairly minor effect on $\beta_\mathrm{eff}$. Only the slope of $\beta_\mathrm{eff}(r)$ in the optically thin part of the disk changes as $T(r)$ varies, in agreement with equation (\ref{eq:beta-surf-0}). 

\subsubsection{Implications for Disk Evolution}
\label{sect:expect}

In order to better understand the implications of the $\beta_\mathrm{eff}$ profiles in Fig.~\ref{fig:beta-eff} for the disk evolution results presented further in \S \ref{sect:results}, we briefly review the impact of cooling on the dynamics of planet-driven density waves and the associated global evolution of the disk, as described in \citet{Miranda-Cooling}. Note that these results are based on a constant-$\beta$ framework.

\begin{figure}
\begin{center}
\includegraphics[width=0.85\textwidth,clip]{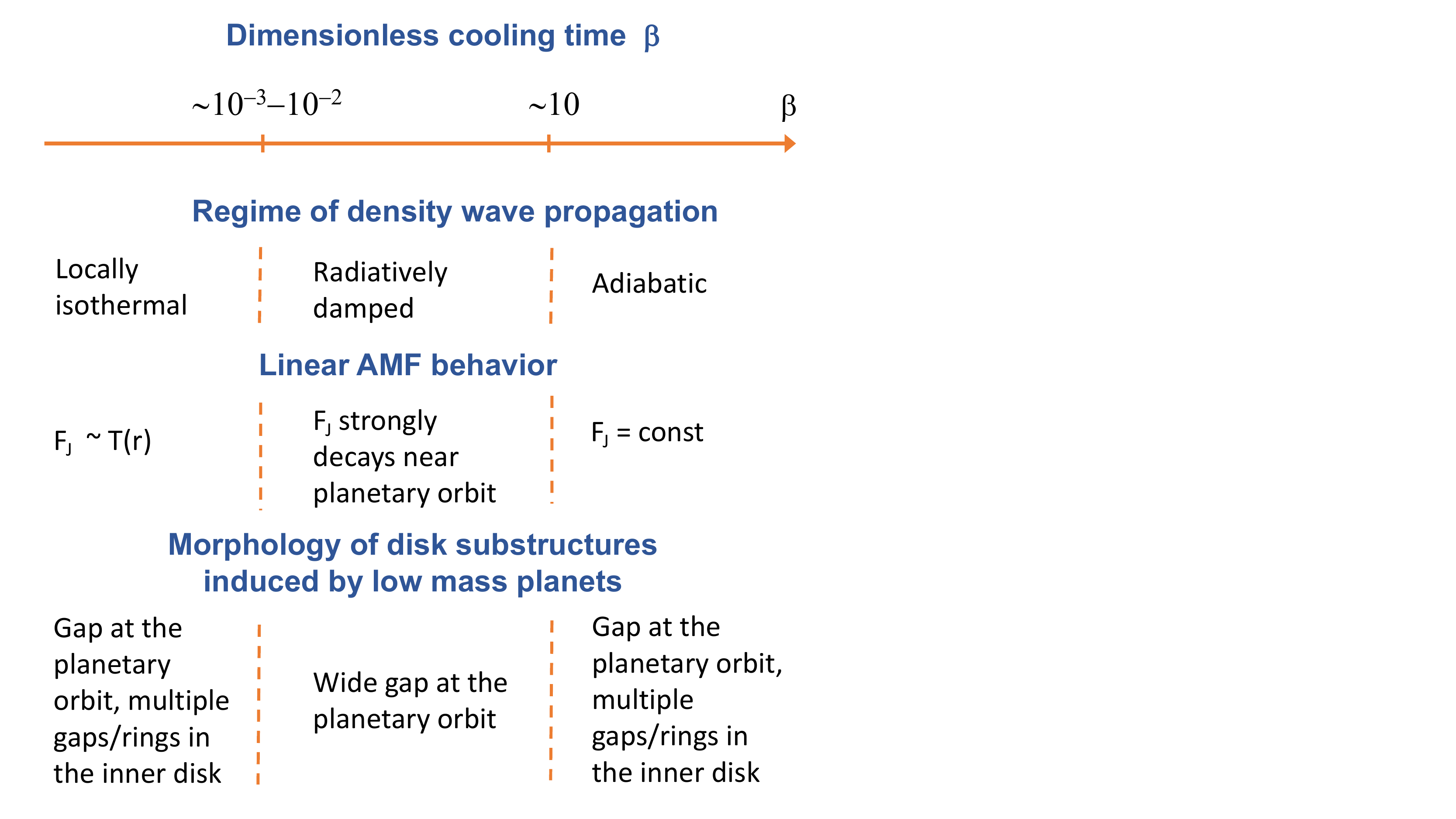}
\caption{Schematic classification of different cooling regimes, parameterized by the dimensionless cooling timescale $\beta$, and their impact on the propagation of the planet-induced density waves and production of different types of annular structures in protoplanetary disks. See \S \ref{sect:expect} for details.}
\label{fig:beta-outcomes}
\end{center}
\end{figure}

The evolution of density waves falls into one of three regimes based on the value of $\beta$, which can be understood through the behavior of the AMF of linear (low-amplitude) waves. These regimes are schematically summarized in Fig.~\ref{fig:beta-outcomes}. For slow cooling, with $\beta \gtrsim 10$, linear waves propagate adiabatically, conserving their AMF. For fast cooling, with $\beta \lesssim 0.01$\footnote{The formal limiting value of $\beta$ for the locally isothermal regime is $\approx h_\mathrm{p}^3$ for linear waves, and roughly an order magnitude larger for moderately nonlinear waves (i.e., launched by a planet whose mass is a moderate fraction of the thermal mass).}, waves propagate in the locally isothermal regime, in which the AMF scales with the local disk temperature \citep{Miranda-ALMA,Miranda-Cooling}. Since typically the temperature decreases with $r$, this means that inward-propagating waves become stronger as they travel inward in this regime, and weaker as they travel outward. Finally, for intermediate values of $\beta\sim (10^{-2}$ -- $10)$, density waves experience strong linear damping as a result of cooling, losing AMF as they propagate. In this regime perturbations with azimuthal wavenumber $m$ experience the strongest damping when $\beta \sim 1/m$.

Gaps and rings emerge in protoplanetary disks as a result of dissipation of the planet-driven density waves, which transfer their angular momentum to the disk material. In both the adiabatic and locally isothermal regimes, the waves dissipate by developing into shocks. This leads to a series of narrow gaps in the disk interior to the planet, each corresponding to the shock locations of different spiral arms into which the density wave splits \citep{Bae2017,Miranda-Spirals}. The structures produced in the adiabatic and locally isothermal regimes are similar in shape, but deeper gaps are formed in the locally isothermal case \citep{Miranda-ALMA}. For intermediate cooling timescales, strong radiative damping transfers the wave AMF to the disk material close to the planetary orbit. As a result, only a single wide gap forms around the orbit of the planet, rather than the series of narrow gaps emerging in adiabatic or locally isothermal disks \citep{Miranda-Cooling,Zhang2020,Ziampras2020b}.

Combining this understanding with the profiles of $\beta_\mathrm{eff}$ in Fig.~\ref{fig:beta-eff}, we would expect density waves to experience strong local radiative damping in our Fiducial disk model for planets with small orbital radii ($r_\mathrm{p} = 10$ au and $20$ au), for which $\beta_\mathrm{eff}\sim 10^{-1}$ -- $1$ for $r<r_\mathrm{p}$. On the other hand, for larger $r_\mathrm{p} = 50$ au and $100$ au, $\beta_\mathrm{eff}\lesssim 10^{-2}$ is small enough that waves may propagate far from the planet in the nearly locally isothermal regime, experiencing minimal cooling-related damping. 

This qualitative conclusion can also be drawn for most of the other disk models shown in Fig.~\ref{fig:beta-eff}, with one exception. In the Cold disk model (Fig.~\ref{fig:beta-eff}(e)), $\beta_\mathrm{eff}\sim (0.1$ -- $10)$ everywhere, for all the different planetary orbital radii considered. Therefore, radiative damping is almost always dominant for this disk model, which should lead to a single gap near the planet and not much structure in the inner disk.

We note that these expectations would be very different if in-plane cooling were neglected. Then cooling would operate on a longer timescale, falling in the range $\beta_\mathrm{surf}\sim (0.1$ -- $10)$ (dashed black curve in each panel of Fig.~\ref{fig:beta-eff}) for most of the disk in all models considered. In this intermediate cooling regime, radiative damping would play a crucial role in wave dissipation and disk evolution almost universally.

We emphasize that these interpretations depend on the detailed, non-trivial profile of $\beta_\mathrm{eff}$, which itself serves only to give an approximate description of the role of cooling on the density waves. It is critical to consider the fact that each Fourier harmonic $\delta e_m$ is thermally relaxed on its own individual timescale $\beta_m$, which is done next.

\section{Numerical Setup and Methods}
\label{sect:setup}

Now we describe the numerical setup and methods used in our calculations of the disk evolution due to planet-driven waves with radiative cooling. We compute a variety of evolutionary and observational diagnostics for each of the disk-planet models listed in Table~\ref{tab:parameters}. The disk temperature and surface density profiles are given by equations~(\ref{eq:T-profile})--(\ref{eq:surfdens-profile}), which represent the initial conditions in our numerical simulations. The dust opacity due to small dust grains is described by the opacity law (\ref{eq:opac-law}). Both cooling from the disk surface and in-plane cooling are considered in all models except for the Surface Cooling Only model, for which it is turned off (see the last column of Table~\ref{tab:parameters}). This allows us to assess the importance of properly including in-plane cooling for understanding density wave dynamics.

\begin{figure*}
\begin{center}
\includegraphics[width=0.99\textwidth,clip]{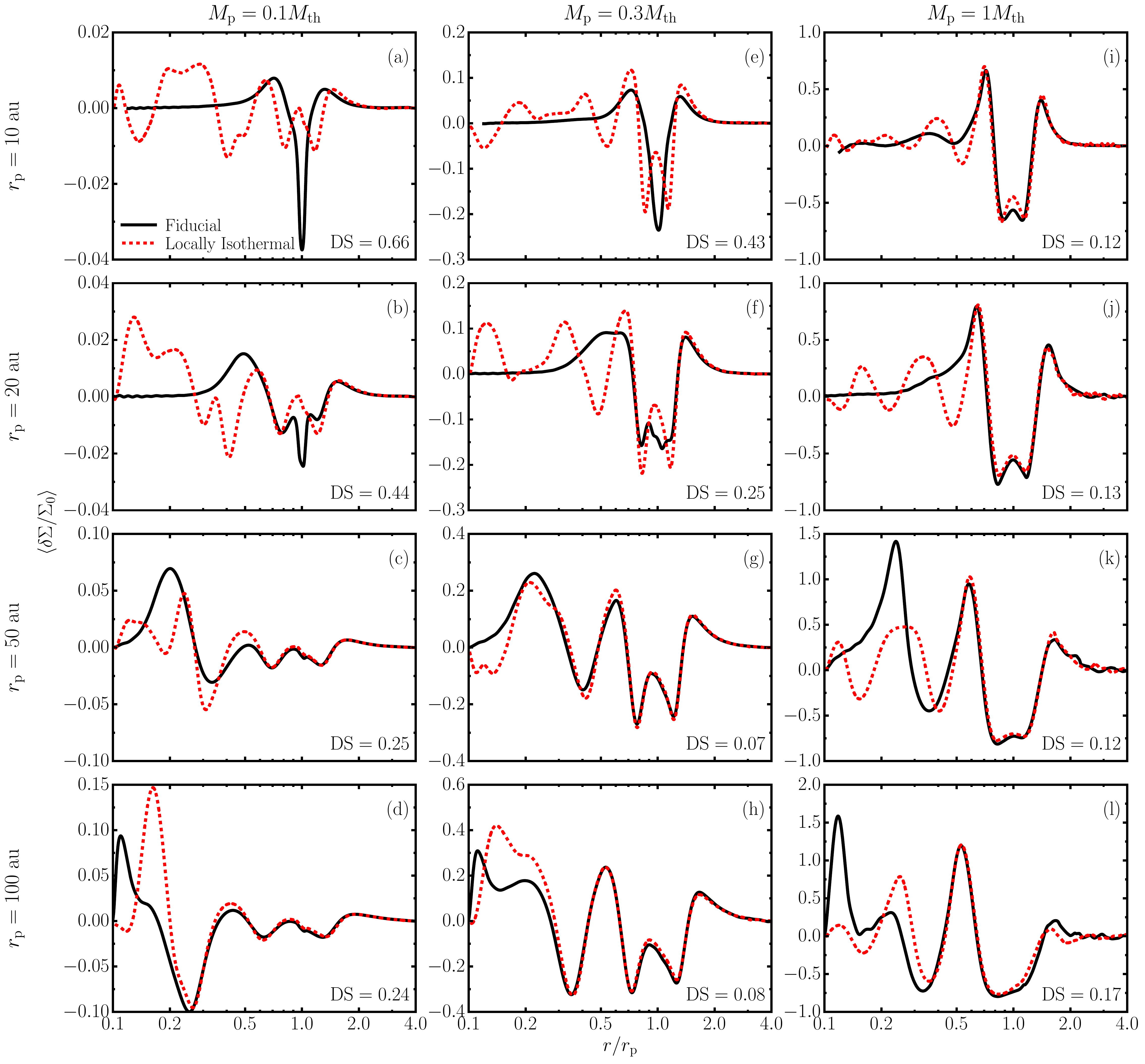}
\caption{Azimuthally averaged gas surface density perturbation $\delta\Sigma = \Sigma - \Sigma_0(r)$, relative to the initial surface density $\Sigma_0(r)$, for planets with different masses $M_\mathrm{p}$ (in terms of the thermal mass $M_\mathrm{th}$, see equation~(\ref{eq:Mth}); different columns) and orbital radii $r_\mathrm{p}$ (different rows) at $500$ orbits. Solid black lines correspond to the results of simulations with cooling, and dashed red lines to the results of locally isothermal simulations. The discrepancy score (see Section~\ref{sect:discrep}) for the pair of profiles is given in the lower right corner of each panel.}
\label{fig:profiles_fiducial}
\end{center}
\end{figure*}

\begin{figure*}
\begin{center}
\includegraphics[width=0.99\textwidth,clip]{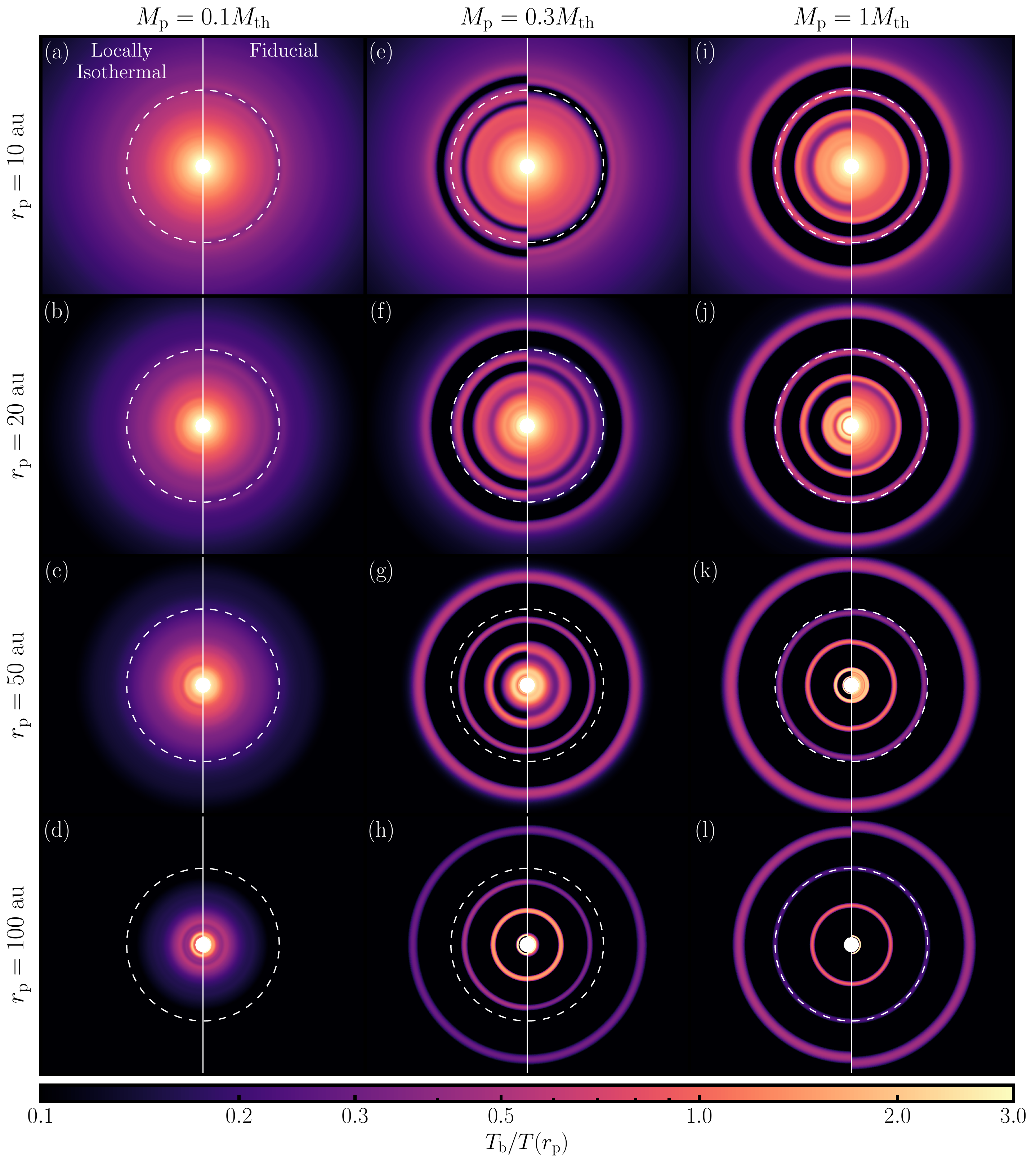}
\caption{Maps of dust continuum emission at $1.25$ mm for $1$ mm dust particles. The brightness temperature $T_\mathrm{b}$ is shown in terms of the gas temperature at the orbital radius of the planet, $T(r_\mathrm{p})$. Different columns correspond to different planet masses $M_\mathrm{p}$, expressed in terms of the thermal mass $M_\mathrm{th}$, and different rows correspond to different planetary orbital radii $r_\mathrm{p}$. Each panel is divided into two subpanels by a vertical white line, with the emission map derived from a locally isothermal hydrodynamical simulation on the left side, and the emission map for a simulation with cooling, using the Fiducial disk model, on the right side. The dashed white circle in each panel indicates the orbit of the planet.}
\label{fig:images_fiducial}
\end{center}
\end{figure*}

\subsection{Linear Calculations}
\label{sect:setup-linear}

As part of our calculations, we compute the linear response of the disk to an orbiting planet. This is used for several different purposes, including calculating the effective cooling timescale (for which we compute the linear $\delta e_m$, see Section~\ref{sect:tc-eff}), validating the treatment of cooling in our numerical simulations (see Appendix~\ref{sect:cooling-test}), and understanding the role of cooling in density wave dissipation (decay of the AMF, see Section~\ref{sect:amf}). We forgo a detailed description of these linear solutions here, and refer the reader to \citet{Miranda-Cooling} for a description of the equations that are solved, and \citet{Miranda-Spirals} for the numerical solution method.

The key modification to the approach described in these previous works is the specific form of the cooling timescale $\beta_m(r)$, which is given by equation~(\ref{eq:beta-m-2}) for each Fourier harmonic of the perturbation. The formalism of \citet{Miranda-Cooling} is easily generalized from the constant-$\beta$ formulation used in that work, as the perturbation equations for each harmonic derived in that study permit an arbitrary radial profile for the cooling timescale.

\subsection{Hydrodynamical Simulations}
\label{sect:hydrosims}

We also carried out a suite of numerical simulations of planet-disk interaction using \textsc{fargo3d} \citep{FARGO3d}. We use a logarithmic (in $r$) grid with a radial extent of $r_\mathrm{in} = 0.08 r_\mathrm{p}$ to $r_\mathrm{out} = 4 r_\mathrm{p}$. Wave damping \citep{deValBorro2006} is applied for $r < 0.1 r_\mathrm{p}$ and $r > 3.6 r_\mathrm{p}$. When presenting the results of our simulations we exclude the inner damping zone, so that $0.1 r_\mathrm{p}$ can be regarded as the effective location of the inner boundary.

The fiducial grid resolution is $N_r \times N_\phi = 1276 \times 2048$, resulting in $32 (h_\mathrm{p}/0.1)$ grid cells per scale height at $r_\mathrm{p}$. In simulations with a thinner or thicker disk, i.e., a smaller or larger value of $h_{50\mathrm{au}}$ (see Table~\ref{tab:parameters}), we use finer or coarser grid, so that the resolution in terms of grid cells per scale height is maintained for a given $r_\mathrm{p}$.

A softening length of $0.6 H_\mathrm{p}$ is applied to the gravitational potential of the planet. Its mass is gradually increased from zero to $M_\mathrm{p}$ over the first $10$ orbits to reduce transient effects. In each simulation we evolve the disk for $500$ orbits.

The key ingredient of our simulations is their inclusion of the radiative effects on density wave propagation. Since the code employed in this study does not include an explicit treatment of radiation transfer, we resort to an effective approach based on the theory developed in \S\ref{sect:cooling}. The details of our implementation of cooling effects are described in Appendix~\ref{sect:cooling-numerical}. There we also test our cooling module (see Appendix~\ref{sect:cooling-test}) by comparing results of simulations with cooling for low-mass planets ($M_\mathrm{p}=0.01 M_\mathrm{th}$) with semi-analytical linear calculations (see \S \ref{sect:setup-linear}), finding good agreement. This validates our implementation of cooling and allows us to extend its use for simulating disks with more massive planets.

For each set of disk/planet parameters, we run two simulations, one with an ideal equation of state (EoS) with $\gamma = 7/5$ and cooling (see Appendix~\ref{sect:cooling-numerical}), and another with a locally isothermal EoS for comparison. We compare the results of the simulations with different thermodynamics, assessing the validity of the locally isothermal approximation for each set of parameters (see Section~\ref{sect:results-iso}). Note that there are two parameters associated with cooling, $\Sigma_{50\mathrm{au}}$ and $\bar{\kappa}_{\mathrm{d},0}$, that are not involved in the locally isothermal simulations. Cooling simulations that differ only by the values of one or both of these parameters can therefore be compared to the same locally isothermal simulation.

\subsection{Dust Dynamics and Submillimeter Emission}

To facilitate comparison with observations, we produce maps of sub-mm dust emission for each simulation using a procedure described in Appendix \ref{sect:dust-maps}. To make the maps we follow the radial motion of large dust grains in the disk as it is perturbed by the dissipation of the planet-driven density waves. This is done by post-processing our hydrodynamical simulations, using an approximate 1D method described in Appendix~\ref{sect:dust-dyn}. 

Our post-processing approach allows us to explore different dust particles sizes with minimal computational cost. The results of a single hydro simulation can be used to quickly generate dust distributions for a variety of particle sizes. Since we perform a systematic exploration of the parameters that affect the disk cooling, with a large number ($\sim 100$) of simulations, the additional computational cost associated with a more self-consistent 2D gas $+$ dust treatment would be prohibitive.

Having determined the radial distribution of the dust, we then compute the maps of sub-mm thermal emission as described in Appendix~\ref{sect:sub-mm}. In most of the maps presented in this paper, we focus on a single particle size ($1$ mm). In Section~\ref{sect:particle-size}, we provide a limited exploration of the effect of varying the particle size. We find that a single dust size is sufficient for characterizing the critical cooling-related effects described in this work.

\section{Results}
\label{sect:results}

We now present our simulation results, focusing first on the gas and dust evolution (\S \ref{sect:results-fiducial}) and AMF behavior (\S \ref{sect:amf}) in the Fiducial disk model, and then moving on to other disk models (\S \ref{sect:pars-dep}). Our main focus will be on structures that form in the disk interior to planetary orbit, which past simulations suggest to be their dominant location. 

\subsection{Disk Structure: Fiducial Disk Model}
\label{sect:results-fiducial}

The disk structure produced by the planet in the Fiducial disk model is illustrated in Figs.~\ref{fig:profiles_fiducial}--\ref{fig:images_fiducial}. The solid black lines in Fig.~\ref{fig:profiles_fiducial} show the profiles of the azimuthally averaged gas surface density perturbation, $\delta\Sigma = \Sigma(r) - \Sigma_0(r)$, relative to the initial surface density profile $\Sigma_0(r)$, in runs with cooling at $500$ orbits, for planets with different masses and orbital radii. The corresponding dust continuum emission maps are shown in right halves of the panels in Fig.~\ref{fig:images_fiducial}. Note that the results for the corresponding locally isothermal simulations are shown as the red dashed lines in Fig.~\ref{fig:profiles_fiducial} and in the left halves of the panels in Fig.~\ref{fig:images_fiducial}; see Section~\ref{sect:results-iso} for a detailed discussion of these results.

The gas surface density profiles in Fig.~\ref{fig:profiles_fiducial} exhibit a varying number of gaps (surface density depletions or minima), as well as rings (surface density enhancements or maxima) between the gaps, depending on $M_\mathrm{p}$ and $r_\mathrm{p}$. The exact number of gaps is somewhat ill-defined, since adjacent minima of $\Sigma$ can to some extent merge into a single gap. In some cases, low-amplitude gap-like structures may be embedded in a larger gap structure. These caveats aside, the number of gaps ranges from one to as many as three or four. Alternatively, we can describe the disks as having less structure when there are fewer gaps, and more structure when there are more gaps, rather than explicitly quantifying the number of gaps.

As a general trend, smaller planet masses or smaller orbital radii result in fewer gaps or less structured disks, while larger planet masses or larger orbital radii result in a larger number of gaps or a more structured disk. The multiplicity of gaps is affected by which dissipation mechanism---cooling or shocks---dominates the evolution of the density waves launched by the planet. Typically, cooling-dominated dissipation produces fewer gaps and shock-dominated dissipation produces more gaps. For low-mass planets in the Fiducial disk model, cooling is dominant for planets with smaller orbital radii and nonlinear dissipation is dominant for larger orbital radii (see Section~\ref{sect:amf}).

For a $0.1 M_\mathrm{th}$ planet, the perturbations to the gas surface density at $500$ orbits (Figs.~\ref{fig:profiles_fiducial}(a)--(d)) are very weak, at the level of a few percent. Correspondingly, the resulting perturbations to the brightness temperature of the dust emission 
(Figs.~\ref{fig:images_fiducial}(a)--(d)) are also weak. They are just barely perceptible relative to the global radial variation of $T_\mathrm{b}$. For the more massive planets ($M_\mathrm{p} = 0.3 M_\mathrm{th}$ and $1 M_\mathrm{th}$), the perturbations are significant enough to produce order unity variations in $T_\mathrm{b}$, modifying the global morphology of the emission.

The dust emission maps largely reflect the same trends illustrated by the gas surface density profiles, with more pronounced secondary gaps (gaps that are not co-orbital with the planet) for larger planet masses and orbital radii. The most notable cases are the ones with $r_\mathrm{p} = 10$ au and $M_\mathrm{p} = 0.1 M_\mathrm{th}$ and $0.3 M_\mathrm{th}$, for which secondary gaps are completely suppressed, and there is only a single wide gap. This is because $\beta_\mathrm{eff}\sim 1$ at this $r_\mathrm{p}$ (see Fig.~\ref{fig:beta-eff}(a)), ensuring rapid deposition of the wave angular momentum into the disk material close to planetary orbit.

In disks with planets at larger orbital radii, the dust ends up concentrated into thinner rings, with wider, more empty gaps between them. The reason for this is the smaller overall gas surface density at larger disk radii, resulting in more mobile dust (large values of $\mathrm{St}$) for a given dust particle size. Dust is therefore able to more efficiently drift into pressure maxima (rings) and evacuate from regions between them (gaps).

\subsection{AMF}
\label{sect:amf}

\begin{figure}
\begin{center}
\includegraphics[width=0.49\textwidth,clip]{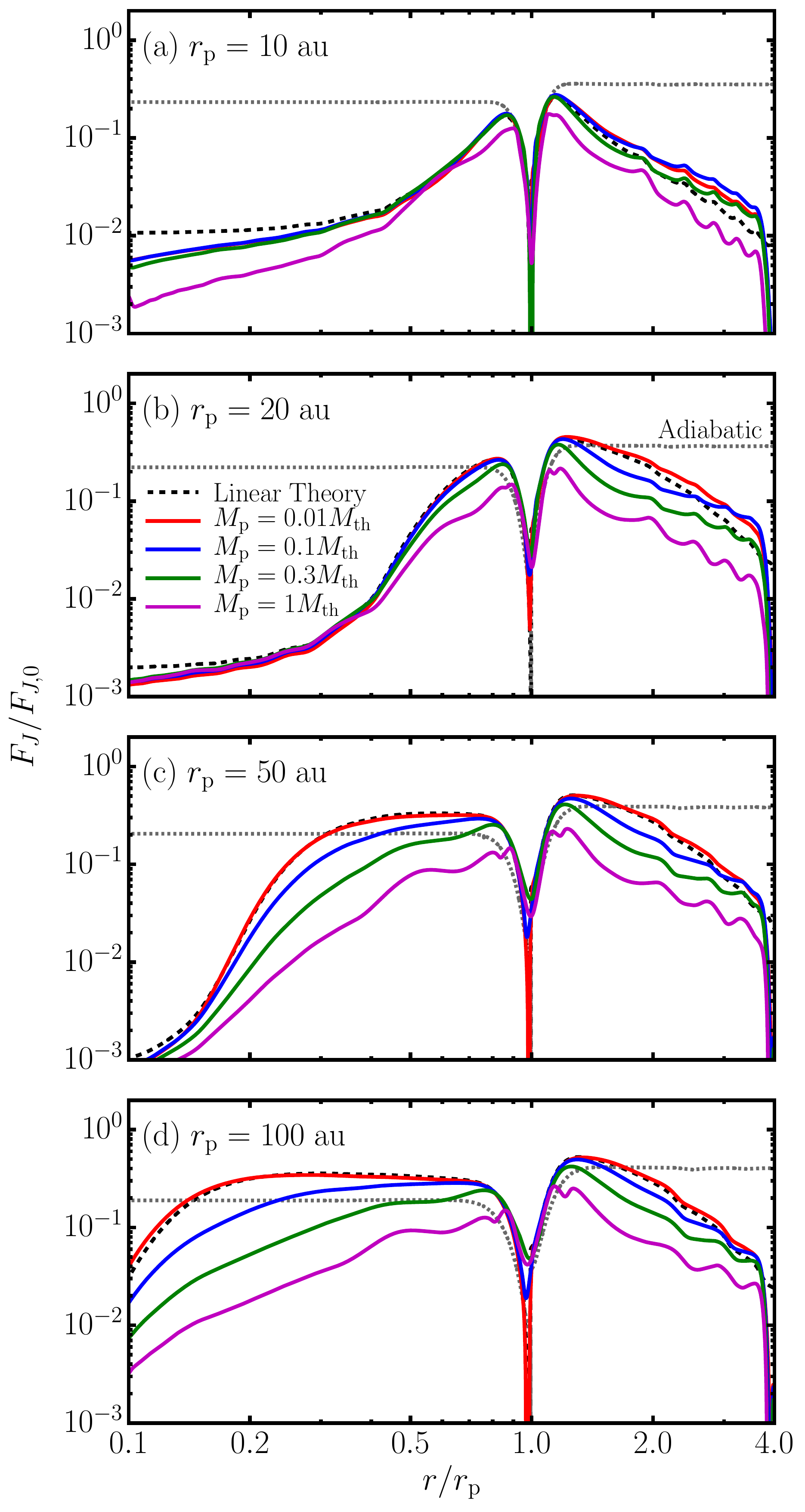}
\caption{Profiles of the angular momentum flux (AMF) $F_J$ (in terms of the characteristic scale $F_{J,0}$, see equation~(\ref{eq:fj0})) for planets with different masses and orbital radii, for the Fiducial disk model. The black dashed line shows the AMF for a low-mass planet computed using linear theory. The solid lines show the AMF from numerical simulations with planets of different masses, ranging from a very low-mass planet, $M_\mathrm{p} = 0.01 M_\mathrm{th}$ (red curves), for which the disk response is approximately linear, to a massive planet with $M_\mathrm{p} = 1 M_\mathrm{th}$ (magenta curves), for which it is strongly nonlinear. The dotted curve in each panel shows the linear AMF profile for an adiabatic (i.e., very slowly cooling) disk.}
\label{fig:amf}
\end{center}
\end{figure}

The AMF of the planet-driven density waves $F_J$ provides an important diagnostic for understanding the wave-driven evolution of the disk \citep{Miranda-Spirals,Miranda-ALMA}. This information is very useful for interpreting the results presented in Section~\ref{sect:results-fiducial}. Profiles of the AMF for planets with different masses and orbital radii, for the Fiducial disk model, are shown in Fig.~\ref{fig:amf}. The AMF is expressed in terms of the characteristic scale \citep{GT80}
\be
\label{eq:fj0}
F_{J,0} = \left(\frac{M_\mathrm{p}}{M_*}\right)^2 h_\mathrm{p}^{-3} \Sigma_\mathrm{p} r_\mathrm{p}^4 \Omega_\mathrm{p}^2,
\ee
associated with the total one-sided torque driving the density waves at Lindblad resonances.

We first examine the profiles of the linear AMF (black dashed curves in Fig.~\ref{fig:amf}), corresponding to the evolution of density waves produced by a very low-mass planet. In the absence of cooling, when the disk is adiabatic, the AMF is constant far from the planet, see the grey dotted lines in Fig.~\ref{fig:amf}. When cooling is present, the AMF falls off---or, if cooling is sufficiently rapid, grows (in the inner disk)---with distance from the planet. The difference between the black and grey curves is indicative of the extent to which cooling affects the wave evolution.

For $r_\mathrm{p} = 10$ au and $20$ au (Fig.~\ref{fig:amf}(a)--(b)), the linear AMF falls off sharply with distance from the planet, indicating that there is strong dissipation due to cooling. Indeed, Fig.~\ref{fig:beta-eff}(a) shows that in these cases $\beta_\mathrm{eff} \sim 0.1$ -- $1$ in the vicinity of the planet. Strong wave damping is typical for a cooling timescale in this range, see \S \ref{sect:expect}.

For planets at larger orbital radii, the linear AMF is either nearly constant ($r_\mathrm{p} = 50$ au; Fig.~\ref{fig:amf}(c)), or slightly grows with distance from the planet ($r_\mathrm{p} = 100$ au; Fig.~\ref{fig:amf}(d)) in the inner disk. This behavior is consistent with the fact that $\beta_\mathrm{eff} \sim 10^{-2}$ near the planet (see Fig.~\ref{fig:beta-eff}(a)), corresponding to the nearly locally isothermal limit. The AMF decreases in the outer disk, as this is the only behavior that can be produced by cooling in the outer disk, no matter how small the cooling timescale is.

The AMF from a numerical simulation with $M_\mathrm{p} = 0.01 M_\mathrm{th}$ follows the linear AMF very well for all of the cases in Fig.~\ref{fig:amf}. This validates our numerical implementation of cooling described in Appendix \ref{sect:cooling-numerical}. We have also verified convergence of these results with respect to resolution using higher resolution runs ($N_\phi = 4096$).

Comparing the AMF profile for more massive planets to the corresponding linear AMF profile allows us to quantify the importance of nonlinear dissipation in the evolution of the planet-driven density waves compared to cooling. For $r_\mathrm{p} = 10$ au and $20$ au, the numerical AMF is similar to the linear AMF even for $M_\mathrm{p} \geq 0.1 M_\mathrm{th}$. This indicates that linear dissipation due to cooling, rather than nonlinear dissipation, is the dominant driver of AMF evolution. But for $r_\mathrm{p} = 50$ au and $100$ au, this is not the case. Instead, the AMF decays substantially faster with the distance from the massive planet than the linear AMF, with larger deviations for more massive planets, indicating that nonlinear evolution plays a significant role in the AMF evolution. Therefore, cooling plays a more substantial role in the wave propagation when massive planets orbit at small radii than it does for larger $r_\mathrm{p}$.

As a rule of thumb, when radiative damping dominates the AMF evolution, the disk substructure is dominated by a single gap, but when nonlinear dissipation is dominant, multiple gaps tend to form \citep{Miranda-Cooling}. Our results demonstrate that cooling is dominant when (i) the (effective) cooling timescale is neither short enough nor long enough for linear waves to propagate without significant linear damping, and (ii) the planet mass is small enough for the waves it produces to be quasi-linear (although for small $r_\mathrm{p}$ cooling dominates even when $M_\mathrm{p}\sim M_\mathrm{th}$, which can be seen in the insensitivity of $F_J(r)$ curves to $M_\mathrm{p}$ variation in Fig.~\ref{fig:amf}(a)). Planets with smaller masses produce density waves with lower amplitudes, which are efficiently damped by radiation for intermediate values of the cooling timescale ($0.01 \lesssim \beta_\mathrm{eff} \lesssim 1$). The cooling timescale is roughly in this range in the inner disk ($r_\mathrm{p} \lesssim 50$ au), see Fig.~\ref{fig:beta-eff}(a). But in the outer disk, $\beta_\mathrm{eff} \lesssim 0.01$, so that density waves are not damped by cooling very efficiently (interior to the orbit of the planet) and propagate nearly in the locally isothermal limit. And for more massive planets (at any orbital radius), the larger amplitude of the density waves enhances the importance of nonlinear dissipation. The AMF profiles (Fig.~\ref{fig:amf}) are therefore consistent with the variation of the gap multiplicity with $M_\mathrm{p}$ and $r_\mathrm{p}$.

Fig.~\ref{fig:amf} also demonstrates that the ``initial amplitude'' of the AMF, i.e., its value just outside the wave excitation zone, varies with $r_\mathrm{p}$. This variation can be quantified by how much the initial AMF with cooling is different from the corresponding value in an adiabatic disk. For $r_\mathrm{p} = 10$ au (Fig.~\ref{fig:amf}(a)), the initial AMF is slightly larger in an adiabatic disk than in a disk with cooling. In this case, the density waves have already experienced significant damping before leaving the wave excitation zone. However, for larger orbital radii (Figs.~\ref{fig:amf}(b)--(d)), the initial AMF is larger in the presence of cooling than in an adiabatic disk. As described in \citet{Miranda-Cooling}, this is the result of the increased torque exerted on the disk when the cooling timescale is short (and the response of the disk is close to locally isothermal). This increased torque is partly responsible for the larger amplitudes of gaps caused by planets at larger radii, as seen in Fig.~\ref{fig:profiles_fiducial}.

\subsection{Dependence on Disk Parameters}
\label{sect:pars-dep}

\begin{figure*}
\begin{center}
\includegraphics[width=0.99\textwidth,clip]{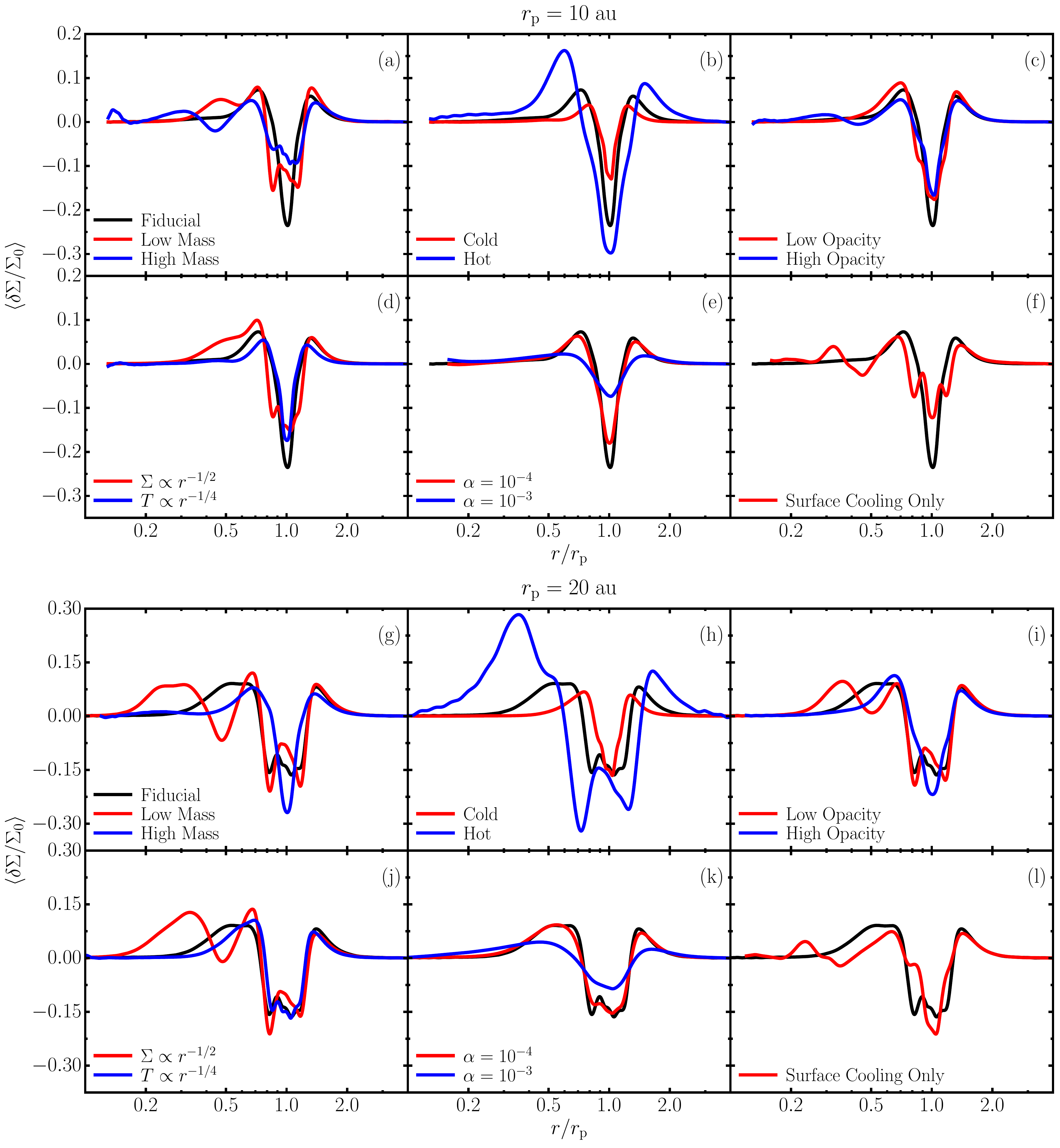}
\caption{Profiles of the gas surface density perturbation (as in Fig.~\ref{fig:profiles_fiducial}) for different disk models and a $0.3 M_\mathrm{th}$ planet with orbital radius $r_\mathrm{p} = 10$ au (panels (a)--(f)) and $20$ au (panels (g)--(l)).}
\label{fig:profiles_10au20au}
\end{center}
\end{figure*}

\begin{figure*}
\begin{center}
\includegraphics[width=0.99\textwidth,clip]{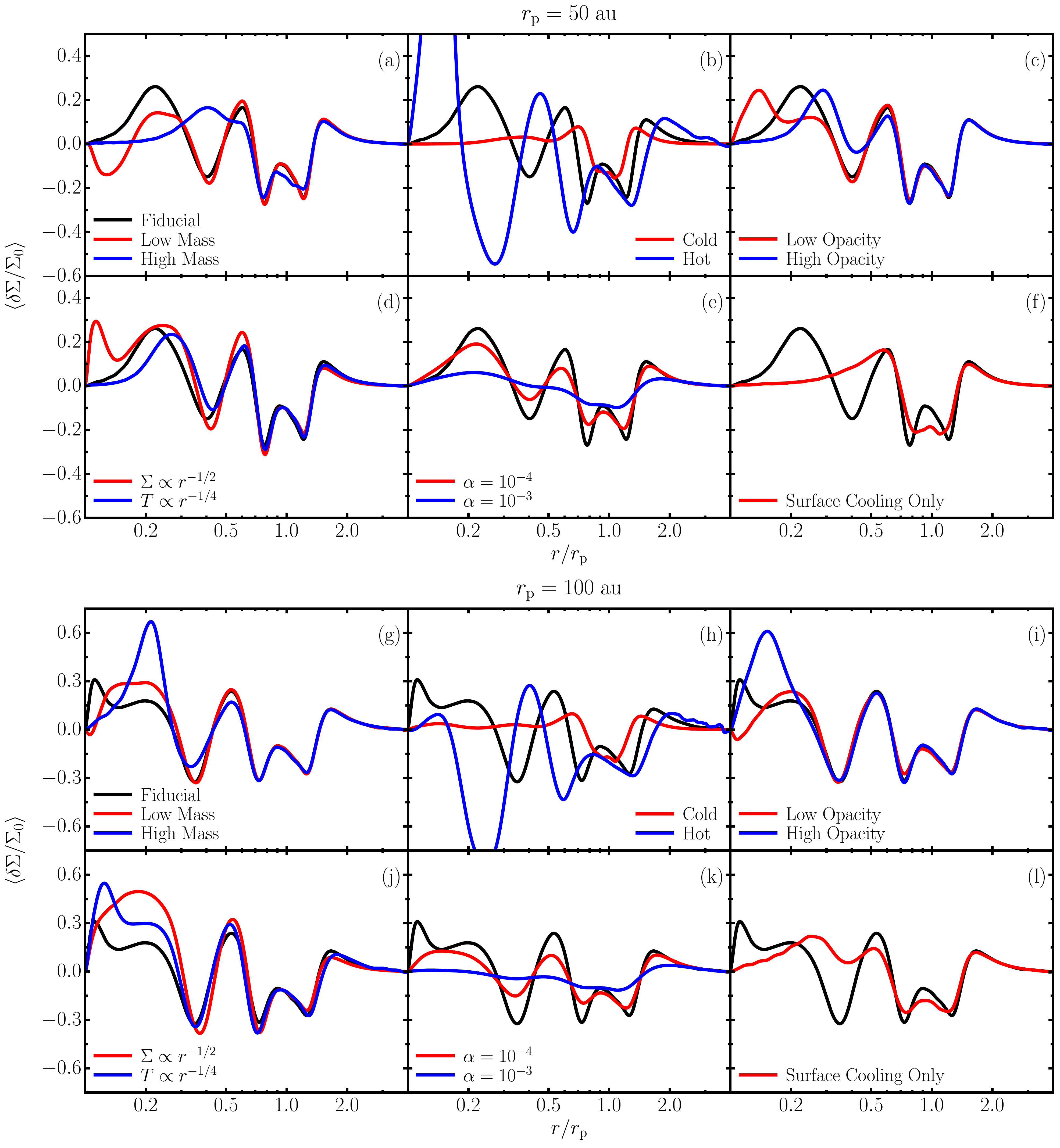}
\caption{Same as Fig.~\ref{fig:profiles_10au20au}, but for $r_\mathrm{p} = 50$ au (panels (a)--(f)) and $100$ au (panels (g)--(l)).}
\label{fig:profiles_50au100au}
\end{center}
\end{figure*}

\begin{figure*}
\begin{center}
\includegraphics[width=0.99\textwidth,clip]{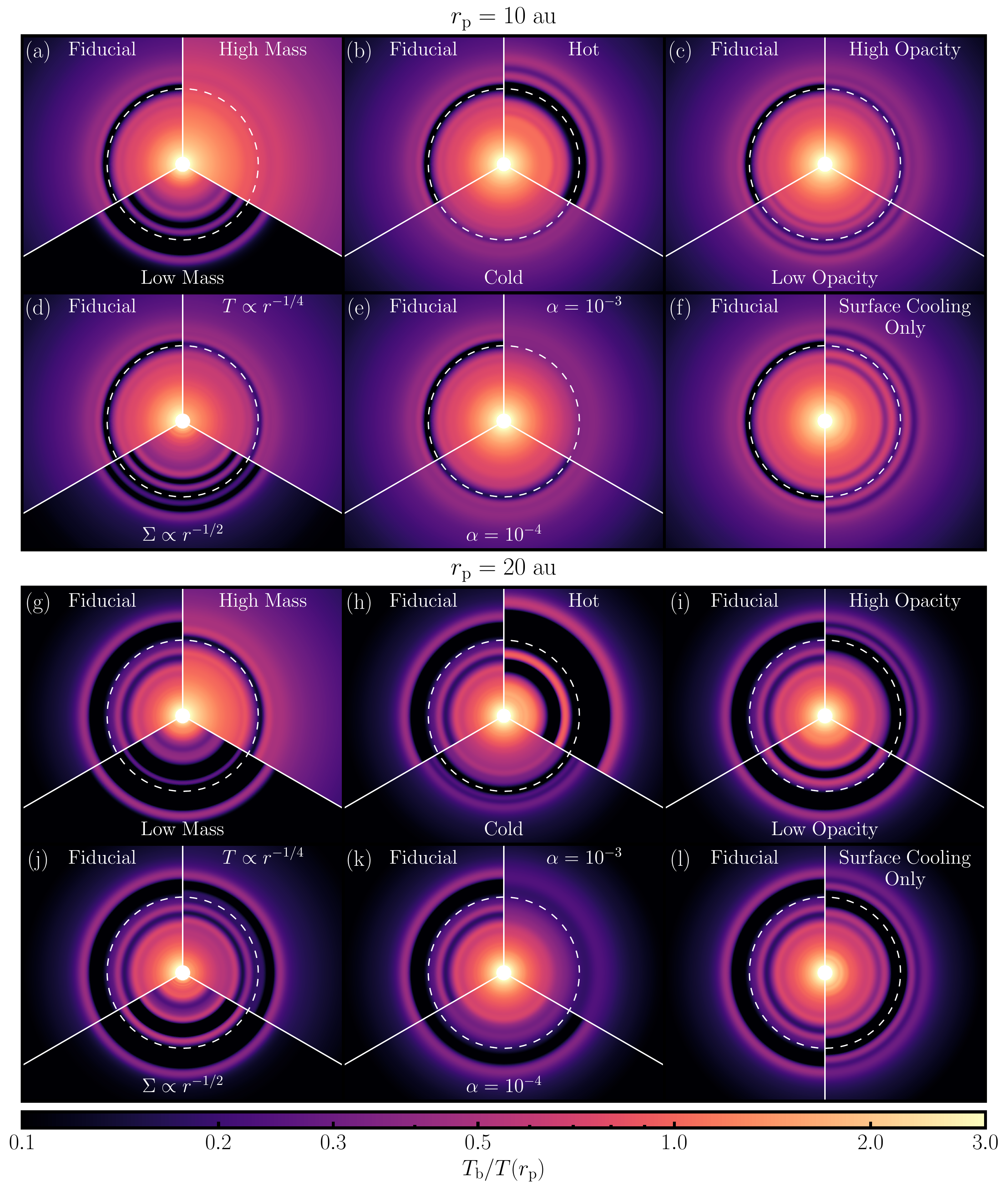}
\caption{Dust continuum emission maps, as in Fig.~\ref{fig:images_fiducial}, for a $0.3 M_\mathrm{th}$ planet with orbital radius $r_\mathrm{p} = 10$ au (panels (a)--(f)) and $20$ au (panels (g)--(l)). In each panel, the solid white radial lines divide the map into results for the Fiducial disk model, and results for one or more disk models with different parameters (see Table~\ref{tab:parameters}), to facilitate direct comparison with the Fiducial model.}
\label{fig:images_10au20au}
\end{center}
\end{figure*}

\begin{figure*}
\begin{center}
\includegraphics[width=0.99\textwidth,clip]{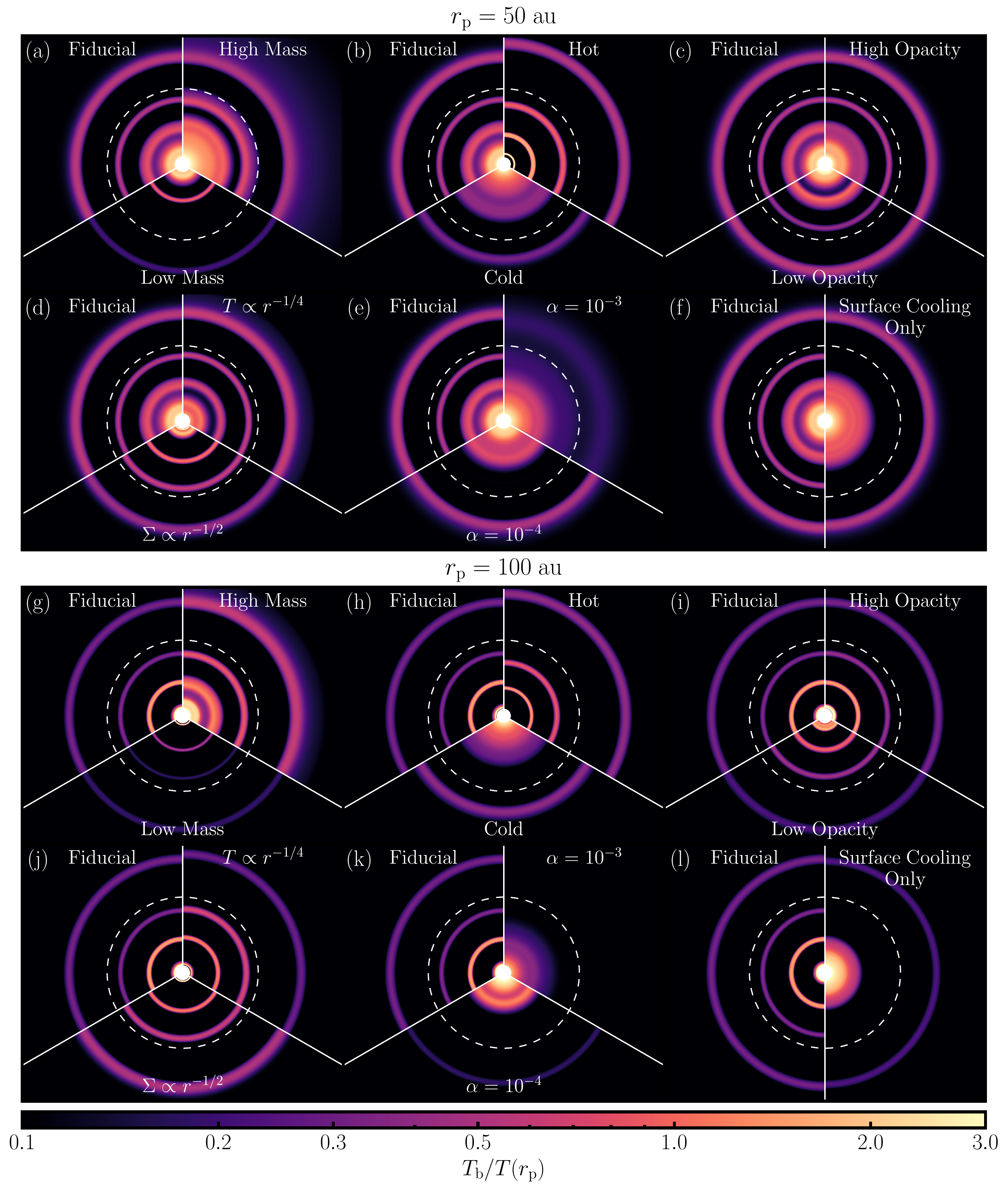}
\caption{Same as Fig.~\ref{fig:images_10au20au}, but for $r_\mathrm{p} = 50$ au (panels (a)--(f)) and $100$ au (panels (g)--(l)).}
\label{fig:images_50au100au}
\end{center}
\end{figure*}

Results for different disk models listed in Table~\ref{tab:parameters} are shown in Figs.~\ref{fig:profiles_10au20au}--\ref{fig:profiles_50au100au} (gas surface density perturbation profiles, analogous to Fig.~\ref{fig:profiles_fiducial}) and Figs.~\ref{fig:images_10au20au}--\ref{fig:images_50au100au} (dust emission maps, as in Fig.~\ref{fig:images_fiducial}). In each panel of these figures, the results (gas surface density or dust emission) are shown for the Fiducial disk model, as well as for other models with different parameters (see Table~\ref{tab:parameters}). In this subsection we describe the variation of the disk morphology associated with varying each of the disk parameters.

\subsubsection{Disk Mass}
\label{sect:M-var}

The effects of varying the disk mass (i.e., the surface density normalization $\Sigma_{50\mathrm{au}}$) are illustrated in panels (a) and (g) of Figs.~\ref{fig:profiles_10au20au}--\ref{fig:images_50au100au}. In the Low Mass model, the radial profile of $\Sigma$ is more structured for small planetary orbital radii $r_\mathrm{p} = 10$ au and $20$ au, as a result of a reduction of the cooling timescale (see Fig.~\ref{fig:beta-eff}(d)), which brings the wave dynamics closer to the locally isothermal regime in both the inner and outer disk. In the High Mass model, there is also more structure relative to the Fiducial model for $r_\mathrm{p} =10$ au. But now this is a result of the cooling timescale being raised into the nearly adiabatic regime in the vicinity of the planet ($\beta_\mathrm{eff}\sim 1$ -- $10$, see Fig.~\ref{fig:beta-eff}(e)), so that damping due to cooling is reduced. 

For $r_\mathrm{p} = 20$ au and $50$ au, the High Mass model produces less structure, i.e., less prominent secondary gaps relative to the Fiducial model. For disks with planets at these radii, $\beta_\mathrm{eff}\sim 0.01$ -- $0.1$, leading to cooling playing a more important role in density wave dissipation. For $r_\mathrm{p} = 100$ au, the disk structure is not too different from the Fiducial model, as in both cases the cooling timescale is short enough to be nearly in the locally isothermal regime.

In addition to the impact on the gas distribution, varying the disk mass makes the dust more or less mobile (by increasing or decreasing $\mathrm{St} \propto \Sigma^{-1}$). This results in more diffuse rings for more massive disks and sharper rings for less massive disks, which is evident in the dust emission maps. 

\subsubsection{Temperature}
\label{sect:T-var}

The effects of varying the disk temperature (or $h_{50\mathrm{au}}$) are shown in panels (b) and (h) of Figs.~\ref{fig:profiles_10au20au}--\ref{fig:images_50au100au}. In the Cold disk model (with $h_{50\mathrm{au}} = 0.07$), the planet clears a single gap in gas for $r_\mathrm{p} = 10$ au and $20$ au. The surface density profiles in these cases are qualitatively similar to the Fiducial model, differing only in the depth and detailed shape of the gap. For $r_\mathrm{p} = 50$ au and $100$ au, the Cold model also results in a single strong gap, with only very weak secondary gaps. This is in contrast to the strong multiple gap structure in the Fiducial model. This is a result of $\beta_\mathrm{eff}\sim 0.1$ -- $1$ throughout the disk for all $r_\mathrm{p}$ in the Cold model (Fig.~\ref{fig:beta-eff}(g)), which leads to density waves being strongly damped by cooling, rather than behaving nearly locally isothermally, and hence being damped by shocks.

In the Hot model, the planet produces one gap for $r_\mathrm{p} = 10$ au, as in the Fiducial and Cold models. But for larger orbital radii, there is additional structure within the gap (for $r_\mathrm{p} = 20$ au) or stronger secondary features ($r_\mathrm{p} = 50$ au and $100$ au). In these cases, the higher temperature results in a shorter cooling timescale almost everywhere in the disk ($\beta_\mathrm{eff}\lesssim 0.01$ for $r_\mathrm{p} = 50$ au and $100$ au, see  (Fig.~\ref{fig:beta-eff}(h)), so that density waves propagate in the locally isothermal regime. Thus, the increase in disk temperature leads to radiative wave damping playing a less important role in the disk evolution.

\subsubsection{Opacity}
\label{sect:kappa-var}

The dust opacity is varied in panels (c) and (i) of Figs.~\ref{fig:profiles_10au20au}--\ref{fig:images_50au100au}. In almost all of the cases with varied opacity, the surface density profile is qualitatively similar to in the Fiducial model. The exception is the Low Opacity model with $r_\mathrm{p} = 20$ au, in which there is a prominent secondary gap, where none is present in the Fiducial model. Overall, varying the opacity (by an order of magnitude) has a fairly minor effect on the planet-driven disk structure. This is because the profiles of $\beta_\mathrm{eff}$ are not very sensitive to the opacity variation, see Fig.~\ref{fig:beta-eff}(b),(c).

\subsubsection{Surface Density and Temperature Profiles}
\label{sect:slopes-var}

Results for disks with different surface density and temperature  profiles are shown in panels (d) and (j) of Figs.~\ref{fig:profiles_10au20au}--\ref{fig:images_50au100au}. The power law indices of the surface density and temperature profiles have a fairly minor effect on the structure of the disk. This is especially true for large $r_\mathrm{p} = 50$ au and $100$ au, as the $\delta\Sigma$ profiles for these models do not differ much from the Fiducial model, except at small radii in the inner disk ($r/r_\mathrm{p} \lesssim 0.4$).

For planets with small orbital radii ($r_\mathrm{p} = 10$ au and $20$ au), the $\delta\Sigma$ profile in the $T \propto r^{-1/4}$ model is fairly similar to the one in the Fiducial model. However, the case with a shallower surface density profile ($\Sigma \propto r^{-1/2}$) shows additional structure (i.e., secondary gaps) as compared to the Fiducial model.

\subsubsection{In-plane cooling}
\label{sect:no-in-plane}

The importance of the in-plane cooling is highlighted in panels (f) and (l) of Figs.~\ref{fig:profiles_10au20au}--\ref{fig:images_50au100au}, where we present results of hypothetical simulations with in-plane cooling artificially turned off. We find that whether or not the in-plane cooling is included has a significant impact on the gas surface density profile and dust emission map, in terms of the multiplicity, locations and amplitudes of rings and gaps, in all cases considered.

For example, for $r_\mathrm{p} = 10$ au, in the Fiducial model (including in-plane cooling), the disk profile is characterized by a single wide gap. However, without in-plane cooling, there is a more complex structure consisting of multiple gaps. This can be understood by comparing the value of the effective cooling timescale between the two cases, see Fig.~\ref{fig:beta-eff}(a). With in-plane cooling (red curve in Fig.~\ref{fig:beta-eff}(a)), $\beta_\mathrm{eff} \sim 1$ in the inner disk ($r < r_\mathrm{p}$), so that linear damping due to cooling dominates the disk evolution. In the absence of in-plane cooling, $\beta_\mathrm{eff}$ is instead equal to $\beta_\mathrm{surf}$ (black dashed curve in Fig.~\ref{fig:beta-eff}(a)), which is $\sim 10$ in the inner disk. In this case, planet-driven density wave behaves nearly adiabatically, so its AMF is not damped by cooling. The wave then has a chance to split into multiple spiral arms that eventually damp nonlinearly far from the planet, giving rise to disk substructure. This trend is also present for the case $r_\mathrm{p} = 20$ au (Fig.~\ref{fig:profiles_10au20au}(l)). Here again the density waves behave closer to adiabatically without in-plane cooling, resulting in a more complex gap structure.

However, the situation is different for larger planetary orbital radii, $r_\mathrm{p} = 50$ au and $100$ au. Here the $\delta\Sigma$ profile with only the surface cooling exhibits less structure than in the Fiducial model that includes in-plane cooling. In these cases, without in-plane cooling we have $\beta_\mathrm{eff} = \beta_\mathrm{surf} \sim 0.1$ in the inner disk, while with in-plane cooling, $\beta_\mathrm{eff} \sim 0.01$ (see the green and magenta curves in Fig.~\ref{fig:beta-eff}(a)). This indicates that, without in-plane cooling, radiative wave damping would be important, leading to less structure in the disk. However, proper inclusion of in-plane cooling pushes the effective cooling timescale down into the nearly locally isothermal regime. As a result, nonlinear dissipation becomes the primary driver of density wave evolution.

\begin{figure}
\begin{center}
\includegraphics[width=0.49\textwidth,clip]{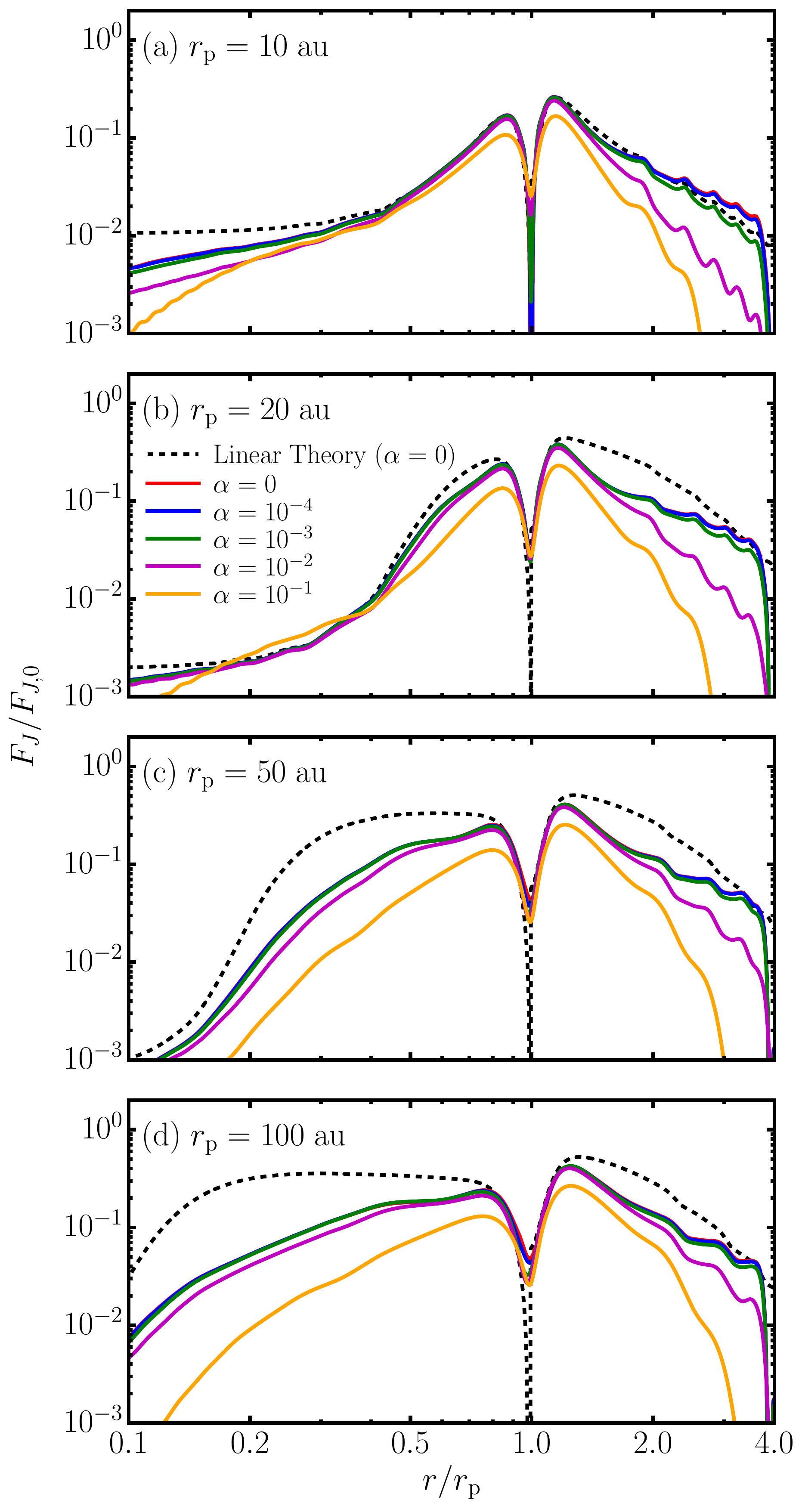}
\caption{AMF profiles for a $0.3 M_\mathrm{th}$ planet with different orbital radii, using the Fiducial disk model with different values of the disk viscosity parameter $\alpha$. The black dashed curve in each panel corresponds to the AMF predicted by linear theory for an inviscid disk.}
\label{fig:amf_alpha}
\end{center}
\end{figure}

\subsubsection{Viscosity}
\label{sect:nu-var}

All of the simulations discussed so far are inviscid, with a viscosity parameter $\alpha = 0$. To check how a non-zero viscosity can modify our results, we also performed several runs with the Fiducial model setup and a finite $\alpha$. The effect of a finite viscosity on the formation of disk substructure is explored in panels (e) and (k) of Figs.~\ref{fig:profiles_10au20au}--\ref{fig:images_50au100au}, which illustrate the results for the $\alpha = 10^{-4}$ and $\alpha = 10^{-3}$ models.

For the case of a small viscosity ($\alpha = 10^{-4}$), the surface density profiles are not qualitatively different than in the Fiducial (inviscid) case. The number of gaps and their locations are not substantially changed. However, the complex structure inside the primary gap that is present in the Fiducial model often gets smeared out with the addition of a small viscosity. Overall, the main effect of this small $\alpha = 10^{-4}$ is to somewhat suppress the depths of the gaps and height of the rings, which is best seen in Figs.~\ref{fig:profiles_50au100au}(e),(k).

But for a larger viscosity ($\alpha = 10^{-3}$), the changes are more dramatic: gap depths are further suppressed (very severely for planets at $r_\mathrm{p} = 50$ and $100$ au), even though their radial locations still remain approximately unchanged. Some hints of the multiple gap structure for planets with larger planetary orbital radii ($r_\mathrm{p} = 50$ and $100$ au) are still evident in the gas surface density profiles, although with a drastically reduced amplitude. However, these perturbations are so weak that they are not discernible in the dust emission maps, which primarily exhibit a single wide gap structure.

In order to interpret these results, we note that introduction of a finite viscosity has two distinct effects on the evolution of the disk. First, it leads to viscous dissipation of density waves \citep{Takeuchi1996}. This dissipation, like the dissipation associated with cooling, is a linear process. Hence in a viscous disk, dissipation generally occurs through three distinct channels---radiative, viscous, and nonlinear. By modifying the wave dissipation, viscosity has the potential to alter the associated gap structure. Second, the viscous spreading of the disk tends to fill in gaps and wipe out rings, thus erasing radial structure. In the classical scenario of gap opening in a viscous disk, this spreading inhibits gap formation for planets below a threshold mass \citep{R02b}.

To disentangle these two ways in which viscosity can affect disk substructure, we examine the effect of a non-zero $\alpha$ on the profile of the wave AMF, which allows us to isolate density wave dissipation from viscous spreading. In Fig.~\ref{fig:amf_alpha} we show the profiles of the wave AMF (as in Fig.~\ref{fig:amf}), taken at $20$ orbits, for a $0.3 M_\mathrm{th}$ planet, for disks with different values of the viscosity parameter $\alpha$. In addition to the cases with $\alpha = 10^{-4}$ and $10^{-3}$, for which the gas profiles and dust emission maps are shown in Figs.~\ref{fig:profiles_10au20au}--\ref{fig:images_50au100au}, two cases with larger values, $\alpha = 10^{-2}$ and $10^{-1}$, are also shown. While these large values of $\alpha$ are not expected to be realized in protoplanetary disks, they are illustrative of the effect of viscous damping on planet-driven density waves.

One can see that the AMF profiles for $\alpha = 10^{-4}$ and $10^{-3}$ show essentially no difference relative to the profiles for $\alpha = 0$. Therefore, we see that viscous damping is completely unimportant for the evolution of the wave AMF---relative to the damping associated with cooling and nonlinear evolution---for $\alpha \leq 10^{-3}$. It is only for much larger values of $\alpha$ ($\gtrsim 10^{-2}$) that the AMF profiles begin to differ substantially from the inviscid profiles. We note that the relative importance of viscous damping will vary with the planet mass (which should be explored in future work), as a result of the varying importance of nonlinear damping (see Fig.~\ref{fig:amf}), with which it competes. However, at least for the case of $M_\mathrm{p} = 0.3 M_\mathrm{th}$ considered here, we can conclude that the differences in the gas and dust profiles seen in Figs.~\ref{fig:profiles_10au20au}--\ref{fig:images_50au100au} are entirely due the second effect described above---the filling in of the gaps due to viscous spreading---rather than due to viscous wave damping.

\subsection{Dust Particle Size}
\label{sect:particle-size}

\begin{figure*}
\begin{center}
\includegraphics[width=0.99\textwidth,clip]{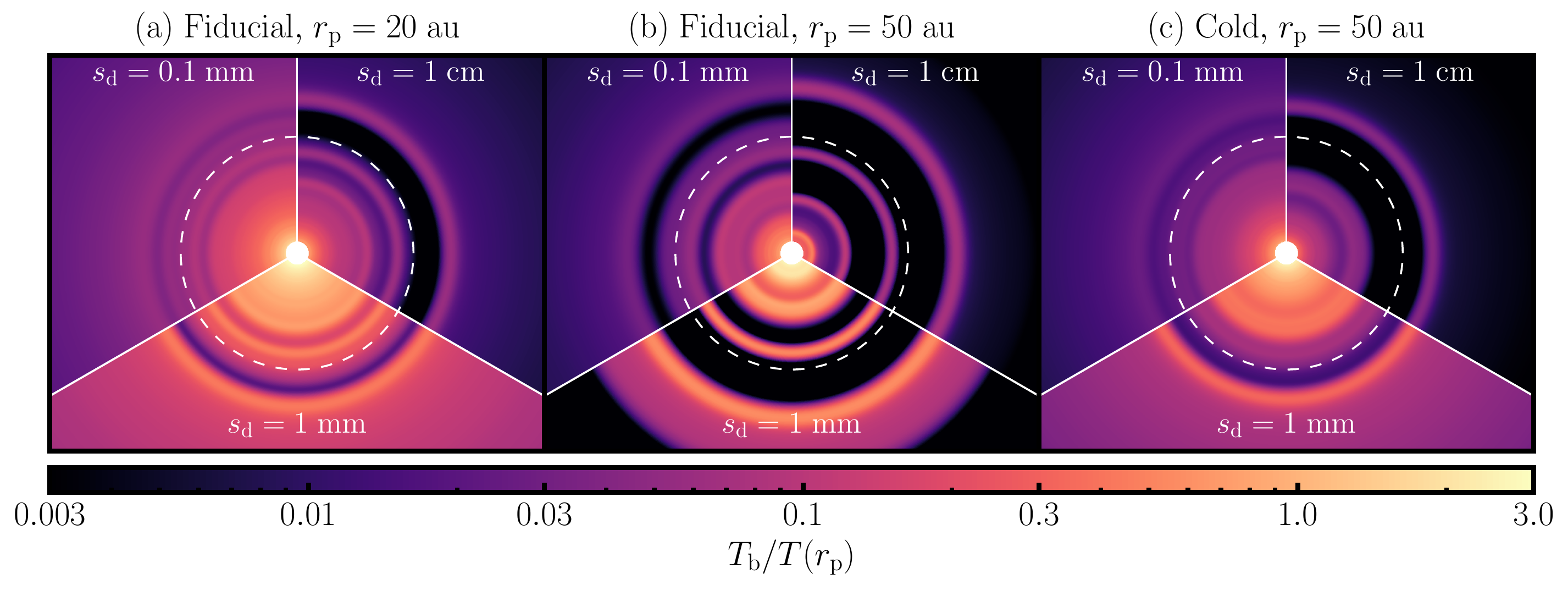}
\caption{Dust emission maps for three different dust particle sizes ($s_\mathrm{d} = 0.1$ mm, $1$ mm, and $1$ cm; as indicated in the different subsections of each panel), for the Fiducial disk model with $r_\mathrm{p} = 20$ au and 50 au (panels (a)--(b)), and for the Cold disk model with $r_\mathrm{p} = 50$ au (panel (c)).}
\label{fig:dust-size}
\end{center}
\end{figure*}

The dust emission maps presented in Sections~\ref{sect:results-fiducial} and \ref{sect:pars-dep} were computed using a single dust particle size of $s_\mathrm{d}=1$ mm. To examine the effect of varying $s_\mathrm{d}$, in Fig.~\ref{fig:dust-size} we present emission maps for three different particle sizes ($0.1$ mm and $1$ cm, in addition to the fiducial $1$ mm) for several disk model/planetary orbital radius combinations. Note that in Fig.~\ref{fig:dust-size}, the dynamic range of $T_\mathrm{b}$, spanning three orders of magnitude, is significantly larger than in the previous emission maps (Figs.~\ref{fig:images_fiducial}, \ref{fig:images_10au20au}--\ref{fig:images_50au100au}). This is necessary in order to highlight the details of the maps for $s_\mathrm{d} = 0.1$ mm and $1$ cm, which are significantly fainter than for $s_\mathrm{d} = 1$ mm. This is the result of the smaller opacity (at a wavelength of $1.25$ mm) for $0.1$ mm and $1$ cm particles (about $4$ and $13$ times smaller than for $1$ mm particles, respectively). 
Fig.~\ref{fig:dust-size} demonstrates that the ring/gap structure for a particular disk model and planetary orbital radius is more or less consistent across different particle sizes. The number of gaps and rings, as well as their approximate locations, do not vary much with $s_\mathrm{d}$. Only the widths of the features vary, with rings becoming narrower and gaps becoming wider (and the contrast of ring/gap pairs increasing correspondingly) for larger particles. Additionally, for larger particles there is a smaller amount of dust that is co-orbital with the planet, as a result of the dust being more mobile and hence less susceptible to trapping in the weak pressure bump near the planet. Aside from these details, there is a great deal of resemblance between the maps for different particle sizes (for the same disk model/planetary orbital radius, i.e., within one panel of Fig.~\ref{fig:dust-size}). This reflects the fact that the distributions of dust with different particles sizes are determined by the same underlying gas distribution in each case.

We also see that qualitative differences between maps for {\it different} disk models and planetary orbital radii can be found for all three particle sizes shown. For example, the suppression of the complex multiple gap structure seen in Fig.~\ref{fig:dust-size}(a)--(b), in favor of a single gap in Fig.~\ref{fig:dust-size}(c), is evident for all of the particle sizes. The emission maps for $1$ mm particles presented in previous subsections are therefore not unique in exhibiting varied morphology across different disk models. This variation is not associated with any details of the dust dynamics, but with differences in the underlying gas distributions resulting from the effects of cooling.

\section{Comparison with Locally Isothermal Simulations}
\label{sect:results-iso}

The majority of existing numerical studies addressing the problem of planet-disk interaction utilize the locally isothermal approximation for treating gas thermodynamics. In this section we address the validity of this approximation in the context of our results, fully accounting for the different forms of radiative transport affecting density wave propagation in disks.

\subsection{Fiducial Disk Model}

We first explore the differences between simulations with cooling and the analogous locally isothermal simulations, using the Fiducial disk model. In Fig.~\ref{fig:profiles_fiducial} we plot the gas surface density profiles, using the solid black lines to show the results for simulations with cooling, and the dashed red lines to show the results of the corresponding locally isothermal simulations.

In all cases considered in Fig.~\ref{fig:profiles_fiducial}, the gas profiles in the locally isothermal simulations exhibit $3$ -- $5$ pairs of rings and gaps. In contrast, the gas profiles in the simulations with cooling exhibit a much more diverse range of morphologies. Most notably, for $r_\mathrm{p} = 10$ au and $20$ au, the multiple gap structure seen in the locally isothermal simulations is strongly suppressed in the simulations with cooling. There is a single gap at the location of the planet, while secondary and higher-order gaps are either absent or very weak.

Two general trends can be observed in Fig.~\ref{fig:profiles_fiducial}. First, discrepancies between the profiles with cooling and the locally isothermal profiles become less pronounced with increasing planetary orbital radius $r_\mathrm{p}$. This is because the cooling timescale---as represented by, e.g., $\beta_\mathrm{eff}$, see Fig.~\ref{fig:beta-eff}(a)---is typically smaller for larger disk radii. In particular, for $r_\mathrm{p} = 50$ au and $100$ au, $\beta_\mathrm{eff} \sim 0.01$ at moderate distances from the planet. As a result, the behavior of density waves in these regions is nearly locally isothermal. For planets in the inner part of the disk, $r_\mathrm{p} = 10$ au and $20$ au, the larger cooling timescale leads to strong linear damping of density waves, resulting in a preference for single gaps.

The second trend illustrated in Fig.~\ref{fig:profiles_fiducial} is the rough convergence of the profiles in the cooling and locally isothermal simulations with increasing planet mass $M_\mathrm{p}$. This arises because, for more massive planets, nonlinear dissipation due to shocks (depositing angular momentum locally, close to the planetary orbit) plays an increasingly dominant role in the density wave evolution and the associated disk evolution (see Fig.~\ref{fig:amf}). This is in contrast to the potentially important role played by linear dissipation associated with cooling for less massive planets. For a $1 M_\mathrm{th}$ planet, the surface density profiles for the cooling and locally isothermal calculations are very similar to one another for $r \gtrsim 0.5 r_\mathrm{p}$. However, discrepancies between the two still emerge at smaller radii. This is because the high-amplitude waves launched by the planet are initially damped predominantly by nonlinear dissipation, but at some large distance from the planet, their amplitude becomes small enough for cooling to take over and become the dominant driver of the wave evolution.

Dust emission maps computed from simulations with cooling and from locally isothermal simulations are shown in the right and left halves, respectively, of each panel in Fig.~\ref{fig:images_fiducial}. Differences between them become less apparent for larger planet masses or larger orbital radii. There are significant differences between the two maps for the two smallest orbital radii, $r_\mathrm{p} = 10$ au and $20$ au, for planets with masses $0.1 M_\mathrm{th}$ and $0.3 M_\mathrm{th}$ (although they may be hard to see because of the intensity scale of these maps). They are less pronounced at these radii for the case with a $1 M_\mathrm{th}$ planet. For the two largest orbital radii, $r_\mathrm{p} = 50$ au and $100$ au, differences between the two emission maps are fairly minor for all three planet masses shown.

The differences between the simulations with cooling and locally isothermal simulations can be understood in terms of the AMF behavior of the planet-driven waves (see Section~\ref{fig:amf}). The locally isothermal approximation represents the limit in which $\beta_\mathrm{eff} \to 0$. In locally isothermal disks, the AMF of linear waves is proportional to the disk temperature far from the planet \citep{Miranda-ALMA}. In the Fiducial disk model, this means that $F_J\propto r^{-1/2}$ in a locally isothermal disk. Therefore, the AMF grows as waves propagate towards smaller radii (see Fig.~1 of \citet{Miranda-ALMA}), and their effect on the inner disk when they dissipate --- the formation of gaps --- is enhanced in locally isothermal disks.

This is to be contrasted with the behavior of the linear AMF in disks with cooling (dashed curves in Fig.~\ref{fig:amf}), which typically decreases with distance from the planet (Fig.~\ref{fig:amf}(d) is an exception, with the AMF exhibiting a very slight rise towards the inner disk since cooling is very fast in this case and thermodynamics approaches the isothermal limit). This implies that, when cooling is not rapid enough for the locally isothermal approximation to be applicable, the waves damp predominantly in the vicinity of the planetary orbit, and therefore fail to open secondary gaps in the inner disk far from the planet.

\subsection{Discrepancy Score and Parameter Exploration}
\label{sect:discrep}

\begin{figure*}
\begin{center}
\includegraphics[width=0.99\textwidth,clip]{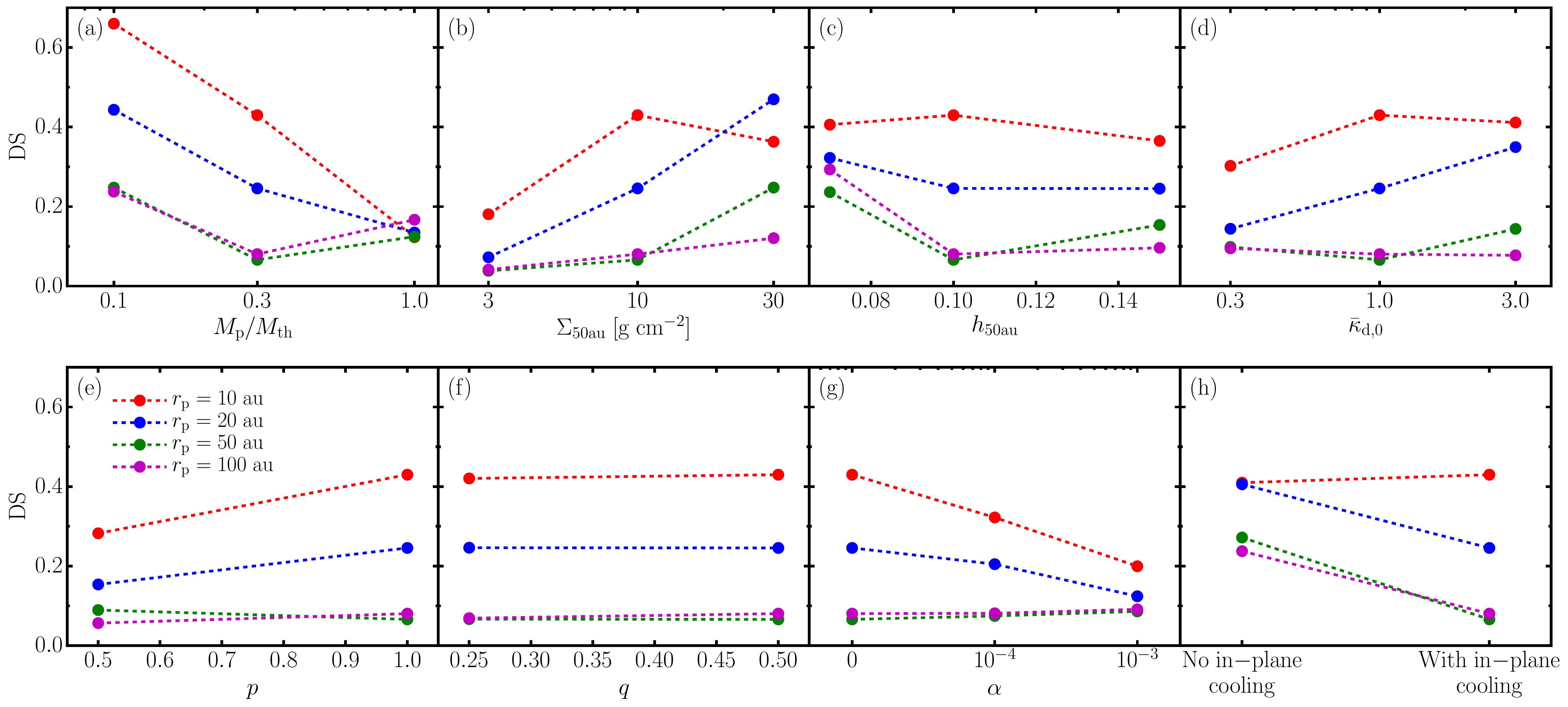}
\caption{The discrepancy score $\mathrm{DS}$ (equation~(\ref{eq:ds})), which approximately describes the validity of the locally isothermal approximation. Values of $\mathrm{DS}$ close to zero indicate that the locally isothermal approximation reproduces the disk structure found in the corresponding simulation with cooling reasonably well, while large values indicate that it does not. The dependence on various disk parameters is explored in the panels (a)--(g). In (h), the effect of turning in-plane cooling on or off is shown.}
\label{fig:discrepancy}
\end{center}
\end{figure*}

It is useful to understand how the degree of discrepancy between simulations with cooling and locally isothermal simulations depends on the parameters of the disk and planet. However, since our grid of simulations consists of $56$ simulations with cooling and $36$ locally isothermal simulations, a detailed comparison for all cases (as in Figs.~\ref{fig:profiles_fiducial}--\ref{fig:images_fiducial}) is impractical and may be difficult to interpret. Instead we chose to quantify the difference in terms of a convenient summary statistic.

We define a ``discrepancy score'' DS that quantifies the difference between $\delta\Sigma_\mathrm{C}(r)$, the (azimuthally averaged) gas surface density perturbation resulting from a simulation with cooling, and $\delta\Sigma_\mathrm{LI}(r)$, the gas surface density perturbation in the corresponding locally isothermal simulation. It is computed according to
\be
\label{eq:ds}
\mathrm{DS} = \frac{\int_{r_1}^{r_2} |\delta\Sigma_\mathrm{C}(r)-\delta\Sigma_\mathrm{LI}(r)|\Sigma_0(r)^{-1}\mathrm{d}r}{\int_{r_1}^{r_2}\left(|\delta\Sigma_\mathrm{C}|+|\delta\Sigma_\mathrm{LI}|\right)\Sigma_0(r)^{-1}\mathrm{d}r}.
\ee
Here $r_1 = 0.1 r_\mathrm{p}$ (just outside the inner wave damping zone) and $r_2 = 2 r_\mathrm{p}$ (beyond which the disk is typically unperturbed) are the radial boundaries of the comparison region, and $\Sigma_0(r)$ is the initial unperturbed surface density. Notice that DS compares profiles of the fractional surface density perturbation $\delta\Sigma/\Sigma$, as shown for different cases in Figs.~\ref{fig:profiles_fiducial} and \ref{fig:profiles_10au20au}--\ref{fig:profiles_50au100au}. Also note that $0 \leq \mathrm{DS} \leq 1$. Small values of DS indicate that the two profiles (cooling and locally isothermal) are very similar, while large values indicate that they differ substantially.

Examples of how well DS describes the difference between the simulations with cooling and locally isothermal simulations are shown in Fig.~\ref{fig:profiles_fiducial}. In each panel the gas surface density perturbation is shown for both simulations, with the corresponding value of $\mathrm{DS}$ given in the lower right corner. Note that the case with $M_\mathrm{p} = 0.1 M_\mathrm{th}$ and $r_\mathrm{p} = 10$ au has the largest score value, $\mathrm{DS} = 0.66$. Correspondingly, the two surface density perturbation profiles bear little resemblance to one another. For $M_\mathrm{p} = 0.3 M_\mathrm{th}$ with $r_\mathrm{p} = 50$ au and $100$ au, $\mathrm{DS} < 0.1$, which is indicative of the close resemblance between the two profiles. Other cases in between these two extremes have intermediate values of $\mathrm{DS}$. We see that DS seems to provide a reasonable measure of the discrepancy between the two profiles (although, obviously, a single numeric metric cannot fully capture the richness of the $\Sigma(r)$ behaviors), and that $\mathrm{DS} \approx 0.1$ -- $0.2$ is the rough threshold separating very similar pairs of profiles to very discrepant pairs.

The validity of the locally isothermal approximation using DS as a metric is explored as a function of the various disk parameters in Fig.~\ref{fig:discrepancy}. Generally, for smaller planetary orbital radii ($r_\mathrm{p} = 10$ au and $20$ au) DS is large ($\gtrsim 0.2$), while for the larger orbital radii ($r_\mathrm{p} = 50$ au and $100$ au), it is comparatively small ($\lesssim 0.2$). In other words, the locally isothermal limit tends to provide a reasonable approximation of the more realistic thermodynamics for larger orbital radii, but a poor approximation for smaller orbital radii.

Moreover, DS tends to exhibit more variation with the different disk parameters for smaller planetary orbital radii, and less variation for larger orbital radii. That is, the validity of the locally isothermal assumption for small $r_\mathrm{p}$ is more sensitive to the details of the disk model than it is for large $r_\mathrm{p}$. For example, for $r_\mathrm{p} = 10$ au and $20$ au, DS decreases significantly as $M_\mathrm{p}$ is increased (Fig.~\ref{fig:discrepancy}(a)), and increases as either $\Sigma_{50\mathrm{au}}$ (Fig.~\ref{fig:discrepancy}(b)), $\bar{\kappa}_{\mathrm{d},0}$ (Fig.~\ref{fig:discrepancy}(d)), or $p$ (Fig.~\ref{fig:discrepancy}(e)) are increased. By comparison, for $r_\mathrm{p} = 50$ au and $100$ au, DS is fairly insensitive to these parameters.

However, there are exceptions to these trends. For example, for the coldest disk with $h_{50\mathrm{au}} = 0.07$ DS is large ($\approx 0.25$ -- $0.4$) for all orbital radii considered (Fig.~\ref{fig:discrepancy}(c)). Also, DS is quite insensitive to the temperature power law index $q$ (see Fig.~\ref{fig:discrepancy}(f)) in all cases.

The variation of DS with the viscosity parameter is explored in Fig.~\ref{fig:discrepancy}(g). For the small orbital radii with large discrepancy scores for an inviscid disk ($\alpha = 0$), the score becomes much smaller for a large viscosity. This is a result of viscous spreading filling in the gaps, and hence erasing the differences in the gap structure resulting from the different thermodynamics (see Section~\ref{sect:nu-var}).

The role of in-plane cooling is illustrated in Fig.~\ref{fig:discrepancy}(h). Except for the case with $r_\mathrm{p} = 10$ au, the inclusion of in-plane cooling significantly reduces the discrepancy score, in comparison to the case in which only surface cooling is considered. For the cases with $r_\mathrm{p} = 50$ au and $100$ au, $\mathrm{DS}$ is fairly large ($\approx 0.2$ -- $0.3$) without in-plane cooling, but drops to $\lesssim 0.1$ when it is included. In other words, for these large orbital radii, the locally isothermal assumption provides a poor approximation for the real thermodynamics when only surface cooling is considered, but is a better approximation when the in-plane cooling is included. This is a result of the substantial reduction in the effective cooling timescale that is brought on by the inclusion of in-plane cooling.

\section{Discussion}
\label{sect:disc}

\subsection{The Importance of In-plane Cooling}

One of the key findings of this work is that the inclusion of in-plane cooling has a significant impact on the details of density wave propagation and gap opening. In-plane cooling has a dominant effect on wave damping simply because the radial length scale of the wave as it propagates rapidly becomes considerably shorter than the vertical disk thickness $H$---the length scale on which surface cooling operates, see Fig.~\ref{fig:spiral_width}. As a result, thermal relaxation of the wave perturbation occurs predominantly {\it along} the disk midplane, rather than normal to it. The importance of in-plane cooling was first noted in \citet{GR01} and our detailed calculations provide the first quantitative confirmation of  their conclusion.

We find rather generally, that for planets at small orbital radii ($r_\mathrm{p}\lesssim 20$ au), inclusion of in-plane cooling leads to a suppression of the multiple gap structure that is present when only radiative cooling from the disk surface is considered, see Fig.~\ref{fig:images_10au20au}(f),(l). On the contrary, for planets with larger orbital radii ($r_\mathrm{p} \gtrsim 50$ au), the inclusion of in-plane cooling has essentially the opposite effect on gap opening. Without in-plane cooling, the planet often tends to open a single gap (due to strong linear radiative damping), but with in-plane cooling, it opens multiple gaps. In the latter case, the resulting disk morphology is fairly similar to the one produced in a locally isothermal simulation. 

As described in \S \ref{sect:no-in-plane}, this variation is naturally explained by the effective cooling timescale acounting for in-plane cooling being $\approx 5$ -- $25$ times shorter than the cooling timescale due to surface cooling alone. This difference is often substantial enough to entirely change the regime in which density waves thermally relax---from flux-preserving, almost adiabatic to strongly damped, or from strongly damped to almost locally isothermal---with obvious implications for the emergence of substructure in the disk. 

Given the disk-planet model, it may not be easy to a priori assess the effect of radiative effects on wave dynamics and disk evolution. While the surface cooling rate can be directly computed as a function of distance as described in \S \ref{sect:cooling-surf}, the calculation of in-plane cooling is a far more involved procedure that requires understanding the characteristics of the density wave in the Fourier domain, see \S \ref{sect:cooling-mid}--\ref{sect:tc-eff}. At the same time, a useful approximation for the in-plane cooling timescale behavior can still be provided by $\beta_{m*}$, defined in Section \ref{sect:beta-mstar} by assuming that certain modes (namely $m=m_*$) dominate the wave dynamics. The results of \S \ref{sect:tc-eff} and Fig.~\ref{fig:beta-eff} show that in many cases $\beta_{m*}(r)$ reproduces the behavior of the more accurate effective cooling timescale $\beta_\mathrm{eff}(r)$ quite well. For this reason we recommend the use of $\beta_{m*}$ given by equations (\ref{eq:beta-m-2})--(\ref{eq:kr-lrad}) with $m=m_*$ for making simple exploratory analytical estimates of the radiative damping timescale for planet-driven density waves.

More generally, our findings emphasize the need for explicitly including radiative effects in numerical studies of planet-disk interaction, something that has been previously brought up in \citet{Miranda-ALMA,Miranda-Cooling}. Whereas cooling may be fast enough for waves launched by planets in the outer disk to propagate almost in the locally isothermal regime, the simulations relying on this thermodynamic simplification may still not be accurate enough to reproduce all details of the observed disk substructures at the quantitative level, as we show in \S \ref{sect:results-iso}. Explicit treatment of radiative effects in numerical studies of planet-disk interaction should become a standard approach when interpreting protoplanetary disk observations.

\subsection{Observational Implications}

Using simulated dust emission maps, we have shown that the multiplicity, widths, separations, and relative contrasts of planet-driven gaps/rings can exhibit significant variation depending on the details of the underlying disk model (see Figs.~\ref{fig:images_10au20au}--\ref{fig:images_50au100au}). The gap structure is sensitive to disk properties such as the temperature, mass, and opacity of the disk, which are often poorly constrained in real observations. Due to the variety of possible morphologies, determining whether or not annular substructures in disks are the signatures of planet-disk interaction, as well as estimating the masses and orbital radii of the putative planets responsible for these structures, may therefore be much more difficult than previously thought. This issue was recently highlighted by \citet{Facchini2020}. Also, this variety of outcomes may complicate the assessment of the role played by other physical processes, e.g., planet migration \citep{Dong2017,Perez2019}, in formation of the disk substructures. In order to make progress in this area, precise quantitative comparisons between observations and numerical simulations would require an improved understanding of the physical properties of protoplanetary disks for the former, and consideration of realistic thermodynamics with radiative effects for the latter.

Our simulated emission maps do not account for the finite spatial resolution in real observations. We may therefore ask whether or not the variety of morphologies seen in our emission maps could be probed in ALMA observations, which achieve a spatial resolution of, e.g., about $5$ au for disks at a characteristic distance of $100$ -- $150$ pc \citep{Andrews-DSHARP}. We found that the thermodynamics details of the disk model have the most substantial impact on the structure of planet-driven gaps for planets with small orbital radii, $\lesssim 20$ au (see Fig.~\ref{fig:images_10au20au}). However, at such separations, a spatial resolution of $\approx 5$ au constitutes a substantial fraction of the disk radius. On the other hand, for larger separations, $r_\mathrm{p} \gtrsim 50$ au, where such a resolution allows for finer characterization of the gap structure, there is somewhat less variation in the structures seen in dust emission with respect to the physical disk parameters (see Fig.~\ref{fig:images_50au100au}).

Despite these limitations, variation of the gap structure with the disk parameters is often significant enough---producing qualitatively different structures---that it should be discernible in real observations. For example, varying the disk temperature has a substantial effect for all the planetary orbital radii we considered. In particular, in our Cold disk model, secondary gaps are strongly suppressed relative to the Fiducial or Hot models. Significant and observationally recognizable differences also arise when the disk mass is varied. 

\subsection{Comparison with Previous Work}

In \S \ref{sect:results-iso} we showed that locally isothermal approximation used in many numerical studies of planet-disk interaction \citep{Bae2017,DongGaps2017,DongGaps2018,Zhang2018} can often lead to a biased representation of the planet-driven evolution of the protoplanetary disk. At the same time, a number of authors have studied this problem using hydrodynamical simulations that do include some treatment of cooling/radiative transfer. These treatments vary in their levels of approximation and complexity, and it is instructive to put them in context of our present results.

In the simplest approximation, one assumes that thermal relaxation occurs with either a (spatially) constant cooling timescale, or a constant value of $\beta$, i.e., a cooling timescale that is a fixed fraction of the orbital timescale (e.g., \citealt{Miranda-Cooling}). 
In particular, \citet{Zhu2015} used this approximation to explore the structure of spiral shocks using both 2D and 3D simulations. They noted the modification of the pitch angle of spirals, as a result of the $\beta$-dependent sound speed \citep{Miranda-Cooling}; however, the damping of density waves by cooling or modification of gap opening was not explored. More recently, \citet{Zhang2020} employed constant-$\beta$ cooling to study the structure of planet-driven spirals and gaps. In agreement with \citet{Miranda-Cooling}, they found that for intermediate cooling timescales, density waves are strongly damped, and that the multiple gap structures produced by a planet are highly suppressed in favor of a single gap. 

At the next level of sophistication, one considers a value for the cooling timescale that is derived from physical processes. Here the cooling timescale will typically vary spatially and have explicit dependence on disk properties such as the surface density. This is the approach that we have taken in this work. Our prescription for the cooling timescale associated with radiative losses from the disk surface (\S \ref{sect:cooling-surf}) represents a linearized version of the cooling law of \citet{Menou2004}, which is often employed in 2D simulations. 

However, our simultaneous inclusion of the physically-motivated prescription for in-plane cooling is an entirely new development. Our use of the WKB approximation for planet-driven density waves to characterize the in-plane radiative transfer (equation~(\ref{eq:beta-mid})) is in some ways similar to \citet{LinYoudin2015}, who considered in-plane cooling in their study of the vertical shear instability (without including surface cooling). However, the application of this machinery to the problem of planet-disk interaction is unique to our work.

Finally, the most self-consistent treatments of radiative effects in simulations directly compute thermal energy transfer along the midplane of the disk (and normal to it in 3D studies), typically resorting to the flux-limited diffusion (FLD) approximation \citep{Levermore1981,Kley1989}. A number of studies have made use of FLD in 2D (e.g., \citealt{DAngelo2003,Kley2008,Pierens2015}), as well as in 3D (e.g., \citealt{Kley2009,Lega2014,Tsuk2015}), to explore the role of radiative effects on disk structure, planetary torques and migration. However, they did not examine the role of radiative effects in the propagation and dissipation of density waves.

Recently, \citet{Ziampras2020a} presented 2D numerical simulations of planet-disk interaction, some of which included radiation transport using FLD. They were interested in the effects of planetary density waves on the global temperature profile of the disk, and found radiative diffusion to be unimportant in terms of its impact on the global disk structure (presumably because in most of their models planet-induced heating was rather moderate). Subsequently, \citet{Ziampras2020b} investigated the structure of gaps opened by planets, however, not including radiative diffusion, i.e., in-plane cooling. Their conclusions are fully consistent with \citet{Miranda-Cooling} and our results with only the surface cooling. However, since in-plane cooling was not included, its effect on density wave dynamics and the formation of the wave-driven structure was missed.

We also note that fully capturing the effect on in-plane cooling in simulations with computationally expensive direct radiation transfer may be rather challenging. The need to resolve radiation transport across the small radial scale of a tightly-wrapped density wave, possibly as short as $0.1H$ (see Fig.~\ref{fig:spiral_width}), requires very high resolution, which may be difficult to achieve even in 2D.  

\subsection{Approximations Used in This Work}

Our study made a number of simplifying assumptions, which we review next. Our treatment of in-plane cooling represents the main improvement made upon existing work. On lengthscales longer than the radiative lengthscale $l_\mathrm{rad}$, radiation is transported along the disk midplane in a diffusive manner. In our semi-analytical calculations (\S \ref{sect:cooling}), rather than explicitly propagating the associated diffusion problem through our mathematical framework (which would result in a fourth-order differential equation for the enthalpy perturbation, see Section \ref{sect:diffusion-limit}), we make use of the local approximation. Perturbations are decomposed into Fourier harmonics, and it is assumed that the lengthscale on which the internal energy of each harmonic varies is small compared to the disk radius. As a result of making this approximation, the diffusion timescale---for a each Fourier harmonic---ends up depending on the lengthscale over which each harmonic varies. This lengthscale is then approximated as $k_m^{-1}$, where $k_m$ the radial wavenumber of the mode, which we estimate using the standard WKB dispersion relation for planet-driven waves. This treatment of in-plane cooling may not be accurate close to the planet, where, e.g., $\beta_\mathrm{eff}$ exhibits a cusp, but in-plane cooling is not critical in this region anyway. 

The implementation of this prescription for in-plane cooling in numerical simulations makes use of the quasi-linearity of the planet-driven density waves, since in our code cooling is applied to each Fourier harmonic of the perturbation separately. For massive planets with $M_\mathrm{p} \gtrsim M_\mathrm{th}$, the WKB approximation, and hence our implementation of cooling in simulations, may begin to break down. However, for massive planets ($M_\mathrm{p} \gtrsim M_\mathrm{th}$), nonlinear effects also play a more important role in the evolution of density waves than cooling does, rendering irrelevant any inaccuracy of the cooling prescription. 

Since our hydrodynamical simulations do not explicitly treat the dust component, we simulated dust emission maps using a 1D treatment in post-processing, neglecting non-axisymmetric effects. We use an artificial smoothing applied to the dust surface density distribution to approximate possible non-axisymmetric effects (see Appendix~\ref{sect:dust-calibration}). Our calculations use dust particles with a single size (approximately representing a maximum dust size), rather than a distribution of sizes as expected in real protoplanetary disks. The resulting dust distributions are used to create raw (i.e., not convolved with an interferometric beam) maps of the submillimeter emission. Therefore, the resulting emission maps primarily serve to explore the observational impact of cooling and varying the disk parameters.

Finally, our simulations are 2D. In 3D, the treatment of cooling becomes considerably more complicated. For example, it is necessary to properly account for the vertical structure of the disk. Additionally, density wave propagation and dissipation may be modified in 3D \citep{Lubow1998,Ogilvie1999,Lee2015}.

\section{Summary}
\label{sect:summary}

We carried out a suite of hydrodynamical simulations of gap opening by planets, including a detailed treatment of radiative physics. In addition to cooling from the disk surface, we also considered transport of radiation induced by density waves along the midplane of the disk. We examined the global structure of gaps and rings produced by planets, using the gas surface density profiles and simulated dust continuum emission maps. We summarize our key results below.

\begin{itemize}

\item In-plane cooling, rather than cooling from the disk surface, plays a dominant role in the evolution of planet-driven density waves. Neglecting in-plane cooling leads to a substantial overestimation of the cooling timescale, typically by more than an order of magnitude.

\item We propose and validate a simple estimate for the (dimensionless) cooling timescale---$\beta_{m*}$ defined in Section \ref{sect:beta-mstar}---that can be used to assess the effect of radiation transport along the disk midplane on density wave propagation.

\item For the effective viscosity expected in protoplanetary disks ($\alpha\lesssim 10^{-3}$), viscous damping plays an insignificant role in dissipating planet-driven density waves. It can become competitive with radiative damping or nonlinear wave dissipation only if $\alpha \gtrsim 10^{-2}$.

\item The number of gaps/rings produced by a planet, as well as their widths, separations, and relative contrasts, depend on the disk temperature, surface density, and opacity, all of which govern the efficiency of cooling for planet-driven density waves. 

\item The locally isothermal assumption represents a poor approximation for describing gap opening by low-mass planets at orbital distances of $\lesssim 50$ au. For planets at larger separations ($\gtrsim 50$ au) or more massive planets ($\gtrsim M_\mathrm{th}$), its performance may improve, but this depends sensitively on the details of the adopted disk model.

\end{itemize}

Our results on the role of radiative effects in wave dynamics and disk evolution can be naturally applied to other astrophysical systems featuring density waves: disks in stellar binaries \citep{Kley-binary,Ju2016}, circumbinary disks \citep{MacFadyen2008,Miranda-CB}, boundary layers of accretion disks \citep{BRS12,BRS13}, and so on.

\acknowledgements

Financial support for this work was provided by NASA via grants 14-ATP14-0059 and 15-XRP15-2-0139. We thank the referee, Ruobing Dong, for a careful reading of the paper and for helpful comments and suggestions, which have helped us to improve the paper.

\appendix

\section{Derivation of the energy equation}
\label{sect:en-eq}

The evolution equation for the internal energy density $e = c_V T$ (per unit mass) in 3D is
\be
\label{eq:e-evol}
\frac{\mathrm{d}}{\mathrm{d}t} \left(\rho e\right) + \gamma \rho e \mathbf{\nabla} \cdot \mathbf{u} = -\mathbf{\nabla} \cdot \left(\mathbf{F} + \mathbf{F}_*\right),
\ee
where $\rho$ is the gas density (so that $\rho e$ is the internal energy per unit volume), $\mathbf{u}$ is the 3D gas velocity, 
$\mathbf{F}$ is the radiative flux associated with the gas, and $\mathbf{F}_*$ is the radiative flux of the central star. The right-hand side represents the energy source term. 

In an axisymmetric ($\partial/\partial\phi \rightarrow 0$) and steady ($\partial/\partial t \rightarrow 0$) state, when the left hand side of equation~(\ref{eq:e-evol}) vanishes, thermal equilibrium requires $\mathbf{F}_0 = -\mathbf{F}_*$, where $\mathbf{F}_0$ is the disk radiative flux in equilibrium. However, the passage of a density wave through the disk disturbs this equilibrium, 
inducing a perturbation $\delta\mathbf{F} = \mathbf{F} - \mathbf{F}_0$ to the radiative flux; this turns the right-hand side of equation (\ref{eq:e-evol}) into $-\mathbf{\nabla} \cdot \delta\mathbf{F}$. 

Integrating equation~(\ref{eq:e-evol}) vertically and making the conventional conversion from 3D to vertically integrated, 2D variables, one obtains
\be
\label{eq:e-evol1}
\frac{\mathrm{d}}{\mathrm{d}t} \left(\Sigma e_2\right) + \gamma \Sigma e_2 \mathbf{\nabla} \cdot \mathbf{u}_2 =  -\int_{-\infty}^\infty \left(\mathbf{\nabla} \cdot\delta\mathbf{F}\right)\mathrm{d}z,
\ee
where $e_2$ and $\mathbf{u}_2$ are the appropriately vertically averaged values of internal energy and 2D velocity in the disk plane. Next we write $\delta\mathbf{F} = \delta\mathbf{F}_\parallel + \delta F_z \hat{\mathbf{z}}$, where $\delta\mathbf{F}_\parallel$ is the component of $\delta\mathbf{F}$ in the disk plane, which allows us to integrate $\partial \delta\mathbf{F}_z/\partial z$ term in the right-hand side of (\ref{eq:e-evol1}). Also, using the 2D continuity equation to eliminate $\mathbf{\nabla} \cdot \mathbf{u}_2$ and introducing the 2D pressure $P = (\gamma-1)\Sigma e_2$, we arrive at 
\be
\label{eq:e-evol2}
\Sigma\left[\frac{\mathrm{d}e_2}{\mathrm{d}t} + P\frac{\mathrm{d}}{\mathrm{d}t}\left(\frac{1}{\Sigma}\right)\right] = -\mathbf{\nabla}_2 \cdot\delta\mathbf{F}_2 - 2\delta F_z(\infty),
\ee
where $\delta\mathbf{F}_2 = \int\delta\mathbf{F}_\parallel\mathrm{d}z$ is the vertically integrated perturbation of the radiative flux along the disk plane, and $\delta F_z(\infty)$ is the deviation of the radiative flux normal to the disk from its unperturbed value $F_z(\infty) = \sigma T_\mathrm{eff}^4$, both measured high above the disk plane (above its photosphere). The first term on the right hand side represents energy transport along the disk plane, while the second term accounts for the energy losses from the disk surface.

To relate $\delta F_z(\infty)$ to various parameters of the disk and their perturbations is, in general, a non-trivial exercise requiring a full consideration of the time-dependent transport of radiation across the vertical extent of the disk. To avoid these complications, we employ a simple but intuitive approximation, in which we relate $\delta F_z(\infty)$ to $F_z(\infty)$ and the energy perturbation $\delta e_2 = e_2-e_{2,0}$ (relative to its unperturbed value $e_{2,0}$) according to
\be  
\delta F_z(\infty)=\frac{\delta e_2}{e_{2,0}}F_z(\infty)=
\frac{\sigma T_\mathrm{eff}^4}{e_{2,0}}\delta e_2.
\label{eq:flux-rel}
\ee  
The explicit form of $\delta\mathbf{F}_2$ is specified in  Appendix \ref{sect:flux}.

Plugging the expression (\ref{eq:flux-rel}) into equation (\ref{eq:e-evol2}), dividing it by $\Sigma$ and dropping the subscript ``2'' from all variables, we finally arrive at the 2D form of the energy equation (\ref{eq:energy-eq}) for a perturbed disk used in this work. 

Note that our transition from the 3D energy equation to its 2D version is accurate up to insignificant factors of order unity (which depend on the detailed vertical structure of the disk) and is roughly as rigorous as the definition of the conventional thin disk variables such as the 2D pressure.

\section{Vertically integrated flux perturbation $\delta\mathbf{F}$}
\label{sect:flux}

In the diffusion limit ($l_e\ll l_\mathrm{rad}$) the radiative flux in 3D is \citep{Rybicki}
\be
\label{eq:flux-diff1}
\mathbf{F} = -\frac{16\sigma T^3}{3\kappa\rho}{\nabla}T=-\eta\rho\nabla e,
\ee
where $\eta$ is defined by equation (\ref{eq:diff-coef}). Note that $\kappa$ is the Rosseland-mean opacity, but for simplicity we use the same opacity in both the diffusive and streaming limits. To get the vertically integrated flux we will assume (as we do everywhere in this paper) that $\eta$ is a constant (i.e., does not depend on $z$) with a value set by $\rho=\rho_\mathrm{mid}$. Integrating (\ref{eq:flux-diff1}) over $z$ we get $\mathbf{F}_2=-\eta\Sigma\nabla e_2$ (again, appropriate averaging of the thermal energy $e$ is assumed here). Taking a perturbation to this equation and dropping the subscript ``2'', we arrive at equation (\ref{eq:flux-diff}).

In the streaming limit, we recall that, most generally, the radiative flux can be expressed through the specific intensity $I_\nu$ as \citep{Rybicki}
\begin{gather}
\label{eq:flux-stream1}
\mathbf{F} = \int\int I_\nu(\mathbf{n})\,\mathbf{n}\,\mathrm{d}\Omega
\mathrm{d}\nu = 4\pi \,\hat{\mathbf{f}}\int H_\nu\mathrm{d}\nu, \\
H_\nu =  \frac{1}{2}\int_{-1}^1 I_\nu\,\mu\,\mathrm{d}\mu,
\end{gather}
where $\hat{\mathbf{f}} = \mathbf{\nabla}T/|\mathbf{\nabla}T|$ is the direction of radiation propagation, and $\mu$ is the cosine of an angle that the unit vector $\mathbf{n}$ makes with $\hat{\mathbf{f}}$ ($\Omega$ is the solid angle). 

In the streaming limit, we can neglect flux absorption and consider only its variation due to disk emissivity. Then, using the Kirchhoff law, the conventional expression for the first moment of the radiation transport equation becomes \citep{Rybicki}
\be  
\frac{\mathrm{d}H_\nu}{\mathrm{d}s} \approx \kappa\rho\, B_\nu,
\label{eq:radtran}
\ee 
where $s$ is the distance along $\hat{\mathbf{f}}$ and $B_\nu$ is the Planck function. 

Taking the divergence of the equation (\ref{eq:flux-stream1}), using equation (\ref{eq:radtran}) and integrating $B_\nu$ over $\nu$ (we take $\kappa$ to be frequency-independent), we get 
\be  
\mathbf{\nabla}\cdot\mathbf{F} \approx 4\kappa\rho\sigma T^4.
\label{eq:flux-stream2}
\ee  
Taking a perturbation to this equation, we integrate it over $z$, finding $\mathbf{\nabla}\cdot\delta\mathbf{F}_2 = 16\kappa\Sigma\sigma T^3\delta T$. Dropping the subscript ``2'', we finally arrive at equation (\ref{eq:flux_stream}).

\section{Implementation of Cooling in Simulations}
\label{sect:cooling-numerical}

In our formulation, the cooling timescale (equation~(\ref{eq:beta-m-2})) has explicit dependence on the azimuthal wavenumber $m$ of the perturbation. For this reason, in our simulations, which do not compute explicit radiation transfer, cooling is implemented using a spectral approach, in which the different Fourier harmonics of the specific internal energy perturbation are evolved independently. The cooling step takes place after the main hydrodynamical update step in \textsc{fargo3d}. Given the specific internal energy $e(r,\phi,t)$, as a function of the polar coordinates $r$ and $\phi$ and at time $t$, we obtain the updated energy due to cooling at time $t + \delta t$, denoted $e(r,\phi,t+\delta t)$, using the following procedure.

\begin{itemize}

\item At each radius $r$, the specific internal energy at $t = 0$, denoted $e_0(r)$, is subtracted from $e(r,\phi,t)$ in order to obtain the specific internal energy perturbation $\delta e(r,\phi,t)$. The perturbation is then decomposed into Fourier harmonics $\delta e_m(r,t)$ by means of a fast Fourier transform (FFT).

\item The evolution of the harmonics is described by equation~(\ref{eq:em-cool}), with the dimensionless cooling timescale $\beta_m$ given by equations~(\ref{eq:beta-m-2})--(\ref{eq:kr-lrad}) (see also equations~(\ref{eq:tau}) and (\ref{eq:beta-surf})--(\ref{eq:ftau})). To compute $\beta_m$, we require the local temperature and surface density as inputs. For these we use the initial temperature profile---assuming the fixed equilibrium temperature distribution set by stellar irradiation---and the numerical azimuthally averaged surface density, $\langle \Sigma(r,\phi,t) \rangle_\phi$. 

\item Each harmonic of the perturbation, up to the maximum azimuthal number $m_\mathrm{max}$ (and including $m = 0$), is then updated using the exact solution to equation~(\ref{eq:em-cool}),
\be
\label{eq:cool-numerical}
\delta e_m(r,t+\delta t) = \exp\left[-\frac{\Omega(r)\delta t}{\beta_m}\right] \delta e_m(r,t).
\ee
For $m > m_\mathrm{max}$, we set $\delta e_m(r,t+\delta t) = 0$.

\item Subsequently, the updated energy perturbation $\delta e(r,\phi,t+\delta t)$ is constructed from the updated Fourier harmonics $\delta e_m(r,t+\delta t)$, using the inverse FFT. Finally, the initial profile $e_0(r)$ is added back to $\delta e(r,\phi,t+\delta t)$, to obtain the final updated energy $e(r,\phi,t+\delta t)$.

\end{itemize}

As described in the third step above, the $\delta e_m$ harmonics are updated according to equation~(\ref{eq:cool-numerical}) for $m < m_\mathrm{max}$, but are set to zero for $m > m_\mathrm{max}$. There are three reasons for this simplification. First, for very large values of $m$ ($m \gg m_*$), the amplitude of $\delta e_m$ (or the Fourier harmonic of any other fluid variable) is very small, and a negligible amount of angular momentum is carried by the harmonic. Second, the cooling timescale for these harmonics is typically very short, especially in the optically thick inner disk where $\beta_m \propto m^{-2}$. Hence, if one of these harmonics acquires an appreciable amplitude, it should be erased very quickly anyway. Finally, the maximum $m$, which can be potentially captured by the simulation, $N_\phi/2$ (i.e., the Nyquist frequency) is very large ($1024$ in our simulations using the fiducial resolution), and updating every harmonic is computationally expensive. In our simulations we choose $m_\mathrm{max} = 8 h_\mathrm{p}^{-1}$, and find that using a larger value has a negligible effect on the results. 

Using this value of $m_\mathrm{max}$, our simulations with cooling incur a moderate slowdown, about $40$ -- $50\%$, relative to the corresponding locally isothermal simulations. In Appendix~\ref{sect:cooling-test}, we present a test of our implementation of cooling. We find that the numerical implementation reproduces the results of linear calculations very closely.

In some of the simulations with cooling, there is a numerical artifact that arises at the boundary of the inner damping zone. It is a result of a mismatch between the short cooling timescale ($\beta \sim 1$) inside the damping zone (associated with the artificially imposed wave damping) and the long damping timescale due to radiative cooling just outside of it. This only occurs when the cooling timescale in the inner disk is large, e.g., $\beta_\mathrm{eff} \gtrsim 10$, which is only the case for small planetary orbital radii (this is never the case for $r_\mathrm{p} \geq 50$ au). The mismatch leads to an artificial density enhancement just outside the damping zone. When this feature is present, the region inside of $r = (0.12$ -- $0.15) r_\mathrm{p}$, where the disk evolution is affected by the artifact, is removed when presenting numerical results.

\subsection{Verification of Numerical Cooling Treatment}
\label{sect:cooling-test}

\begin{figure}
\begin{center}
\includegraphics[width=0.49\textwidth,clip]{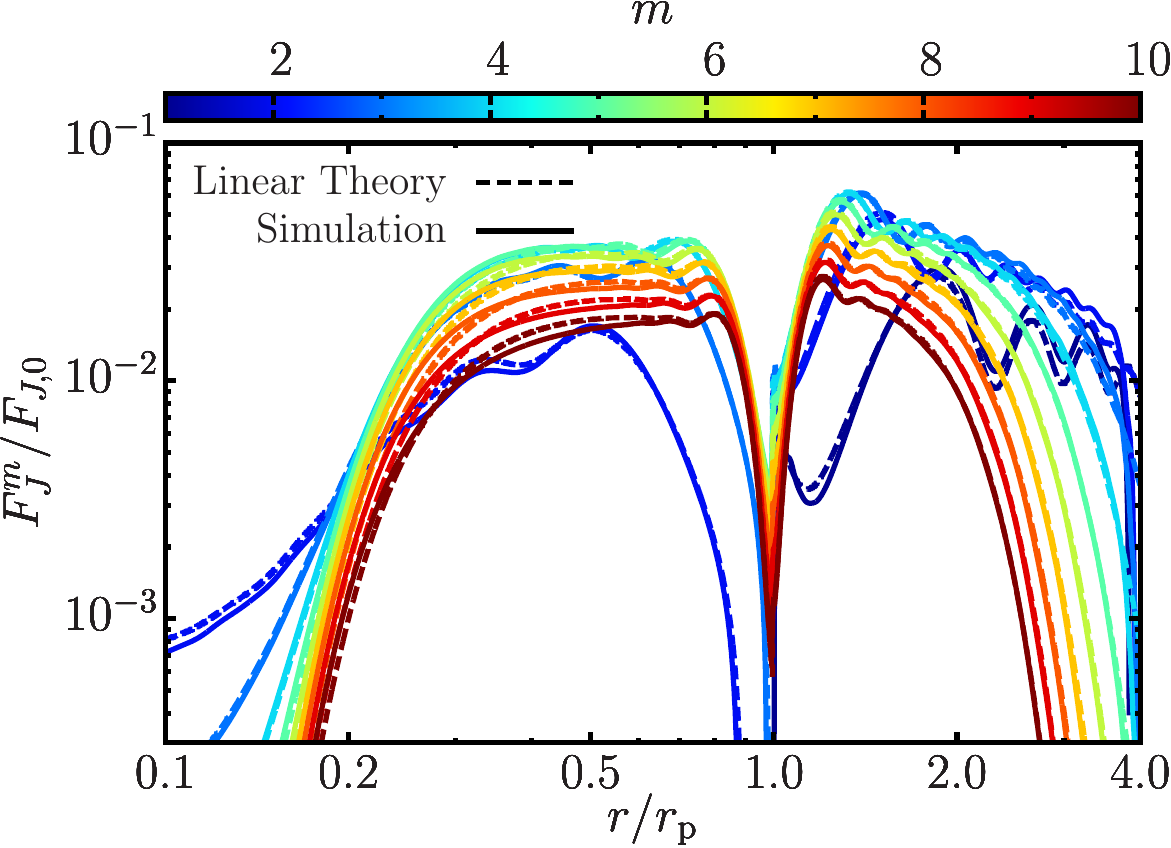}
\caption{Fourier harmonics of AMF, $F_J^m$, for waves driven by a planet with orbital radius $r_\mathrm{p} = 50$ au, using the Fiducial disk model. The dashed lines represent the results of linear perturbation calculations, and the solid lines the results of numerical simulation with a low-mass ($0.01 M_\mathrm{th}$) planet.}
\label{fig:amf-m}
\end{center}
\end{figure}

Now we test our implementation of cooling in numerical simulations by comparing in Fig.~\ref{fig:amf-m} the profiles of the Fourier harmonics of the AMF for the waves produced by a $M_\mathrm{p}=0.01 M_\mathrm{th}$ planet in a simulation with cooling implemented as described above to the predictions from linear theory. Here we have used the Fiducial disk model, with the planet located at $r_\mathrm{p} = 50$ au. Overall, one can see excellent agreement between the numerical results and linear theory. Most importantly, the damping rates of the different harmonics (i.e., the rates at which the different $F_J^m$ components fall off with distance from the planet) are accurately reproduced in the numerical simulation. We also found similarly good agreement for different planetary orbital radii and disk models. This validates our numerical treatment of cooling.

We also experimented with several alternative implementations of cooling, in which the total internal energy perturbation (rather than its Fourier harmonics) is relaxed with single cooling timescale, which is intended to produce approximately the correct behavior for all of the harmonics of a perturbation. This would bypass the need to decompose the internal energy perturbation $\delta e$ into Fourier harmonics, which is computationally convenient. Possible choices for this timescale include $\beta_\mathrm{eff}$ (equation~(\ref{eq:beta-eff})), as well as other similarly constructed quantities, e.g., the averages of $\beta_m$ weighted by either $|\delta e_m|$ or $|\delta e_m|^2$. Such ``single effective $\beta$'' methods generally do not adequately reproduce the linear behavior of the $F_J^m$ components. A common failure of these methods is an overestimation of the damping rate for harmonics with small values of $m$ (e.g., $\lesssim 3$). For these harmonics the cooling timescale $\beta_m$ (equation~(\ref{eq:beta-m-2})), set primarily by in-plane cooling, is usually much longer than for higher order harmonics (in the diffusive regime; see Fig.~\ref{fig:beta-m}). Therefore, using an effective cooling timescale that is weighted towards higher azimuthal numbers (which tend to have larger amplitudes) generally produces incorrect behavior for the low-$m$ harmonics. In our simulations we therefore use the spectral method described in Section~\ref{sect:cooling-numerical}.

\section{Dust Emission Maps}
\label{sect:dust-maps}

\subsection{Dust Dynamics}
\label{sect:dust-dyn}

Dust evolution is treated by post-processing our hydrodynamical simulations, using an approximate 1D method similar to the approach used in \citet{Miranda-ALMA}. However, in this work we treat dust as Lagrangian particles rather than as a pressureless fluid. In this approach, it is not necessary to mitigate the large density contrasts that arise at the outer boundary in a fluid-like treatment.

The positions of $2 \times 10^5$ dust particles are initialized at $t = 0$ following the distribution $\Sigma_\mathrm{d}(r) = \eta_\mathrm{d}\Sigma(r)$, where $\eta_\mathrm{d} = 0.01$ is the initial dust-to-gas ratio. At each timestep, their positions are updated using the dust radial velocity \citep{TL02}
\be
\label{eq:vd}
u_{r,\mathrm{d}}(r) = \frac{1}{1+\mathrm{St}^2}\left(\bar{u}_r + \frac{\mathrm{St}}{\langle\Sigma\rangle_\phi\Omega_\mathrm{K}}\frac{\mathrm{d}\langle P\rangle_\phi}{\mathrm{d}r}\right).
\ee
Here $\mathrm{St}$ is the Stokes number of the particle, $\langle \Sigma \rangle_\phi$ and $\langle P \rangle_\phi$ are the azimuthally averaged gas surface density and pressure, and $\bar{u}_r = \langle\Sigma u_r\rangle_\phi/\langle\Sigma\rangle_\phi$ is the effective 1D gas radial velocity. The fluid variables in equation~(\ref{eq:vd}) are extracted from the hydrodynamical simulation at a frequency $20/t_\mathrm{p}$ ($t_\mathrm{p} = 2\pi/\Omega_\mathrm{p}$) and are linearly interpolated from the radial hydrodynamical grid to the particle position. Particles are dropped if they leave the grid, and so the total dust mass can decrease with time. However, typically only a few percent of the dust mass is lost.

The Stokes number of a particle is
\be
\mathrm{St} = \frac{\pi\rho_\mathrm{d}s_\mathrm{d}}{2\Sigma} = 0.016~\rho_{\mathrm{d},1} s_\mathrm{d,mm} \Sigma_{10}^{-1},
\ee
where $\rho_{\mathrm{d},1} = \rho_\mathrm{d}/(1~\mathrm{g}~\mathrm{cm}^{-3})$ and $s_{\mathrm{d,mm}} = s_\mathrm{d}/(1~\mathrm{mm})$ are the scaled bulk density and particle size. For most of our calculations, we a choose a particle size of $s_\mathrm{d}=1$ mm for the dust particles (the effects of varying the particle size are explored in Section~\ref{sect:particle-size}). Since we consider a fixed particle size and the gas surface density decreases with $r$, initially $\mathrm{St}$ increases with $r$, so that the particles are more mobile, i.e., drift faster, at larger radii. In the Fiducial disk model, the initial Stokes number at $r_\mathrm{p}$, denoted $\mathrm{St}_0$, ranges from $0.008$ to $0.08$ for $1$ mm particles.

In some of our simulations (several cases with $r_\mathrm{p} = 100$ au and different disk models), the dust is so mobile (with $\mathrm{St}_0 \gtrsim 0.1$) that it all drifts inward and is lost to the central star (i.e., leaves the inner edge of the grid) before appreciable pressure bumps in the gas can be built up to trap it. In these cases we hold the dust particles fixed for the first several hundred orbits of the simulation, allowing pressure bumps to be built up, before they are released to drift freely, ensuring that all the dust is not lost through the inner boundary. In these cases, the final results are not sensitive to the exact time at which the dust particles are released, as long as it is $\gtrsim 200$ orbits.

\subsection{Calibration of 1D Dust Surface Density}
\label{sect:dust-calibration}

In our 1D treatment of dust dynamics, the particles accumulated in a gas pressure maximum tend to collapse into unphysically narrow radial structures. This is the result of neglecting several different effects in our 1D treatment: explicit diffusion, and the stirring of dust by non-axisymmetric density waves. Both of these effects would serve to smear out the radial dust distribution. Without accounting for them, the dust distribution near a pressure maximum resembles a $\delta$-function. The width of a the peak is therefore underestimated, and the concentration is overestimated.

\begin{figure*}
\begin{center}
\includegraphics[width=0.99\textwidth,clip]{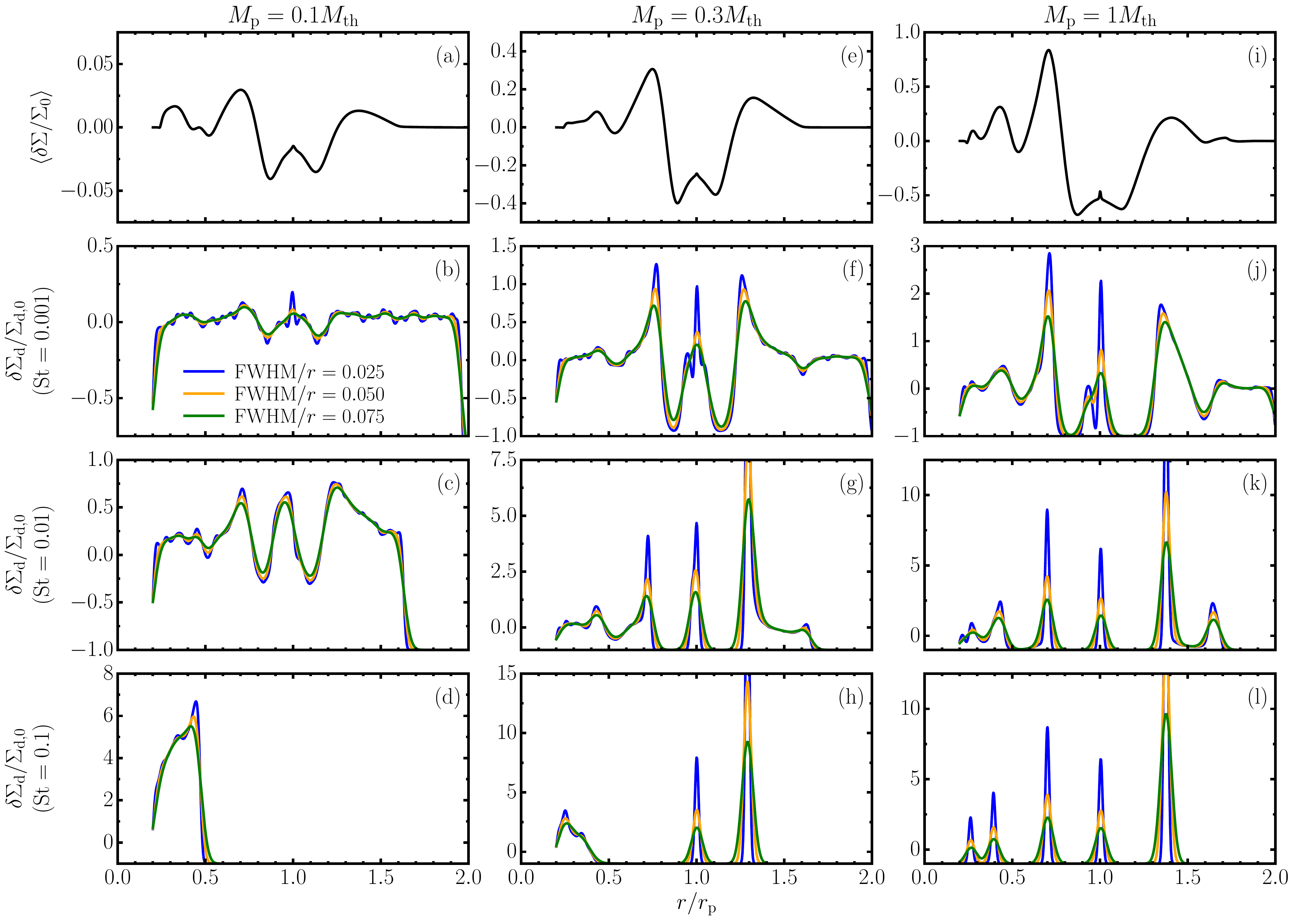}
\caption{Gas surface density perturbation (relative to the initial profile; top row) and dust surface density perturbation, for different sized dust particles (parameterized by the initial Stokes number at $r_\mathrm{p}$), for different planet masses $M_\mathrm{p}$ (in terms of the thermal mass). The dust surface density is convolved in $\ln (r)$ with a Gaussian profile with different widths, described by the FWHM (different colored curves), in order to approximate 2D effects not captured in our 1D calculation of the dust evolution.}
\label{fig:dust}
\end{center}
\end{figure*}

In order to approximate the neglected stirring effects, we artifically smooth the dust density distribution produced in our calculations. The degree of smoothing is calibrated against the dust surface density profiles presented in \citet{Zhang2020}, which come from hydrodynamical simulations with a self-consistent 2D treatment of dust dynamics. 

To perform this calibration we carried out a dedicated set of numerical simulations following the ``B17'' setup described in \citet{Zhang2020}. The parameters of these runs are aspect ratio $h_\mathrm{p} = 0.07$, temperature power law index $q = 0$ (i.e., the temperature profile is globally isothermal), surface density power law index $p = 1$, and viscosity parameter $\alpha = 5 \times 10^{-5}$. The smoothing length for the planetary potential is $0.1 H_\mathrm{p}$ (substantially smaller than  the value used in our other runs, $0.6 H_\mathrm{p}$). The radial extent of the disk is $[0.2 r_\mathrm{p}, 2 r_\mathrm{p}]$, the boundaries of the inner and outer wave damping zones are $0.24 r_\mathrm{p}$ and $1.6 r_\mathrm{p}$, and the size of the grid is $N_r \times N_\phi = 752 \times 2048$ (approximately two times smaller than that of \citealt{Zhang2020}).

We ran these simulations with planet masses $M_\mathrm{p} = 0.1 M_\mathrm{th}$, $0.3 M_\mathrm{th}$, and $1 M_\mathrm{th}$, each for $500$ orbits, and then performed 1D dust evolution calculations. The final dust surface density $\Sigma_\mathrm{d}(r)$ was convolved with a Gaussian profile in $\ln(r)$, so that the width of the Gaussian relative to the radius $r$ is constant. The convolution is parameterized by the full width at half maximum (FWHM) of the Gaussian ($\mathrm{FWHM} \approx 2.36 \sigma$). We chose $\mathrm{FWHM}/r = 0.025, 0.05,$ and $0.075$.

The profiles of the gas surface density and convolved dust surface density for three different dust sizes are shown in Fig.~\ref{fig:dust}. Here the dust size is parameterized by $\mathrm{St}_0$, the Stokes number at $r = r_\mathrm{p}$ and $t = 0$. Fig.~\ref{fig:dust} can be compared with Fig.~8 of \citet{Zhang2020} (specifically, their case with $Q = 100$, which is closest to our simulations, in which the disk self-gravity is not included). Our gas surface density profiles reproduce $\Sigma(r)$ in \citet{Zhang2020} very closely. 

We find that the dust profiles convolved with $\mathrm{FWHM}/r = 0.025$ and $\mathrm{FWHM}/r = 0.05$ produce peaks that are too narrow and whose amplitudes are too large. But convolution with $\mathrm{FWHM}/r = 0.075$ reproduces the correct widths and amplitudes of the peaks quite well. The agreement is not as good for $M_\mathrm{p} = 0.1 M_\mathrm{th}$, however, only a small subset of the simulations in this paper use such a small planet mass.

Therefore, we find that smoothing our 1D distributions with $\mathrm{FWHM}/r = 0.075$ provides an adequate approximation of 2D effects neglected in our 1D calculations. We use this value of the Gaussian FWHM for producing all of our dust maps.

\subsection{Submillimeter Emission}
\label{sect:sub-mm}

Continuum emission maps are computed from the dust surface distributions using a procedure that closely follows that of \citet{Zhang2018}. However, we simplify the process by considering only a single particle size, rather than a size distribution.

We choose an observing frequency of $\nu_\mathrm{obs} = 240$ GHz or wavelength $\lambda_\mathrm{obs} = 1.25$ mm, as in the DSHARP survey. We make use the DSHARP opacity, tabulated as a function of particle size and wavelength, from \citet{Birnstiel2018}. For $1$ mm particles and a wavelength of $1.25$ mm, the absorption opacity is $\kappa_\nu = 2.2$ cm$^2$ g$^{-1}$ (per unit dust mass). Note that $\kappa_\nu$ is the opacity that controls the emission at frequency $\nu$, and should not be confused with $\kappa_\mathrm{d}$, the frequency-averaged opacity of smaller dust grains that mediates disk cooling (see equation~(\ref{eq:opac-law})).

The optical depth is $\tau_\nu(r) = \kappa_\nu\Sigma_\mathrm{d}(r)/2$. For the Fiducial disk model, the optical depth at $t = 0$ is $\tau_\nu = 0.11~r_{50}^{-1}$. Using the optical depth $\tau_\nu(r)$ of the dust distribution at $500$ orbits, we compute the brightness temperature of the dust emission,
\be
T_\mathrm{b}(r) = [1 - \mathrm{e}^{-\tau_\nu(r)}] T(r).
\ee
To bring our emission maps to the same scale, the brightness temperature is normalized by the disk temperature at $r_\mathrm{p}$,
\be
T(r_\mathrm{p}) = 43~M_{*,1} h_{50\mathrm{au},0.1}^2 r_{\mathrm{p},50}^{-q}~\mathrm{K},
\ee
where $r_{\mathrm{p},50} = r_\mathrm{p}/(50~\mathrm{au})$ and $h_{50\mathrm{au},0.1} = h_{50\mathrm{au}}/0.1$.

In the Rayleigh-Jeans limit, the intensity of the emission is proportional to the brightness temperature, $I_\nu(r) \propto T_\mathrm{b}(r)$. The peak blackbody emission wavelength for $T(r)$ is $\lambda_\mathrm{peak}(r) = 0.067~h_{50\mathrm{au},0.1}^{-2} r_{50}^{1/2}~\mathrm{mm}$. For the Fiducial disk model, we therefore have $\lambda_\mathrm{peak} \ll \lambda_\mathrm{obs}$, even at large radii (hundreds of au), and so the Rayleigh-Jeans limit is always applicable. Only for the Cold disk model, the Rayleigh-Jeans limit becomes only marginally applicable far in the outer disk.

\bibliographystyle{apj}
\bibliography{references}

\end{document}